\documentclass[11pt,a4paper]{article}
\pdfoutput=1

\usepackage{jheppub}

\usepackage{amsmath}
\usepackage{bbm}
\usepackage{float}
\usepackage{graphicx}	
\usepackage{hyperref}
\usepackage{mathtools}
\usepackage{cancel}

\usepackage[normalem]{ulem}

\usepackage{multirow}
\usepackage{rotating}
\usepackage{subcaption}

\def\lsim{\raise0.3ex\hbox{$\;<$\kern-0.75em\raise-1.1ex
\hbox{$\sim\;$}}}
\def\gsim{\raise0.3ex\hbox{$\;>$\kern-0.75em\raise-1.1ex
\hbox{$\sim\;$}}}

\def\thetitle{ 
Symmetry in neutrino oscillation in matter with non-unitarity \\
 \vspace{- 6mm}
}
\title{\thetitle}
\hypersetup{pdftitle={\thetitle}}
\author{Hisakazu Minakata}
\affiliation{
Center for Neutrino Physics, Department of Physics, Virginia Tech, Blacksburg, Virginia 24061, USA \\
}

\emailAdd{hisakazu.minakata@gmail.com}
\date{\today}

\abstract{Recently we have developed a method called ``Symmetry Finder'' (SF) for hunting the reparametrization symmetry in the three-neutrino system in matter. Here, we apply SF to the Denton {\it et al.} (DMP) perturbation theory extended by including unitarity violation (UV), a possible low-energy manifestation of physics beyond the $\nu$SM. Implementation of UV into the SF framework yields the additional two very different constraints, which nonetheless allow remarkably consistent solutions, the eight DMP-UV symmetries. Treatment of one of the constraints, the genuine non-unitary part, leads to the key identity which entails the UV $\alpha$ parameter transformation only by rephasing, which innovates the invariance proof of the Hamiltonian. The quantum mechanical nature of the symmetry dictates the both $\nu$SM and UV variables to transform jointly, through which the response of the two sectors are related to reveal their interplay. Thus, the symmetry can serve for a tool for diagnostics, probing the interrelation between the $\nu$SM and a low-energy description of new physics. Problem of SF symmetry in vacuum is revisited to complete eight symmetries akin to DMP's. 
 } 

\newcommand{\Dmsqren}{\Delta \lambda_{ \text{ren} }}

\begin{document} 

\maketitle

\section{Introduction} 
\label{sec:introduction} 

Symmetry positions in a deepest place in our understanding of quantum mechanics and quantum field theory~\cite{Itzykson:1980rh,Coleman:1985rnk}. Recently we have studied the reparametrization symmetry in neutrino oscillations in matter~\cite{Minakata:2021dqh,Minakata:2022yvs,Minakata:2021goi}. We have constructed a machinery called ``Symmetry Finder'' (SF) for a systematic search for symmetries in the neutrino system in matter. SF has been successfully applied to the Denton~{\it et al.}~(DMP)~\cite{Denton:2016wmg} and the solar-resonance perturbation (SRP)~\cite{Martinez-Soler:2019nhb} theories to uncover the ``twin''~\cite{Minakata:2022pyr} eight reparametrization symmetries of the 1-2 state exchange type~\cite{Minakata:2021dqh,Minakata:2022yvs}, and the sixteen 1-3 state exchange symmetries~\cite{Minakata:2021goi} in the helio-perturbation theory~\cite{Minakata:2015gra}. Enrichment of the symmetries in matter environment~\cite{Wolfenstein:1977ue} is emphasized~\cite{Minakata:2022yvs,Minakata:2022pyr}.\footnote{
We remark that in this manuscript citation of the articles that are devoted to the SF symmetry discussions includes the ones which appeared after version 1 of e-Print:~2206.06474. This is to make our description of the Rep symmetry up to date, and to provide a less incomplete list of the references on the SF symmetry within the scope of the present author. } 
For our frequent usage, we denote the reparametrization symmetry discussed in these references as the ``Rep symmetry'' throughout this paper. 

Despite the above mentioned successes in symmetry hunting, we still do not know the whole picture of the Rep symmetry. In particular, nature of the Rep symmetry has been the subject of scrutiny since the SF search has been started. Invariance of the oscillation probability under a Rep symmetry transformation implies that there is another way of parametrizing the equivalent solution of the theory. It suggests a different characterization of the Rep symmetry from the ones discussed in refs.~\cite{Itzykson:1980rh,Coleman:1985rnk}. Then, the imminent questions would be: 
\begin{itemize}

\item 
What {\em is} the Rep symmetry? 
Under which raison d'~\^etre does it exist? 

\item
In which way can a symmetry merely reparametrizing the same physics be useful to understand the system? 

\end{itemize}
While it is not practical to squarely tackle these questions at this moment, we present our emphasis of quantum nature of the SF symmetry in matter in section~\ref{sec:quantum-nature}, and some of the remaining questions in section~\ref{sec:conclusion}. We would like to remind the readers that, in view of state of the art summary of our search for the Rep symmetry given in ref.~\cite{Minakata:2022yvs}, we are still in a phase of ``experimental searches'' for the Rep symmetry. 
In this paper, we will fill a remaining hole by presenting six new SF symmetries in vacuum to complement to the existing two ones~\cite{Parke:2018shx} to complete the eight 1-2 exchange symmetries which is very similar to those of DMP and SRP. 

The principal purpose of this paper is to examine the theory based on the $\nu$SM but extended by including non-unitarity. Hereafter, the ``$\nu$SM'' is a shorthand notation for the neutrino-mass-embedded Standard Model. Incorporation of non-unitarity, or unitarity violation (UV),\footnote{
We are aware that in physics literatures UV usually means ``ultraviolet''. But, in this paper UV is used as an abbreviation for ``unitarity violation'' or ``unitarity violating''.  }
is one of the promising ways to discuss physics beyond the $\nu$SM~\cite{Antusch:2006vwa,Escrihuela:2015wra,Blennow:2016jkn,Fong:2016yyh,Fong:2017gke}. It turned out that inclusion of the extra ingredients, non-unitarity and the associated UV sector, into the theory facilitates to have an insight toward answering the above utility question.  
For concreteness, we examine in this paper the DMP-UV perturbation theory formulated in ref.~\cite{Minakata:2021nii}. The DMP theory~\cite{Denton:2016wmg}, the unique ``globally valid''~\cite{Minakata:2022yvs} framework after the Agarwalla {\it et al.} perturbation theory~\cite{Agarwalla:2013tza}, and its extensions have been studied in a variety of contexts, e.g., in ref.~\cite{Parke:2019vbs,Minakata:2020oxb,Parke:2019jyu}. 
Our investigation will reveal the characteristically new two features (visit section~\ref{sec:SF-solution-UV} for more about the both items): 
\begin{itemize}

\item
All the eight $\nu$SM DMP symmetries~\cite{Minakata:2021dqh} have UV extensions, with extremely simple transformation property $\widetilde{\alpha} \rightarrow \text{Rep(X)}~\widetilde{\alpha}~\text{Rep(X)}^{\dagger}$ of the non-unitarity variable $\widetilde{\alpha}$, where Rep(X) stands for the rephasing matrix, diagonal one with entries $\pm 1 = e^{ \pm i k \pi }$ ($k=0,1$). 

\item 
The identity which will be born out from the study of transformation property of the UV sector variables has an impact on our understanding of the Rep symmetry. 

\end{itemize}

In a concurrently progressing work~\cite{Minakata:2022yvs}, the SF symmetry search has also been undertaken by using the helio-UV theory~\cite{Martinez-Soler:2018lcy}, a UV extension of the helio-perturbation theory~\cite{Minakata:2015gra}. 
Despite the parallelism between the SF treatments in these two UV extended theories we try to make our presentation in this paper self-contained. This is preferable because of the difference between the structure of symmetries, the eight 1-2 in DMP-UV and the sixteen 1-3 symmetries in the helio-UV theories. Nonetheless we try to make our description in this paper complementary to that of ref.~\cite{Minakata:2022yvs}. 

To the author's knowledge, the first discussion of the Rep symmetry in neutrino oscillation is given by Fogli {\it et al.} in ref.~\cite{Fogli:1996nn}. In broader contexts the symmetries in theories of neutrino mixing and the closely related topics had been investigated from various point of views, for which we can give only a very limited list~\cite{Gluza:2001de,deGouvea:2000pqg,Fogli:2001wi,Altarelli:2010gt,Fogli:1996pv,Minakata:2001qm,Minakata:2010zn,Coloma:2016gei,deGouvea:2008nm,Zhou:2016luk}. We hope that our approach contributes to further advance the field. 

There exists ample reason to consider theories with non-unitarity from physics point of view. It is a possible low-energy manifestation of new physics beyond the $\nu$SM at high (or possibly low) scale. The most well-known example of high-scale UV is given by the active three neutrino sector in the seesaw model of neutrino masses~\cite{Minkowski:1977sc,Yanagida:1979as,Gell-Mann:1979vob,Glashow:1979nm,Mohapatra:1979ia}. With three right-handed heavy neutrinos, the whole active-sterile $6 \times 6$ system is unitary, but low-energy description of $3 \times 3$ active neutrino subsystem is not. When the UV effects coexist with the $\nu$SM there arise many new features in the three neutrino system. Most notably, the UV effects could affect determination of the standard mixing parameters. Disentanglement of the UV effect from the $\nu$SM ones is then required, the topics being extensively discussed, e.g., in refs.~\cite{Escrihuela:2015wra,Miranda:2016wdr,Ge:2016xya,Dutta:2016vcc,Abe:2017jit,Rout:2017udo,Li:2018jgd}. The other side of the coin is the correlations between the $\nu$SM and UV parameters~\cite{Escrihuela:2015wra,Miranda:2016wdr,Abe:2017jit,Martinez-Soler:2018lcy,Martinez-Soler:2019noy}, which could enhance the detection capability of the UV effect. It would be nice if our symmetry study can shed light on the question of how the UV effect could be distinguished from the $\nu$SM one. For an incomplete list of the additional references on non-unitarity, see e.g., refs.~\cite{Fernandez-Martinez:2007iaa,Goswami:2008mi,Antusch:2009pm,Antusch:2009gn,Antusch:2014woa,Fernandez-Martinez:2016lgt,Parke:2015goa,Ellis:2020hus,Coloma:2021uhq}. 
We note that even more generic framework of adding ``non-standard interactions'' (NSI)~\cite{Wolfenstein:1977ue} is vigorously investigated as a possible description of physics beyond the $\nu$SM. See e.g. refs.~\cite{Ohlsson:2012kf,Miranda:2015dra,Farzan:2017xzy,Proceedings:2019qno} for reviews of NSI, and refs.~\cite{Davidson:2003ha,Antusch:2008tz,Biggio:2009nt,Esteban:2018ppq} for constraints on NSI. 

In section~\ref{sec:quantum-nature}, we introduce SF, discuss its quantum nature, and present a full set of the SF symmetry in vacuum which is derived in Appendix~\ref{sec:vacuum-SF}. In sections~\ref{sec:3nu-non-unitarity}, \ref{sec:DMP-UV}, and~\ref{sec:Vmatrix-method}, we set up the basics for our analysis framework of the UV extended DMP, which can be skipped for the readers familiar to the theory including the $V$ matrix computation. Section~\ref{sec:SF-DMP-UV} provides the pillar of our symmetry search, the SF formulation of the DMP-UV perturbation theory, through which the readers can move on to the solutions given in sections~\ref{sec:SF-solution-UV} and~\ref{sec:SF-solution-EV}. 
The readers who are really knowledgeable to our SF formalism~\cite{Minakata:2021dqh,Minakata:2022yvs,Minakata:2021goi} may want to start from Table~\ref{tab:DMPUV-symmetry}, and then go directly to section~\ref{sec:SF-solution-UV} which allows them to understand all essentials. 
For most of the readers, however, the author recommends to go through the whole manuscript because it is written in a self-contained way. Nature of the DMP-UV symmetries as a Hamiltonian symmetry is revealed in section~\ref{sec:hamiltonian-symmetry}. We conclude in section~\ref{sec:conclusion} with discussions of some of the remaining issues including the questions of $\nu$SM - UV inter-sector communications, how big is the symmetry, and a possible extension of the Rep symmetry to cover the double beta decay observable. 

\section{Reparametrization (Rep) symmetry: New progress in understanding and its quantum nature}
\label{sec:quantum-nature} 

A framework for systematic search for symmetry, which we call ``Symmetry Finder'' (SF), started from the one in vacuum~\cite{Parke:2018shx}. The underlying principle of SF in vacuum as well as in matter is very simple. Suppose that the flavor basis state (i.e., wave function) $\nu$ allows the two different expressions by the mass eigenstate $\bar{\nu}$ in vacuum as~\cite{Minakata:2021dqh}  
\begin{eqnarray} 
&&
\nu = U (\theta_{23}, \theta_{13}, \theta_{12}, \delta) \bar{\nu} 
= U (\theta_{23}^{\prime}, \theta_{13}^{\prime}, \theta_{12}^{\prime}, \delta^{\prime}) \bar{\nu}^{\prime}, 
\label{SF-eq-general}
\end{eqnarray}
where $U \equiv U_{\text{\tiny MNS}}$ denotes the usual unitary $\nu$SM flavor mixing matrix~\cite{Maki:1962mu}, see section~\ref{sec:3nu-non-unitarity} for its explicit expressions.  
The quantities with ``prime'' imply the transformed ones, and $\bar{\nu}^{\prime}$ may involve eigenstate exchanges and/or rephasing of the wave functions. 
If it is in matter the $U$ matrix in eq.~\eqref{SF-eq-general} must be replaced by the one, so called the $V$ matrix~\cite{Minakata:1998bf}, see section~\ref{sec:Vmatrix-method}. The mixing angles and the CP phase are often elevated to the matter-dressed variables, as it is the case of $\theta_{12}$ and $\theta_{13}$ in DMP. 
As eq.~\eqref{SF-eq-general}, or its matter version, represents the unique flavor state by the two different sets of the physical parameters, it implies a symmetry. This is nothing but the key statement of SF~\cite{Parke:2018shx,Minakata:2021dqh,Minakata:2021goi,Minakata:2022yvs}. 

\subsection{Reparametrization (Rep) symmetry in vacuum revisited} 
\label{sec:symmetry-vacuum} 

In a series of our previous and concurrently progressed works on the SF symmetries in matter reported in refs.~\cite{Minakata:2021dqh,Minakata:2022yvs,Minakata:2021goi,Minakata:2022pyr}, we have always referred only Symmetry IA- and IB-vacuum in our nomenclature as the 1-2 exchange Rep symmetries in vacuum~\cite{Parke:2018shx}. If it has left an impression that there exist only the two symmetries in vacuum, this is neither correct nor what it was meant. While the statement itself is useful as it was done in the context of heuristic discussion of the SF method, we prefer to straighten out the problem of SF symmetry in vacuum.  
In Appendix~\ref{sec:vacuum-SF} we conduct an explicit reanalysis of the SF equation to show that there exist the eight 1-2 state exchange symmetries in vacuum, which we call Symmetry X-vacuum (X=IA, IB, $\cdot \cdot \cdot $, IVB). Here, we only present the resulting Rep symmetries in Table~\ref{tab:SF-symmetry-vacuum} below.

\begin{table}[h!]
\vglue -0.2cm
\begin{center}
\caption{The eight SF symmetries in vacuum, Symmetry X-vacuum where X=IA, IB, $\cdot \cdot \cdot $, IVB. The symmetries in the upper and lower row correspond to the upper and the lower solutions in Table~\ref{tab:SF-solutions}. $s_{12}$ etc. are the simplified notations for $\sin \theta_{12}$ etc..
}
\label{tab:SF-symmetry-vacuum}
\vglue 0.2cm
\begin{tabular}{c|c}
\hline 
Symmetry & 
Vacuum mixing parameter transformation 
\\
\hline 
\hline 
IA & 
$m^2_{1} \leftrightarrow m^2_{2}$, $c_{12} ^{\prime} = - s_{12}$, $s_{12} ^{\prime} = c_{12}$ \\ 
& $m^2_{1} \leftrightarrow m^2_{2}$, $c_{12} ^{\prime} = s_{12}$, $s_{12} ^{\prime} = - c_{12}$ \\
\hline
IB & 
$m^2_{1} \leftrightarrow m^2_{2}$, $c_{12} ^{\prime} = s_{12}$, $s_{12} ^{\prime} = c_{12}$, $\delta^{\prime} = \delta + \pi$ \\
& $m^2_{1} \leftrightarrow m^2_{2}$, $c_{12} ^{\prime} = - s_{12}$, $s_{12} ^{\prime} = - c_{12}$, $\delta^{\prime} = \delta + \pi$ \\
\hline 
IIA & 
$m^2_{1} \leftrightarrow m^2_{2}$, $c_{12} ^{\prime} = s_{12}$, $s_{12} ^{\prime} = c_{12}$, $s_{23} ^{\prime} = - s_{23}$ \\
& $m^2_{1} \leftrightarrow m^2_{2}$, $c_{12} ^{\prime} = - s_{12}$, $s_{12} ^{\prime} = - c_{12}$, $s_{23} ^{\prime} = - s_{23}$ \\
\hline 
IIB & 
$m^2_{1} \leftrightarrow m^2_{2}$, $c_{12} ^{\prime} = - s_{12}$, $s_{12} ^{\prime} = c_{12}$, $s_{23} ^{\prime} = - s_{23}$, $\delta^{\prime} = \delta + \pi$ \\ 
& $m^2_{1} \leftrightarrow m^2_{2}$, $c_{12} ^{\prime} = s_{12}$, $s_{12} ^{\prime} = - c_{12}$, $s_{23} ^{\prime} = - s_{23}$, $\delta^{\prime} = \delta + \pi$ \\
\hline 
IIIA & 
$m^2_{1} \leftrightarrow m^2_{2}$, $c_{12} ^{\prime} = s_{12}$, $s_{12} ^{\prime} = c_{12}$, $s_{13} ^{\prime} = - s_{13}$ \\ 
& $m^2_{1} \leftrightarrow m^2_{2}$, $c_{12} ^{\prime} = - s_{12}$, $s_{12} ^{\prime} = - c_{12}$, $s_{13} ^{\prime} = - s_{13}$ \\
\hline 
IIIB & 
$m^2_{1} \leftrightarrow m^2_{2}$, $c_{12} ^{\prime} = - s_{12}$, $s_{12} ^{\prime} = c_{12}$, $s_{13} ^{\prime} = - s_{13}$, $\delta^{\prime} = \delta + \pi$  \\ 
& $m^2_{1} \leftrightarrow m^2_{2}$, $c_{12} ^{\prime} = s_{12}$, $s_{12} ^{\prime} = - c_{12}$, $s_{13} ^{\prime} = - s_{13}$, $\delta^{\prime} = \delta + \pi$ \\
\hline 
IVA & 
$m^2_{1} \leftrightarrow m^2_{2}$, $c_{12} ^{\prime} = - s_{12}$, $s_{12} ^{\prime} = c_{12}$, $s_{23} ^{\prime} = - s_{23}$, $s_{13} ^{\prime} = - s_{13}$ \\ 
& $m^2_{1} \leftrightarrow m^2_{2}$, $c_{12} ^{\prime} = s_{12}$, $s_{12} ^{\prime} = - c_{12}$, $s_{23} ^{\prime} = - s_{23}$, $s_{13} ^{\prime} = - s_{13}$ \\
\hline 
IVB & 
$m^2_{1} \leftrightarrow m^2_{2}$, $c_{12} ^{\prime} = s_{12}$, $s_{12} ^{\prime} = c_{12}$, $s_{23} ^{\prime} = - s_{23}$, $s_{13} ^{\prime} = - s_{13}$, $\delta^{\prime} = \delta + \pi$ \\ 
& $m^2_{1} \leftrightarrow m^2_{2}$, $c_{12} ^{\prime} = - s_{12}$, $s_{12} ^{\prime} = - c_{12}$, $s_{23} ^{\prime} = - s_{23}$, $s_{13} ^{\prime} = - s_{13}$, $\delta^{\prime} = \delta + \pi$ \\
\hline 
\end{tabular}
\end{center}
\vglue -0.4cm 
\end{table}

In fact, the eight Rep symmetries in vacuum have identical structure with those of the eight symmetries in DMP, as one recognizes by comparing Table~\ref{tab:SF-symmetry-vacuum} with Table~\ref{tab:DMPUV-symmetry} (first three columns only for $\nu$SM DMP) in section~\ref{sec:SF-DMP-UV}. We will also see that the solutions for the phases in the SF equation are identical to those of DMP in Table~\ref{tab:SF-solutions}. These features may suggest the picture of vacuum symmetry as a vacuum limit of the DMP symmetry. Nonetheless, in Appendix~\ref{sec:vacuum-SF} where we show that Symmetry X-vacuum is the Hamiltonian symmetry, we will see that the vacuum symmetry is {\em not} a simple vacuum limit of the DMP symmetry. 

If we elevate the system to the one diagonalized exactly in matter assuming the uniform density~\cite{Zaglauer:1988gz}, it implies the existence of eight symmetries, Symmetry X-ZS (X=IA, IB, $\cdot \cdot \cdot $, IVB). See ref.~\cite{Minakata:2021dqh} for the definition of the ZS symmetry including replacement of the vacuum variables by the matter-dressed ones. 


\subsection{Reparametrization symmetry as a vacuum-matter hybrid symmetry} 
\label{sec:hybrid-symmetry} 

Since our SF search for the symmetry has been initiated, nature of the Rep symmetry has been the subject of scrutiny. In this and the next section~\ref{sec:Q-dynamical} we shall try to give a partial answer to the question: ``what is the Rep symmetry?''. 

Let us start from a typical, and a very frequently raised question: 
Isn't it merely a part of the reparametrization of neutrino mixing matrix? If this is essentially the case, the probability is trivially invariant, and we learn nothing new but what is well known. However, a glance over a few examples of the DMP Rep symmetries in Table~\ref{tab:DMPUV-symmetry}, again with the nomenclature of ref.~\cite{Minakata:2021dqh}, reveals incongruity with the ``mere reparametrization'' picture: 
\begin{eqnarray} 
\text{IIA-DMP:} 
&& 
\theta_{23} \rightarrow - \theta_{23}, 
\hspace{4mm} 
\theta_{12} \rightarrow- \theta_{12}, 
\hspace{4mm} 
\lambda_{1} \leftrightarrow \lambda_{2}, 
\hspace{4mm}
c_{\psi} \rightarrow \pm s_{\psi}, 
\hspace{4mm}
s_{\psi} \rightarrow \pm c_{\psi}, 
\nonumber \\
\text{IIIA-DMP:} 
&&
\theta_{13} \rightarrow - \theta_{13},
\hspace{4mm}
\theta_{12} \rightarrow- \theta_{12}, 
\hspace{4mm}
\lambda_{1} \leftrightarrow \lambda_{2}, 
\hspace{4mm}
\phi \rightarrow - \phi, 
\hspace{4mm}
c_{\psi} \rightarrow \pm s_{\psi}, 
\hspace{4mm}
s_{\psi} \rightarrow \pm c_{\psi}, 
\nonumber \\
\text{IVA-DMP:} 
&& 
\theta_{23} \rightarrow - \theta_{23},
\hspace{4mm}
\theta_{13} \rightarrow- \theta_{13}, 
\hspace{4mm}
\lambda_{1} \leftrightarrow \lambda_{2}, 
\hspace{4mm}
\phi \rightarrow - \phi, 
\hspace{4mm}
c_{\psi} \rightarrow \mp s_{\psi}, 
\hspace{4mm}
s_{\psi} \rightarrow \pm c_{\psi},
\nonumber \\
\label{Symmetry-II-IVA}
\end{eqnarray}
In eq.~\eqref{Symmetry-II-IVA}, $\theta_{ij}$ ($ij = 1,2,3$) is the mixing angle in vacuum, and $\phi$ and $\psi$ denote, respectively, $\theta_{13}$ and $\theta_{12}$ in matter~\cite{Denton:2016wmg}, often denoted as the ``matter-dressed'' $\theta_{13}$ and $\theta_{12}$ in this paper. $s_{\phi} \equiv \sin \phi$ and $s_{\psi} \equiv \sin \psi$, etc. 
These matter-dressed variables are the ``dynamical variables'' that arise when  (dominant part of) the Hamiltonian is diagonalized. As emphasized in ref.~\cite{Minakata:2021dqh}, a dynamical symmetry is the symmetry that has no hint in the original Hamiltonian of the system, but the one which indeed arises after the system is solved. 

Therefore, our Rep symmetry can be characterized as the vacuum- and dynamical-variables hybrid symmetry. The crucial feature is that the vacuum variables' transformations are determined in tight correlation with the matter variables' ones, and the correlated choice of the vacuum and matter variables pair is different in each symmetry. Unless symmetry respects the correlation, it never leaves the probability invariant. 
Thus, our characterization of the Rep symmetry is based on the fact that the three-generation neutrino system exists with the lepton flavor mixing~\cite{Maki:1962mu} and in matter environment~\cite{Wolfenstein:1977ue} whose effect is treated as a background. Within this treatment neutrino evolution is enriched by the resonant or other enhanced flavor transformations~\cite{Wolfenstein:1977ue,Mikheyev:1985zog,Barger:1980tf,Smirnov:2016xzf}, which entails a variety of the Rep symmetries, as stressed in ref.~\cite{Minakata:2022pyr}. 

\subsection{Quantum nature of the Rep symmetry}
\label{sec:Q-dynamical} 

In all the eight DMP symmetries~\cite{Minakata:2021dqh}, see Table~\ref{tab:DMPUV-symmetry}, the eigenvalue exchange $\lambda_{1} \leftrightarrow \lambda_{2}$ is involved where $\lambda_{i} / 2E$ denote the eigenvalues of the Hamiltonian of the eigenstate $\nu_{i}$ ($i=1,2,3$)~\cite{Denton:2016wmg}.\footnote{
In this paper, the state $\nu_{1}$ denotes the one with the largest $\nu_{e}$ component. The state $\nu_{2}$ is the one that is separated from the state $\nu_{1}$ by the mass squared difference $m^2_{2} - m^2_{1} \equiv \Delta m^2_{ \text{solar} } \simeq 7.5 \times 10^{-5}$ eV$^2 > 0$. } 
But none of them involves the vacuum mass exchange $m^2_{1} \leftrightarrow m^2_{2}$. It may be puzzling that the 1-2 state exchange is not accompanied by the mass exchange, but this is what the solution of the SF equation dictates. We should note here that the vacuum mass exchange is indeed involved in all the SF symmetries in vacuum as tabulated in Table~\ref{tab:SF-symmetry-vacuum}. 

We argue below that absence of the vacuum mass exchange in the DMP symmetries comes out naturally given the fact that our theory is based on quantum mechanics, and the state space on which the Rep symmetry generator acts has no $m_{i}$ dependence. 
We span the physical state space by the eigenbases of the diagonalized Hamiltonian $\bar{H}^{(0)}$ in eq.~\eqref{barH-0th-1st}. Since we work under the approximation that neutrinos undergo no inelastic scattering, no absorption, and do not decay, the state space we construct is a direct product of the one that has definite neutrino energy $E$. Then, our system has only the three component states, the neutrino flavor states $[\nu_{e}, \nu_{\mu}, \nu_{\tau}]^T$, or the mass eigenstates $[\nu_{1}, \nu_{2}, \nu_{3}]^T$, which is analogous to the spin 1 system. In this analogy our treatment of the three neutrino system looks similar to the quantum theory of angular momentum. The flavor mixing matrix $U$ which acts on the three-component state is nothing but the matrix representation of the ``rotation'' operator which is composed of the Gell-Mann matrix~\cite{Itzykson:1980rh}, instead of the Pauli matrix for a spin $\frac{1}{2}$, two-component system. 

But, in our case, unlike the case of angular momentum, the ``rotation angles'' themselves undergo the transformation. That is, the vacuum and matter-dressed mixing angles transform at the same time when the state exchange is performed. In such quantum mechanical formulation, the mixing matrix $U \equiv U_{\text{\tiny MNS}}$~\cite{Maki:1962mu} and the $V$ matrix~\cite{Minakata:1998bf}, see eq.~\eqref{V-matrix-def} for the definition, must be elevated to the operator-valued $\bf{U}$ and $\bf{V}$ matrices, respectively. Or in other word, the $U$ and $V$ matrices are the matrix representations of $\bf{U}$ and $\bf{V}$, respectively, which act on the unique ground state. To generate the mixing angles' transformations, such as $c_{\psi} \rightarrow \mp s_{\psi}$ and $s_{\psi} \rightarrow \pm c_{\psi}$, there must be the generators to produce the Rep symmetry transformations by acting onto the operator $\bf{V}$ matrix. 

The $V$ matrix, the matrix representation of $\bf{V}$, will be calculated to first order in section \ref{sec:Vmatrix-method}. Then, we confirm that the $V$ matrix has no $m_{i}$ dependence, and hence the masses  do not transform under the SF symmetry transformations in DMP. In vacuum, on the other hand, $m^2_{i}$ can transform because the state space is spanned by the eigenstate of the vacuum Hamiltonian $H_{ \text{vac} } \equiv U~\text{diag} (m^2_{1}/2E, m^2_{2}/2E, m^2_{3}/2E )~U^{\dagger}$ with the eigenvalues to be elevated to the operators. Due to the difference in the state space structure it is unlikely that the vacuum SF symmetry can be regarded as the vacuum limit of the DMP symmetry. This point will be revisited in Appendix~\ref{sec:vacuum-SF}. 

\subsection{Utility of the Rep symmetry?}
\label{sec:utility} 

As raised in section~\ref{sec:introduction}, the symmetry we discuss here implies reparameterization of the same physics, which may sound that no new physical insights will be gained by studying it. Despite that the suspect may be legitimate, we would like to show in this paper that this naive view is not supported by the results obtained by our treatment. In a nutshell and in simple terms, it appears that the Rep symmetry sheds light on interplay between $\nu$SM and UV sectors of the theory. 

In the seesaw model of neutrino masses~\cite{Minkowski:1977sc,Yanagida:1979as,Gell-Mann:1979vob,Glashow:1979nm,Mohapatra:1979ia}, the connection between the low-mass active neutrinos and high-scale new physics with the heavy sterile states is explicitly visible. But the connection is model dependent. In the usual treatment of non-unitarity~\cite{Antusch:2006vwa} the high-energy sector is integrated out, which makes the interplay between the $\nu$SM and UV sectors of the theory model-independent. But, then, the inter-connection between the two sectors ceases to be visible, or it becomes rather obscured. Is it still possible to probe the interplay between the $\nu$SM and UV sectors of the theory in a model-independent manner? 

Our approach to answering this question is through the symmetry. As we discussed above the Rep symmetry resides deep in quantum mechanics, see also ref.~\cite{Minakata:2020oxb}, which of course governs universally both the $\nu$SM and UV sectors of the theory. Then, it is conceivable that the quantum-mechanics-rooted Rep symmetry reveals how the $\nu$SM and UV sectors of the theory are mutually correlated, or communicate to each other. This is nothing but what we will see in sections~\ref{sec:SF-solution-UV} and~\ref{sec:SF-solution-EV}. We even bravely speculate on a picture of inter-sector communications between the two sectors through the phases in section~\ref{sec:conclusion}. 

\section{Three active-neutrino system with non-unitary flavor mixing matrix}
\label{sec:3nu-non-unitarity} 

This section is to define the system of three active neutrinos propagating under the influence of non-unitary flavor mixing matrix and the matter potential. In our formalism the three active neutrino evolution in matter in the presence of non-unitary flavor mixing is based on the Schr\"odinger equation in the vacuum mass eigenstate basis~\cite{Blennow:2016jkn,Fong:2017gke}, the ``check basis'', 
\begin{eqnarray}
i \frac{d}{dx} \check{\nu} = 
\frac{1}{2E} 
\left\{  
\left[
\begin{array}{ccc}
0 & 0 & 0 \\
0 & \Delta m^2_{21} & 0 \\
0 & 0 & \Delta m^2_{31} \\
\end{array}
\right] + 
N^{\dagger} \left[
\begin{array}{ccc}
a - b & 0 & 0 \\
0 & -b & 0 \\
0 & 0 & -b \\
\end{array}
\right] N 
\right\} 
\check{\nu} 
\equiv 
\check{H} 
\check{\nu}. 
\label{evolution-check-basis} 
\end{eqnarray}
In eq.~\eqref{evolution-check-basis}, which also defines the check-basis Hamiltonian $\check{H}$, $N$ denotes the $3 \times 3$ non-unitary flavor mixing matrix which relates the flavor neutrino states to the vacuum mass eigenstates as 
\begin{eqnarray}
\nu_{\alpha} = N_{\alpha i} \check{\nu}_{i}. 
\label{N-def}
\end{eqnarray}
In eq.~\eqref{N-def} and hereafter, the subscript Greek indices $\alpha$, $\beta$, or $\gamma$ run over $e, \mu, \tau$, and the Latin indices $i$, $j$ run over the mass eigenstate indices $1,2,$ and $3$. $E$ is neutrino energy and $\Delta m^2_{ji} \equiv m^2_{j} - m^2_{i}$. The usual phase redefinition of neutrino wave function is done to leave only the mass squared differences. Notice, however, that doing this phase redefinition or not (see eq.~\eqref{H-LHS}) does not affect our symmetry discussion in this paper. 

The functions $a(x)$ and $b(x)$ in eq.~(\ref{evolution-check-basis}) denote the
Wolfenstein matter potentials~\cite{Wolfenstein:1977ue} due to charged current (CC) and neutral current (NC) reactions, respectively, 
\begin{eqnarray} 
a(x) &=&  
2 \sqrt{2} G_F N_e E \approx 1.52 \times 10^{-4} \left( \frac{Y_e \rho}{\rm g\,cm^{-3}} \right) \left( \frac{E}{\rm GeV} \right) {\rm eV}^2, 
\nonumber \\
b(x) &=& \sqrt{2} G_F N_n E = \frac{1}{2} \left( \frac{N_n}{N_e} \right) a, 
\label{matt-potential}
\end{eqnarray}
where $G_F$ is the Fermi constant. $N_e$ and $N_n$ are the electron and neutron number densities in matter. $\rho$ and $Y_e$ denote, respectively, the matter density and number of electrons per nucleon in matter. These quantities except for $G_F$ are, in principle, position dependent. 

We use so called the $\alpha$ parametrization~\cite{Escrihuela:2015wra} for the non-unitary flavor mixing matrix $N$, 
\begin{eqnarray} 
N 
&=& 
\left( \bf{1} - \alpha \right) U 
= 
\left\{ 
\bf{1} - 
\left[ 
\begin{array}{ccc}
\alpha_{ee} & 0 & 0 \\
\alpha_{\mu e} & \alpha_{\mu \mu}  & 0 \\
\alpha_{\tau e}  & \alpha_{\tau \mu} & \alpha_{\tau \tau} \\
\end{array}
\right] 
\right\}
U, 
\label{alpha-matrix-def}
\end{eqnarray}
where $U \equiv U_{\text{\tiny MNS}}$ denotes the usual unitary $\nu$SM mixing matrix~\cite{Maki:1962mu}. The $\alpha$ parametrization originates in refs.~\cite{Schechter:1980gr,Okubo:1961jc}. In fact, the definitions of $U$ and $\alpha$ matrices are convention dependent~\cite{Martinez-Soler:2018lcy}. In eq.~\eqref{alpha-matrix-def} we have used the Particle Data Group (PDG) convention~\cite{ParticleDataGroup:2022pth}. 

For convenience of our discussion, following ref.~\cite{Minakata:2021dqh}, we use the solar (SOL) convention for the $U$ and $\alpha$ matrices. They are defined by the phase redefinition of those of the PDG convention: 
\begin{eqnarray} 
U_{\text{\tiny SOL}} 
&=&
\left[
\begin{array}{ccc}
1 & 0 &  0  \\
0 & e^{ - i \delta} & 0 \\
0 & 0 & e^{ - i \delta} \\ 
\end{array}
\right] 
U_{\text{\tiny PDG}} 
\left[
\begin{array}{ccc}
1 & 0 &  0  \\
0 & e^{ i \delta} & 0 \\
0 & 0 & e^{ i \delta} \\
\end{array}
\right] 
=
\left[
\begin{array}{ccc}
1 & 0 &  0  \\
0 & c_{23} & s_{23} \\
0 & - s_{23} & c_{23} \\
\end{array}
\right] 
\left[
\begin{array}{ccc}
c_{13}  & 0 & s_{13} \\
0 & 1 & 0 \\
- s_{13} & 0 & c_{13} \\
\end{array}
\right] 
\left[
\begin{array}{ccc}
c_{12} & s_{12} e^{ i \delta}  &  0  \\
- s_{12} e^{- i \delta} & c_{12} & 0 \\
0 & 0 & 1 \\
\end{array}
\right] 
\nonumber \\
&\equiv& 
U_{23} (\theta_{23}) U_{13} (\theta_{13}) U_{12} (\theta_{12}, \delta), 
\nonumber \\
\alpha_{\text{\tiny SOL}} 
&=& 
\left[
\begin{array}{ccc}
1 & 0 &  0  \\
0 & e^{ - i \delta} & 0 \\
0 & 0 & e^{ - i \delta} \\
\end{array}
\right] 
\alpha 
\left[
\begin{array}{ccc}
1 & 0 &  0  \\
0 & e^{ i \delta} & 0 \\
0 & 0 & e^{ i \delta} \\
\end{array}
\right] 
\equiv
\left[ 
\begin{array}{ccc}
\widetilde{\alpha}_{ee} & 0 & 0 \\
\widetilde{\alpha}_{\mu e} & \widetilde{\alpha}_{\mu \mu}  & 0 \\
\widetilde{\alpha}_{\tau e}  & \widetilde{\alpha}_{\tau \mu} & \widetilde{\alpha}_{\tau \tau} \\
\end{array}
\right], 
\label{U-alpha-SOL-def} 
\end{eqnarray}
with the obvious notations $s_{ij} \equiv \sin \theta_{ij}$ etc., $\delta$ for the lepton analogue of the quark CP violating Kobayashi-Maskawa (KM) phase $\delta$~\cite{Kobayashi:1973fv}, and $\widetilde{\alpha}_{\beta \gamma}$ for the SOL convention $\alpha$ parameters which we use.\footnote{
The other useful form of the $U$ matrix is the one with the atmospheric (ATM) convention used e.g., in refs.~\cite{Minakata:2015gra,Denton:2016wmg,Martinez-Soler:2018lcy} in which the phase factor $e^{ \pm i \delta}$ is attached to $s_{23}$ as opposed to $s_{12}$ in the SOL convention.  }
The second line in eq.~\eqref{U-alpha-SOL-def} defines the rotation matrices in the 2-3, 1-3, and 1-2 spaces in order.

\section{DMP-UV perturbation theory in a nutshell} 
\label{sec:DMP-UV} 

To present our formulation of the DMP-UV perturbation theory, an extension of DMP to include non-unitarity, we rely on the basic formulation given in a previous paper~\cite{Minakata:2021nii}, which will be referred to for details. On the other hand, we have to go beyond the treatment of ref.~\cite{Minakata:2021nii} for the following three reasons: 
(1) While we remain in the first-order treatment in most part of this paper, we have to keep the second-order UV term in the Hamiltonian when we attempt to give an all-order proof of the symmetries in section~\ref{sec:hamiltonian-symmetry}. 
(2) In the SF framework we need the $V$ matrix method~\cite{Minakata:1998bf}, which will be formulated for our theory in section~\ref{sec:Vmatrix-method}. 
(3) We take the eigenvalue-renormalized basis for the eigenstate in matter, which is slightly different from the one in ref.~\cite{Minakata:2021nii}. 

The DMP-UV perturbation theory has two kind of the expansion parameters, $\epsilon$ and the UV $\widetilde{\alpha}$ parameters. $\epsilon$ is defined as 
\begin{eqnarray} 
&&
\epsilon \equiv \frac{ \Delta m^2_{21} }{ \Delta m^2_{ \text{ren} } }, 
\hspace{10mm}
\Delta m^2_{ \text{ren} } \equiv \Delta m^2_{31} - s^2_{12} \Delta m^2_{21},
\label{epsilon-Dm2-ren-def}
\end{eqnarray}
where $\Delta m^2_{ \text{ren} }$ is the ``renormalized'' atmospheric $\Delta m^2$ used in ref.~\cite{Minakata:2015gra}. 
The same quantity is known as the effective $\Delta m^2_{ \text{ee} }$ in the $\nu_{e} \rightarrow \nu_{e}$ channel in vacuum~\cite{Nunokawa:2005nx}. While we prefer usage of $\Delta m^2_{ \text{ren} }$ in the context of the present paper, the question of which symbol should be appropriate to use here is under debate~\cite{Minakata:2015gra}. 
The other expansion parameters are the $\widetilde{\alpha}_{\beta \gamma}$ parameters defined in eq.~\eqref{U-alpha-SOL-def} which represent the UV effects. 

We start from the tilde-basis Hamiltonian 
\begin{eqnarray} 
&&
\widetilde{H} 
\equiv 
( U_{13} U_{12} ) \check{H} ( U_{13} U_{12} )^{\dagger} 
= 
\widetilde{H} _{ \nu\text{SM}} 
+ \widetilde{H}_\text{ UV }^{(1)} 
+ \widetilde{H}_\text{ UV }^{(2)}, 
\label{tilde-H}
\end{eqnarray}
where  
\begin{eqnarray} 
&&
\hspace{-5mm}
\widetilde{H} _{ \nu\text{SM}} 
= 
\frac{ \Delta m^2_{ \text{ren} } }{ 2E } 
\left\{ 
\left[
\begin{array}{ccc}
\frac{ a(x) }{ \Delta m^2_{ \text{ren} }} + s^2_{13} + \epsilon s^2_{12} & 
0 & c_{13} s_{13} \\
0 & \epsilon c^2_{12} & 0 \\
c_{13} s_{13}  & 0 & c^2_{13} + \epsilon s^2_{12} \\
\end{array}
\right] 
+ 
\epsilon c_{12} s_{12} 
\left[
\begin{array}{ccc}
0 & 
c_{13} e^{ i \delta} & 
0 \\
c_{13} e^{ - i \delta} & 
0 & 
- s_{13} e^{ - i \delta} \\
0 & 
- s_{13} e^{ i \delta} & 
0 \\
\end{array}
\right] 
\right\}. 
\nonumber \\
\label{tilde-H-SM}
\end{eqnarray}
The UV part has the first and second order terms in the $\widetilde{\alpha}$ parameters 
\begin{eqnarray} 
&&
\widetilde{H}_\text{ UV }^{(1)} = 
\frac{b}{2E} 
U_{23}^{\dagger} A U_{23}, 
\nonumber \\
&& 
\widetilde{H}_\text{ UV }^{(2)} = 
- \frac{b}{2E} 
U_{23}^{\dagger} A^{(2)} U_{23},
\label{tilde-H-UV}
\end{eqnarray}
where 
\begin{eqnarray}
A &\equiv&
\left[ 
\begin{array}{ccc}
2 \widetilde{\alpha}_{ee} \left( 1 - \frac{ a (x) }{ b (x) } \right) & 
\widetilde{\alpha}_{\mu e}^* & 
\widetilde{\alpha}_{\tau e}^* \\
\widetilde{\alpha}_{\mu e} & 
2 \widetilde{\alpha}_{\mu \mu}  & 
\widetilde{\alpha}_{\tau \mu}^* \\
\widetilde{\alpha}_{\tau e}  & 
\widetilde{\alpha}_{\tau \mu} & 
2 \widetilde{\alpha}_{\tau \tau} \\
\end{array}
\right], 
\nonumber \\
A^{(2)} 
&\equiv& 
\left[
\begin{array}{ccc}
\widetilde{\alpha}_{ee}^2 \left( 1 - \frac{ a (x) }{ b (x) } \right) 
+ |\widetilde{\alpha}_{\mu e}|^2 + |\widetilde{\alpha}_{\tau e}|^2 & 
\widetilde{\alpha}_{\mu e}^* \widetilde{\alpha}_{\mu \mu} 
+ \widetilde{\alpha}_{\tau e}^* \widetilde{\alpha}_{\tau \mu} & 
\widetilde{\alpha}_{\tau e}^* \widetilde{\alpha}_{\tau \tau} \\
\widetilde{\alpha}_{\mu e} \widetilde{\alpha}_{\mu \mu} 
+ \widetilde{\alpha}_{\tau e} \widetilde{\alpha}_{\tau \mu}^* & 
\widetilde{\alpha}_{\mu \mu}^2 + |\widetilde{\alpha}_{\tau \mu}|^2 & 
\widetilde{\alpha}_{\tau \mu}^* \widetilde{\alpha}_{\tau \tau} \\
\widetilde{\alpha}_{\tau e} \widetilde{\alpha}_{\tau \tau} & 
\widetilde{\alpha}_{\tau \mu} \widetilde{\alpha}_{\tau \tau} & 
\widetilde{\alpha}_{\tau \tau}^2 \\
\end{array}
\right]. 
\label{A-A2-def}
\end{eqnarray}
In $\widetilde{H} _{ \nu\text{SM}}$ in eq.~\eqref{tilde-H-SM}, the rephasing to remove the NC potential is understood~\cite{Minakata:2021nii}. For a consistent nomenclature $A$ must carry the superscript as $A^{(1)}$, but for simplicity of the expressions we omit it throughout this paper. 
In what follows we keep omitting the superscript $^{(1)}$ for many of the quantities in first order in the $\widetilde{\alpha}$ parameters, because our treatment will be free from the second order terms apart from section~\ref{sec:hamiltonian-symmetry}. 

\subsection{The renormalized eigenvalue (bar) basis} 
\label{sec:bar-basis} 

We use the two successive rotations in the 1-3 and 1-2 spaces with the angles $\phi$ and $\psi$, the matter-dressed $\theta_{13}$ and $\theta_{12}$, respectively, to diagonalize $\widetilde{H} _{ \nu\text{SM}}$, see ref.~\cite{Denton:2016wmg}. The diagonalized basis is denoted as the ``bar basis'' with the Hamiltonian
\begin{eqnarray} 
\bar{H} 
&=&
U_{12}^{\dagger} (\psi, \delta) U^\dagger_{13}(\phi) 
\widetilde{H} 
U_{13}(\phi) U_{12} (\psi, \delta) 
= 
\bar{H}^{(0)}_{ \nu\text{SM}} + \bar{H}^{(1)}_{ \nu\text{SM}} 
+ \bar{H}^{(1)}_{ \nu\text{UV}} + \bar{H}^{(2)}_{ \nu\text{UV}}.
\label{barH-4terms}
\end{eqnarray}
The notations are such that $U_{12} (\psi, \delta)$, for example, implies $U_{12} (\theta_{12}, \delta)$ with replacement of $\theta_{12}$ by $\psi$. The $\nu$SM part of $\bar{H}$ takes the form in the SOL convention as 
\begin{eqnarray} 
&& 
\bar{H}^{(0)}_{ \nu\text{SM}} + \bar{H}^{(1)}_{ \nu\text{SM}} 
= 
\frac{1}{2E} 
\left[
\begin{array}{ccc}
\lambda_{1}^{ \nu\text{SM} } & 0 & 0 \\
0 & \lambda_{2}^{ \nu\text{SM} } & 0 \\
0 & 0 & \lambda_{3}^{ \nu\text{SM} } \\
\end{array}
\right] 
+ 
\epsilon c_{12} s_{12} 
\sin ( \phi - \theta_{13} ) \frac{ \Delta m^2_{ \text{ren} } }{2E} 
\left[
\begin{array}{ccc}
0 & 0 & - s_\psi \\
0 & 0 & c_\psi e^{ - i \delta} \\
- s_\psi & c_\psi e^{ i \delta} & 0 
\end{array}
\right],
\nonumber \\
\label{barH-nuSM} 
\end{eqnarray}
where the first and second terms in eq.~\eqref{barH-nuSM} defines $\bar{H}^{(0)}_{ \nu\text{SM}}$ and $\bar{H}^{(1)}_{ \nu\text{SM}}$, respectively. $\lambda_{i}^{ \nu\text{SM} } / 2E$ denote the zeroth-order eigenvalues of the $\nu$SM Hamiltonian given in ref.~\cite{Denton:2016wmg}. The UV part of the bar-basis Hamiltonian reads 
\begin{eqnarray} 
&& 
\bar{H}^{(1)}_{ \text{UV}} 
= \frac{b}{2E} G, 
\hspace{10mm} 
\bar{H}^{(2)}_{ \text{UV}} 
= - \frac{b}{2E} G^{(2)}. 
\label{barH-UV} 
\end{eqnarray}
where the $G$ matrices are the 2-3, 1-3 and 1-2 rotated $A$ matrices in eq.~\eqref{A-A2-def}: 
\begin{eqnarray} 
G &=& 
U_{12} (\psi, \delta)^{\dagger} U^\dagger_{13}(\phi) 
U_{23}^{\dagger} (\theta_{23}) 
A 
U_{23} (\theta_{23}) 
U_{13}(\phi) U_{12} (\psi, \delta), 
\nonumber \\
G^{(2)} &=& 
U_{12} (\psi, \delta)^{\dagger} U^\dagger_{13}(\phi) 
U_{23}^{\dagger} (\theta_{23}) 
A^{(2)} 
U_{23} (\theta_{23}) 
U_{13}(\phi) U_{12} (\psi, \delta). 
\label{G-G2-def}
\end{eqnarray} 

To discuss the Rep symmetry, for simplicity, we take the renormalized bar basis in which the diagonal elements of the $G$ matrices are absorbed into the eigenvalues of $\bar{H}^{(0)}$: 
\begin{eqnarray} 
&&
\lambda_{i} 
= \lambda_{i}^{ \nu\text{SM} } + b G_{ii} - b G^{(2)}_{ii} ~~~~~ (i=1,2,3). 
\label{ren-eigenvalues}
\end{eqnarray}
The off-diagonal parts of the $G$ and $G^{(2)}$  matrices are left as they are in $\bar{H}^{(1)}_{ \text{UV}}$ and $\bar{H}^{(2)}_{ \text{UV}}$, and are treated as perturbation. Notice that since the matter-dressed mixing angles are defined through the process of  diagonalizing the $\nu$SM part, $\bar{H}^{(0)}_{ \nu\text{SM}} + \bar{H}^{(1)}_{ \nu\text{SM}}$, the shifted eigenvalues do not affect them. 

From now on until section~\ref{sec:hamiltonian-symmetry}, we concentrate on the first order term in the Hamiltonian and parametrize the $G$ matrix as 
\begin{eqnarray} 
&&
G = 
D H D^{\dagger}
=
\left[
\begin{array}{ccc}
H_{11} & e^{ i \delta} H_{12} & H_{13} \\
e^{ - i \delta} H_{21} & H_{22} & e^{ - i \delta} H_{23} \\
H_{31} & e^{ i \delta} H_{32} & H_{33} \\
\end{array}
\right], 
\label{Hmatrix-def}  
\end{eqnarray}
where $D \equiv \text{diag} (e^{ i \delta}, 1, e^{ i \delta} )$, with the diagonal elements untouched, $G_{ii} = H_{ii}$. The explicit expressions of $H_{ij}$ are given in Appendix~\ref{sec:H-elements}. 
We have introduced the $H$ matrix for two reasons: 
(1) Overall $e^{ \pm i \delta }$ factors in the elements $G_{12}$ and $G_{23}$ are taken out to prevent proliferation of hidden $\delta$ in the expressions of the probability. 
(2) It makes the $\delta$ dependences of the $\nu$SM and UV parts of the SF equation more coherent. 

Then, by using the renormalized eigenvalues, the bar-basis Hamiltonian is given to first order in the DMP-UV expansion as 
\begin{eqnarray} 
\bar{H} 
&=&
\frac{1}{2E} 
\left[
\begin{array}{ccc}
\lambda_{1} & 0 & 0 \\
0 & \lambda_{2} & 0 \\
0 & 0 & \lambda_{3} \\
\end{array}
\right] 
+
\epsilon c_{12} s_{12} 
\sin ( \phi - \theta_{13} ) \frac{\Delta m^2_{ \text{ren} }}{2E} 
\left[
\begin{array}{ccc}
0 & 0 & - s_\psi \\
0 & 0 & c_\psi e^{ - i \delta} \\
- s_\psi & c_\psi e^{ i \delta} & 0 
\end{array}
\right] 
\nonumber \\
&+&
\frac{b}{2E} 
\left[
\begin{array}{ccc}
0 & e^{ i \delta} H_{12} & H_{13} \\
e^{ - i \delta} H_{21} & 0 & e^{ - i \delta} H_{23} \\
H_{31} & e^{ i \delta} H_{32} & 0 \\
\end{array}
\right]. 
\label{barH-0th-1st}
\end{eqnarray}
Hereafter we denote the first term of eq.~\eqref{barH-0th-1st} as $\bar{H}^{(0)}$, and the second and third as $\bar{H}^{(1)}$ which are treated as perturbation. 
After identifying the unperturbed and perturbed Hamiltonian, there is a standard route to compute the $S$ matrix and the oscillation probability. This task is carried out explicitly in the $\nu_{\mu} \rightarrow \nu_{e}$ channel in ref.~\cite{Minakata:2021nii} to first order in the DMP-UV expansion. 
As in ref.~\cite{Minakata:2021nii} we use the uniform matter density approximation until section~\ref{sec:hamiltonian-symmetry}.

\section{$V$ matrix method} 
\label{sec:Vmatrix-method} 

Symmetry Finder (SF)~\cite{Minakata:2021dqh,Minakata:2021goi,Minakata:2022yvs} is formulated by using the $V$ matrix formalism~\cite{Minakata:1998bf}, and for this reason we construct it for the DMP-UV perturbation theory. If we have the expression of the flavor eigenstate in terms of the mass eigenstate basis in matter as 
\begin{eqnarray}
\nu = V \bar{\nu}, 
\label{V-matrix-def}
\end{eqnarray}
the oscillation probability can readily be calculated as 
\begin{eqnarray}
P(\nu_\beta \rightarrow \nu_\alpha) 
&=&
\delta_{\alpha \beta}
-4\sum_{j>i} \mbox{Re}[V_{\alpha i} V_{\beta i}^* V_{\alpha j}^* V_{\beta j}]
\sin^2 \frac{( \lambda_j - \lambda_i) x}{4E} 
\nonumber\\
&-&
2\sum_{j>i} \mbox{Im}[V_{\alpha i} V_{\beta i}^* V_{\alpha j}^* V_{\beta j}]
\sin \frac{( \lambda_j - \lambda_i) x}{2E}.
\label{probability-matter}
\end{eqnarray}
The $V$ matrix method has been utilized for this purpose, e.g., in refs.~\cite{Minakata:2015gra,Denton:2016wmg}. 

\subsection{$V$ matrix at the zeroth and first orders}
\label{sec:Vmatrix-0th-1st} 

However, since we deal with the theory with non-unitary mixing matrix a proper care is needed to calculate the $V$ matrix. We recall that the flavor eigenstate $\nu_{\alpha}$ is related with the vacuum mass eigenstate (denoted as the check basis) as  
$\nu_{\alpha} 
= N_{\alpha i} \check{\nu}_{i} 
= \left\{ ( 1 - \widetilde{\alpha} ) U \right\}_{\alpha i} \check{\nu}_{i}$, see eqs.~\eqref{N-def} and \eqref{alpha-matrix-def}. 
From eqs.~\eqref{tilde-H} and \eqref{barH-4terms}, the relation between the bar and the check bases are given by 
$\bar{H} = U_{12}^{\dagger} (\psi, \delta) U^\dagger_{13}(\phi) U_{13} U_{12} \check{H} U_{12}^{\dagger} U_{13}^{\dagger} U_{13}(\phi) U_{12} (\psi, \delta)$. Here, $U_{13}$ and $U_{12}$ without arguments implies the rotation matrices in vacuum, see eq.~\eqref{U-alpha-SOL-def}. Therefore, the relation between the check-basis and bar-basis states is 
\begin{eqnarray}
\check{\nu} 
= 
U_{12}^{\dagger} U_{13}^{\dagger} U_{13} (\phi) U_{12} (\psi, \delta) 
\bar{\nu}.
\label{bar-check-state}
\end{eqnarray}
Then, the flavor state is connected to the bar-basis state as 
\begin{eqnarray} 
&& 
\nu 
= 
( 1 - \widetilde{\alpha} ) U \check{\nu} 
=
( 1 - \widetilde{\alpha} ) 
U_{23} (\theta_{23}) U_{13}(\phi) U_{12} (\psi, \delta) \bar{\nu}, 
\label{flavor-check}
\end{eqnarray}
which gives the zeroth-order and the first-order ``genuine UV part'' of the $V$ matrix given, respectively, as 
\begin{eqnarray} 
&& 
V^{(0)} = U_{23} (\theta_{23}) U_{13} (\phi) U_{12} (\psi, \delta),
\nonumber \\
&&
V^{(1)}_{ \text{UV} } 
= 
- \widetilde{\alpha} U_{23} (\theta_{23}) U_{13}(\phi) U_{12} (\psi, \delta) 
= 
- \widetilde{\alpha} V^{(0)}. 
\label{Vmatrix-0th-1st-UV}
\end{eqnarray} 
Hereafter we simply use the subscript ``UV'' to indicate the genuine UV part. 

When the perturbation $\bar{H}^{(1)}$ is switched on, the other type of the first-order term in the $V$ matrix, the EV (UV-induced but unitary evolution) part $V^{(1)}_{ \text{EV} }$, is generated. For more about these concepts, the ``genuine UV'' and ``unitary evolution'' parts, see ref.~\cite{Martinez-Soler:2018lcy}. They contribute to the probability in the two quite different forms as we will see in section~\ref{sec:probability-Vmatrix}, the feature also observed in the $S$ matrix computation in ref.~\cite{Minakata:2021nii}. 
To compute the EV part of the first-order correction to the $V$ matrix we take out the prefactor $( 1 - \widetilde{\alpha} )$ because the effect of this $\widetilde{\alpha}$ produces the second order effect. Then, the method for obtaining the first-order correction for the $V$ matrix is identical to the one used in computing the first-order correction to the wave function in quantum mechanics. When we write $\bar{\nu}_{i} = \bar{\nu}_{i}^{(0)} + \bar{\nu}_{i}^{(1)}$, the first-order correction can be calculated as 
\begin{eqnarray}
\bar{\nu}_{i}^{(1)} 
= 
\sum_{j\neq i} \frac{ 2E \bar{H}^{(1)}_{ij} }{ \lambda_i - \lambda_j } 
\bar{\nu}_{j}^{(0)} 
\equiv 
W_{ij} \bar{\nu}_{j}^{(0)}, 
\label{nu-bar-first-order}
\end{eqnarray}
where we have defined the $W$ matrix. Since $\bar{H}$ in eq.~\eqref{barH-0th-1st} has the two first order terms, one from the $\nu$SM and the other due to the EV effect, the $W$ matrix can be written in the form of addition of these terms, $W = W_{ \nu\text{SM} } + W_{ \text{EV} }$, and their explicit forms are given in eq.~\eqref{W-DMP-UV}. 
A short note for clarification of eq.~\eqref{nu-bar-first-order} is in Appendix~\ref{sec:QM-1st-order}. 

Then, the energy eigenstate calculated to first order can be written, using the $W$ matrix defined in eq.~\eqref{nu-bar-first-order}, as 
\begin{eqnarray} 
\bar{\nu} 
&=& 
\bar{\nu}^{(0)} + \bar{\nu}^{(1)} 
= ( 1 + W ) \bar{\nu}^{(0)} 
\nonumber \\
&=&
\left( 1 + W_{ \nu\text{SM} } + W_{ \text{EV} } \right) 
\left[ U_{23} (\theta_{23}) U_{13} (\phi) U_{12} (\psi, \delta) \right]^{\dagger} \nu. 
\label{1st-order-bar-flavor}
\end{eqnarray}
where we have used the zeroth order relation $\nu = V^{(0)} \bar{\nu}^{(0)} = [U_{23} (\theta_{23}) U_{13} (\phi) U_{12} (\psi, \delta)] \bar{\nu}^{(0)}$. Inverting this relation and adding the contribution from the genuine unitary part in eq.~\eqref{Vmatrix-0th-1st-UV}, we obtain the expression of the flavor state by the mass eigenstate in matter to first order as 
\begin{eqnarray} 
\nu 
&=& 
\left[ 
- \widetilde{\alpha} V^{(0)} 
+ V^{(0)} \left( 1 + W_{ \nu\text{SM} } + W_{ \text{EV} } \right) ^{\dagger} 
\right] 
\bar{\nu} 
\equiv V \bar{\nu},
\label{flavor-bar-state-final} 
\end{eqnarray}
which defines the $V$ matrix to first order in the DMP-UV expansion, and in eq.~\eqref{flavor-bar-state-final} we have used the expression of the zeroth-order $V$ matrix in eq.~\eqref{Vmatrix-0th-1st-UV}. 

\subsection{$V$ matrix to first order: Summary}
\label{sec:Vmatrix-summary} 

For the convenience of formulating the SF equation we rewrite the expression of the flavor state~\eqref{flavor-bar-state-final} with use of the $V$ matrix in the following form: 
\begin{eqnarray} 
&& 
\left[
\begin{array}{c}
\nu_{e} \\
\nu_{\mu} \\
\nu_{\tau} \\
\end{array}
\right] 
=  
U_{23} (\theta_{23}) U_{13} (\phi) U_{12} (\psi, \delta)
\biggl\{
1 + \mathcal{W}_{ \nu\text{SM} } ^{(1)} 
+ \mathcal{W}_{ \text{EV} } ^{(1)} 
- \mathcal{Z} _{ \text{UV} } ^{(1)} 
\biggr\} 
\left[
\begin{array}{c}
\nu_{1} \\
\nu_{2} \\
\nu_{3} \\
\end{array}
\right], 
\label{Vmatrix-for-SF}
\end{eqnarray}
where 
\begin{eqnarray} 
\mathcal{W}_{ \nu\text{SM} } ^{(1)} 
( \theta_{23}, \delta, \phi, \psi; \lambda_{1}, \lambda_{2} ) 
\equiv W_{ \nu\text{SM} }^{\dagger} 
&=&
\epsilon c_{12} s_{12} \sin ( \phi - \theta_{13} )
\left[
\begin{array}{ccc}
0 & 0 & 
- s_\psi \frac{ \Dmsqren }{ \lambda_3 - \lambda_1 } \\
0 & 0 & 
c_\psi e^{ - i \delta} \frac{ \Dmsqren }{ \lambda_3 - \lambda_2 } \\
s_\psi \frac{ \Dmsqren }{ \lambda_3 - \lambda_1 } & 
- c_\psi e^{ i \delta} \frac{ \Dmsqren }{ \lambda_3 - \lambda_2 } & 0 \\
\end{array}
\right], 
\nonumber \\
\mathcal{W}_{ \text{EV} } ^{(1)} 
( \theta_{23}, \delta, \phi, \psi; \lambda_{1}, \lambda_{2} ) 
\equiv W_{ \text{EV} }^{\dagger} 
&=&
\left[
\begin{array}{ccc}
0 & e^{ i \delta} H_{12} \frac{ b }{ \lambda_2 - \lambda_1 }  & 
H_{13} \frac{ b }{ \lambda_3 - \lambda_1 } \\
- e^{ - i \delta} H_{21} \frac{ b }{ \lambda_2 - \lambda_1 } & 0 & 
H_{23} e^{ - i \delta} \frac{ b }{ \lambda_3 - \lambda_2 } \\
- H_{31} \frac{ b }{ \lambda_3 - \lambda_1 } & 
- H_{32} e^{ i \delta} \frac{ b }{ \lambda_3 - \lambda_2 } & 0 \\
\end{array}
\right]. 
\label{W-DMP-UV} 
\end{eqnarray}
That is, we label the mass eigenstate basis, the bar-basis state, $\nu_{i}$ ($i=1,2,3$) to make the state label more explicit in our discussion of the Rep symmetry which involves the state exchange. $W_{ \nu\text{SM} }$ and $W_{ \text{EV} }$ are calculated by using eq.~\eqref{nu-bar-first-order}. Notice that the $H$ matrix is hermitian. 

In eq.~\eqref{Vmatrix-for-SF}, we give $V^{(0)}$ in eq.~\eqref{flavor-bar-state-final} the explicit form as in eq.~\eqref{Vmatrix-0th-1st-UV}, and move it to the front position in the right-hand side of eq.~\eqref{Vmatrix-for-SF}. 
For this purpose we have introduced $\mathcal{Z} _{ \text{UV} } ^{(1)}$ with the definition 
\begin{eqnarray} 
&&
\widetilde{\alpha} V^{(0)} 
= 
V^{(0)} \mathcal{Z} _{ \text{UV} } ^{(1)}, 
\label{Z-def}
\end{eqnarray}
which can be readily solved for $\mathcal{Z} _{ \text{UV} } ^{(1)}$ as 
\begin{eqnarray}
&&
\mathcal{Z} _{ \text{UV} } ^{(1)} ( \theta_{23}, \delta, \phi, \psi; \widetilde{\alpha}_{\beta \gamma} ) 
= 
\left( V^{(0)} \right)^{\dagger} \widetilde{\alpha} V^{(0)}. 
\label{Z-sol}
\end{eqnarray} 

\subsection{Computing the oscillation probability using the $V$ matrix method}
\label{sec:probability-Vmatrix} 

Having obtained the $V$ matrix as in eq.~\eqref{flavor-bar-state-final}, it is straightforward to compute the oscillation probability by using the formula~\eqref{probability-matter}. We restrict ourselves into the zeroth- and first-order terms of the oscillation probability. In the UV extensions of the perturbative formulations of neutrino oscillation in matter using the $S$ matrix method~\cite{Martinez-Soler:2018lcy,Martinez-Soler:2019noy,Minakata:2021nii}, the probability can be written to first order in the DMP-UV expansion as 
\begin{eqnarray} 
P(\nu_\beta \rightarrow \nu_\alpha) 
= 
P(\nu_\beta \rightarrow \nu_\alpha)^{(0)}_{ \nu\text{SM} } 
+ P(\nu_\beta \rightarrow \nu_\alpha)^{(1)} _{ \nu\text{SM} } 
+ P(\nu_\beta \rightarrow \nu_\alpha)^{(1)} _{ \text{EV} } 
+ P(\nu_\beta \rightarrow \nu_\alpha)^{(1)} _{ \text{UV} }, 
\nonumber \\
\label{probability-0th-1st}
\end{eqnarray}
where the first two terms denote the standard $\nu$SM contributions. $P(\nu_\beta \rightarrow \nu_\alpha)^{(1)} _{ \text{EV} }$ is the contribution from the EV part, the UV-driven but describing the unitary evolution effect. 
$P(\nu_\beta \rightarrow \nu_\alpha)^{(1)} _{ \text{UV} }$ is the genuine non-unitary contribution which violates unitarity at the $S$ matrix and the probability levels~\cite{Martinez-Soler:2018lcy}. 

The first two terms, $P(\nu_\beta \rightarrow \nu_\alpha)^{(0)}_{ \nu\text{SM} }$ and $P(\nu_\beta \rightarrow \nu_\alpha)^{(1)} _{ \nu\text{SM} }$, are fully calculated in ref.~\cite{Denton:2016wmg}. See also arXiv v3 of ref.~\cite{Minakata:2020oxb} for the less abstract expressions in all the relevant channels. Therefore, we just concentrate into the UV related parts, $P(\nu_\beta \rightarrow \nu_\alpha)^{(1)} _{ \text{EV} }$ and $P(\nu_\beta \rightarrow \nu_\alpha)^{(1)} _{ \text{UV} }$, in this paper. In our $V$ matrix formulation, $P(\nu_\beta \rightarrow \nu_\alpha)^{(1)} _{ \text{UV} }$ comes from $V_{ \text{UV} }$ in eq.~\eqref{Vmatrix-0th-1st-UV}, and $P(\nu_\beta \rightarrow \nu_\alpha)^{(1)} _{ \text{EV} }$ from $V^{(0)} W_{ \text{EV} }^{\dagger}$ term of eq.~\eqref{flavor-bar-state-final}.

The computed results of $P(\nu_\mu \rightarrow \nu_e)^{(1)} _{ \text{EV} }$ and $P(\nu_\mu \rightarrow \nu_e)^{(1)} _{ \text{UV} }$ are given in Appendices~\ref{sec:EV-part} and \ref{sec:UV-part}, respectively. One can readily see that the genuine non-unitary part $P(\nu_\mu \rightarrow \nu_e)^{(1)}_{ \text{UV} }$ is identical with the corresponding formula in ref.~\cite{Minakata:2021nii}. For the EV part, $P(\nu_\mu \rightarrow \nu_e)^{(1)}_{ \text{EV} }$ agrees with the one in ref.~\cite{Minakata:2021nii} apart from that our expression in Appendix~\ref{sec:UV-probabilities} misses the $(b x) / 2E$ term, eq.~(4.9) in ref.~\cite{Minakata:2021nii}. But, there is no problem because the $(b x) / 2E$ term shows up if we expand the renormalized eigenvalues \eqref{ren-eigenvalues} by $b G_{ii} = b H_{ii}$, as will be shown in eq.~\eqref{bx-correction} in Appendix~\ref{sec:bx-term}. 

\section{Symmetry Finder for the DMP-UV perturbation theory} 
\label{sec:SF-DMP-UV} 

\subsection{Symmetry Finder (SF) equation}
\label{sec:SFeq} 

We embody SF in eq.~\eqref{SF-eq-general} and the associated machinery, the SF equation, in the DMP-UV perturbation theory. We define another state physically equivalent to the one in eq.~\eqref{Vmatrix-for-SF}: 
\begin{eqnarray} 
&& 
\hspace{-6mm}
F \left[
\begin{array}{c}
\nu_{e} \\
\nu_{\mu} \\
\nu_{\tau} \\
\end{array}
\right] 
= 
F 
U_{23} (\theta_{23}) U_{13} (\phi) U_{12} (\psi, \delta)
R^{\dagger} R 
\biggl\{
1 + \mathcal{W}_{ \nu\text{SM} } ^{(1)} 
+ \mathcal{W}_{ \text{EV} } ^{(1)} 
- \mathcal{Z} _{ \text{UV} } ^{(1)} 
\biggr\} 
R^{\dagger} R 
\left[
\begin{array}{c}
\nu_{1} \\
\nu_{2} \\
\nu_{3} \\
\end{array}
\right], ~~
\label{SFeq-ansatz-DMPUV}
\end{eqnarray}
with use of the flavor state rephasing matrix $F$ and the generalized 1-2 state exchange matrix $R$ parametrized as\footnote{
In ref.~\cite{Minakata:2021dqh}, we have used the notation $G$ for the $R$ matrix in eq.~\eqref{F-R-def}. We use here the symbol $R$ not to confuse it to the $G$ matrix in eq.~\eqref{G-G2-def}. }
\begin{eqnarray} 
&& 
F \equiv 
\left[
\begin{array}{ccc}
e^{ i \tau } & 0 & 0 \\
0 & e^{ i \sigma } & 0 \\
0 & 0 & 1 \\
\end{array}
\right],
\hspace{8mm}
R \equiv 
\left[
\begin{array}{ccc}
0 & - e^{ i ( \delta + \alpha) } & 0 \\
e^{ - i ( \delta + \beta) } & 0 & 0 \\
0 & 0 & 1 \\
\end{array}
\right], 
\label{F-R-def}
\end{eqnarray}
respectively, where $\tau$, $\sigma$, $\alpha$, and $\beta$ denote the arbitrary phases. Since the rephasing of the states does not change its physical content, the states defined by eqs.~\eqref{Vmatrix-for-SF} and \eqref{SFeq-ansatz-DMPUV} are physically equivalent to each other. 

Now, we introduce the SF equation, the DMP version of eq.~\eqref{SF-eq-general}. If the flavor state~\eqref{SFeq-ansatz-DMPUV} can be written in a form of the same state but with the transformed parameters, it implies existence of symmetry. The concrete form of the statement reads: 
\begin{eqnarray} 
&&
\left[
\begin{array}{ccc}
e^{ i \tau } & 0 & 0 \\
0 & e^{ i \sigma } & 0 \\
0 & 0 & 1 \\
\end{array}
\right]
\left[
\begin{array}{c}
\nu_{e} \\
\nu_{\mu} \\
\nu_{\tau} \\
\end{array}
\right] 
= 
\left[
\begin{array}{ccc}
1 & 0 &  0  \\
0 & c_{23} & s_{23} e^{ i \sigma } \\
0 & - s_{23} e^{ - i \sigma } & c_{23} \\
\end{array}
\right] 
\left[
\begin{array}{ccc}
c_{\phi} & 0 & s_{\phi} e^{ i \tau } \\
0 & 1 & 0 \\
- s_{\phi} e^{ - i \tau } & 0 & c_{\phi} \\
\end{array}
\right] 
\nonumber \\
&\times&
F U_{12} (\psi, \delta) R^{\dagger} R 
\biggl\{
1 + \mathcal{W}_{ \nu\text{SM} }^{(1)} (\Phi; \lambda_{1}, \lambda_{2})
+ \mathcal{W}_{ \text{EV} }^{(1)} (\Phi, \widetilde{\alpha}; \lambda_{1}, \lambda_{2})
- \mathcal{Z} _{ \text{UV} }^{(1)} (\Phi, \widetilde{\alpha})
\biggr\} 
R^{\dagger} R 
\left[
\begin{array}{c}
\nu_{1} \\
\nu_{2} \\
\nu_{3} \\
\end{array}
\right] 
\nonumber \\
&=& 
\left[
\begin{array}{ccc}
1 & 0 &  0  \\
0 & c_{23}^{\prime} & s_{23}^{\prime} \\
0 & - s_{23}^{\prime} & c_{23}^{\prime} \\
\end{array}
\right] 
\left[
\begin{array}{ccc}
c_{\phi}^{\prime} & 0 & s_{\phi}^{\prime} \\
0 & 1 & 0 \\
- s_{\phi}^{\prime} & 0 & c_{\phi}^{\prime} \\
\end{array}
\right] 
U_{12} ( \psi^{\prime}, \delta + \xi) 
\nonumber \\
&\times&
\biggl\{
1 + \mathcal{W}_{ \nu\text{SM} }^{(1)} (\Phi^{\prime}; \lambda_{2}, \lambda_{1})
+ \mathcal{W}_{ \text{EV} }^{(1)} (\Phi^{\prime}, \widetilde{\alpha}^{\prime}; \lambda_{2}, \lambda_{1})
- \mathcal{Z} _{ \text{UV} }^{(1)} (\Phi^{\prime}, \widetilde{\alpha}^{\prime})
\biggr\} 
\left[
\begin{array}{c}
- e^{ i ( \delta + \alpha) } \nu_{2} \\
e^{ - i ( \delta + \beta) } \nu_{1} \\
\nu_{3} \\
\end{array}
\right]. 
\label{SFeq-DMPUV} 
\end{eqnarray}
That is, if we find a solution of the SF equation~\eqref{SFeq-DMPUV} we identify a Rep symmetry in the DMP-UV theory. In eq.~\eqref{SFeq-DMPUV}, $\Phi$ denotes the collective representation of all the parameters involved, $\theta_{23}$, $\phi$, $\psi$, $\delta$, $\theta_{12}$, and $\theta_{13}$, and $\Phi^{\prime}$ their transformed ones. $\widetilde{\alpha}^{\prime}$ is for the transformed $\widetilde{\alpha}$, the collective notation for the $\widetilde{\alpha}_{\beta \gamma}$ parameters, and $\delta^{\prime} = \delta + \xi$. The SF equation is valid to first order in the DMP-UV perturbation theory. 

Clarifying remarks are in order on the structure of the SF equation and the relationship between the DMP and DMP-UV theories: 
(1) Due to the perturbative formulation of the SF equation it can be decomposed into the zeroth and the first order parts, which will be denoted as the first and second conditions, respectively. The former contains only the $\nu$SM variables. 
(2)  The three first-order entities in the second condition, 
$\mathcal{W}_{ \nu\text{SM} }^{(1)} (\Phi^{\prime}; \lambda_{1}, \lambda_{2})$, 
$\mathcal{W}_{ \text{EV} }^{(1)} (\Phi, \widetilde{\alpha}; \lambda_{1}, \lambda_{2})$, and $\mathcal{Z} _{ \text{UV} }^{(1)} (\Phi, \widetilde{\alpha})$, are independent from each other due to the differences in their variable dependences. One can confirm this property by the explicit treatment of the second condition on the UV parts in sections~\ref{sec:2nd-condition-EV} and~\ref{sec:2nd-condition-UV}, in addition to the $\nu$SM part which is worked out in ref.~\cite{Minakata:2021dqh}. Therefore, the second condition decomposes into the three independent equations. See eq.~\eqref{2nd-eq-DMPUV}. 

\subsection{The first condition}
\label{sec:1st-condition}

By eliminating all the first-order terms in the SF equation~\eqref{SFeq-DMPUV},
we obtain the first condition. We look for the solution under the ansatz $s_{23} e^{ i \sigma } = s_{23}^{\prime}$ and $s_{\phi} e^{ i \tau } = s_{\phi}^{\prime}$. Apparently we have no other choice within the present SF formalism. The ansatz implies that the possible values of $\tau$ and $\sigma$ are restricted to integer multiples of $\pi$. 

Then the first condition takes the form 
\begin{eqnarray} 
&&
F U_{12} (\psi, \delta) R^{\dagger} = U_{12} ( \psi^{\prime}, \delta + \xi), 
\label{1st-eq-DMPUV}
\end{eqnarray}
where we recall that 
\begin{eqnarray} 
&&
U_{12} (\psi, \delta) 
= 
\left[
\begin{array}{ccc}
c_{\psi} & s_{\psi} e^{ i \delta}  &  0  \\
- s_{\psi} e^{- i \delta} & c_{\psi} & 0 \\
0 & 0 & 1 \\
\end{array}
\right]. 
\label{U12-def}
\end{eqnarray}
It is shown~\cite{Minakata:2021dqh} that the DMP first condition, which is identical to ours, can be reduced to 
\begin{eqnarray} 
&&
c_{\psi^{\prime}} 
= - s_{\psi} e^{ - i ( \alpha - \tau ) } 
= - s_{\psi} e^{ i ( \beta + \sigma ) }, 
\hspace{8mm}
s_{\psi^{\prime}} 
= c_{\psi} e^{ i ( \beta + \tau - \xi ) } 
= c_{\psi} e^{ - i ( \alpha - \sigma - \xi ) }.
\label{1st-condition}
\end{eqnarray}
First of all, the first condition implies that up to the phase factor, $c_{\psi}$ transforms to $s_{\psi}$ and vice versa. This is consistent with the property of the symmetry we are looking for, which involves the 1-2 state exchange.
We note that under the above restriction of $\tau$ and $\sigma$ being integer multiples of $\pi$, eq.~\eqref{1st-condition} implies that all the rest of the phase parameters, $\xi$, $\alpha$, and $\beta$, must also be integer multiples of $\pi$. All the solutions of the first condition are obtained in ref.~\cite{Minakata:2021dqh}, and they are summarized in Table~\ref{tab:SF-solutions}. 
The universal nature of the first three columns of Table~\ref{tab:SF-solutions} over the DMP, SRP, and the helio-perturbation theories is pointed out in ref.~\cite{Minakata:2022yvs}.

\vspace{3mm}
\begin{table}[h!]
\vglue -0.2cm
\begin{center}
\caption{All the solutions of the first condition~\eqref{1st-condition} for Symmetry X where X = IA, IB, $\cdot \cdot \cdot $, IVB, and the rephasing matrix Ref(X)~\eqref{Rep-II-III-IV} are tabulated. 
The labels ``upper'' and ``lower'' imply the upper and lower sign in the corresponding rows in Table~\ref{tab:DMPUV-symmetry}. }
\label{tab:SF-solutions}
\vglue 0.2cm
\begin{tabular}{c|c|c|c}
\hline 
Symmetry & 
$\tau, \sigma, \xi$ & 
$\alpha, \beta$ & 
Rep(X) 
\\
\hline 
\hline 
IA & 
$\tau = \sigma = 0$, $\xi = 0$ & 
$\alpha = \beta = 0$ (upper) & diag(1,1,1)  \\ 
& &
$\alpha = \pi, \beta = - \pi$ (lower) & \\
\hline
IB & 
$\tau = \sigma = 0$, $\xi = \pi$ & 
$\alpha = \pi, \beta = - \pi$ (upper) & same as IA \\
& & $\alpha = \beta = 0$ (lower) & \\
\hline 
IIA & 
$\tau = 0, \sigma = - \pi$, $\xi = 0$ & 
$\alpha = \pi, \beta = 0$ (upper) & diag(1,-1,1) \\
& & $\alpha = 0, \beta = \pi$ (lower) & \\
\hline 
IIB & 
$\tau = 0, \sigma = - \pi$, $\xi = \pi$ & 
$\alpha = 0, \beta = \pi$ (upper) & same as IIA \\ 
& & $\alpha = \pi, \beta = 0$ (lower) & \\
\hline 
IIIA & 
$\tau = \pi, \sigma = 0$, $\xi = 0$ & 
$\alpha = 0, \beta = \pi$ (upper) & diag(-1,1,1) \\ 
& & $\alpha = \pi, \beta = 0$ (lower) & \\
\hline 
IIIB & 
$\tau = \pi, \sigma = 0$, $\xi = \pi$ & 
$\alpha = \pi, \beta = 0$ (upper) & same as IIIA \\ 
 & & 
$\alpha = 0, \beta = \pi$ (lower) &  \\
\hline 
IVA & 
$\tau = \sigma = \pi$, $\xi = 0$ & 
$\alpha = \pi, \beta = - \pi$ (upper) & diag(-1,-1,1) \\ 
& &
$\alpha = \beta = 0$ (lower) & \\
\hline 
IVB & 
$\tau = \sigma = \pi$, $\xi = \pi$ & 
$\alpha = \beta = 0$ (upper) & same as IVA \\ 
& &
$\alpha = \pi, \beta = - \pi$ (lower) & \\
\hline 
\end{tabular}
\end{center}
\vglue -0.4cm 
\end{table}
\begin{table}[h!]
\vglue 0.1cm
\begin{center}
\caption{All the reparametrization symmetries of the 1-2 state exchange type in the DMP-UV perturbation theory are summarized by giving the transformation properties of the $\nu$SM and UV $\widetilde{\alpha}$ variables. In the text we refer them as ``Symmetry IA-DMP-UV'', etc. }
\label{tab:DMPUV-symmetry}
\vglue 0.2cm
\begin{tabular}{c|c|c|c}
\hline 
Symmetry & 
Vacuum parameter transf. & 
Matter parameter transf. & 
UV parameter transf. 
\\
\hline 
\hline 
IA & 
none & 
$\lambda_{1} \leftrightarrow \lambda_{2}$, 
$c_{\psi} \rightarrow \mp s_{\psi}$, 
$s_{\psi} \rightarrow \pm c_{\psi}$ &
none 
\\
\hline 
IB & 
$\theta_{12} \rightarrow - \theta_{12}$, 
$\delta \rightarrow \delta + \pi$. & 
$\lambda_{1} \leftrightarrow \lambda_{2}$, 
$c_{\psi} \rightarrow \pm s_{\psi}$, 
$s_{\psi} \rightarrow \pm c_{\psi}$ & 
none
\\
\hline
IIA & 
$\theta_{23} \rightarrow - \theta_{23}$, 
$\theta_{12} \rightarrow - \theta_{12}$. & 
$\lambda_{1} \leftrightarrow \lambda_{2}$, 
$c_{\psi} \rightarrow \pm s_{\psi}$, 
$s_{\psi} \rightarrow \pm c_{\psi}$ & 
$\widetilde{\alpha}_{\mu e} \rightarrow - \widetilde{\alpha}_{\mu e}$, 
$\widetilde{\alpha}_{\tau \mu} \rightarrow - \widetilde{\alpha}_{\tau \mu}$ 
\\
\hline 
IIB & 
$\theta_{23} \rightarrow - \theta_{23}$, 
$\delta \rightarrow \delta + \pi$. & 
$\lambda_{1} \leftrightarrow \lambda_{2}$, 
$c_{\psi} \rightarrow \mp s_{\psi}$, 
$s_{\psi} \rightarrow \pm c_{\psi}$ &
same as IIA
\\
\hline 
IIIA & 
$\theta_{13} \rightarrow - \theta_{13}$, 
$\theta_{12} \rightarrow - \theta_{12}$. & 
$\lambda_{1} \leftrightarrow \lambda_{2}$, 
$\phi \rightarrow - \phi$ & 
$\widetilde{\alpha}_{\mu e} \rightarrow - \widetilde{\alpha}_{\mu e}$, 
$\widetilde{\alpha}_{\tau e} \rightarrow - \widetilde{\alpha}_{\tau e}$
\\ 
 & & 
$c_{\psi} \rightarrow \pm s_{\psi}$, 
$s_{\psi} \rightarrow \pm c_{\psi}$ & 
\\
\hline 
IIIB & 
$\theta_{13} \rightarrow - \theta_{13}$, 
$\delta \rightarrow \delta + \pi$. & 
$\lambda_{1} \leftrightarrow \lambda_{2}$, 
$\phi \rightarrow - \phi$ & 
same as IIIA
\\ 
 & & 
$c_{\psi} \rightarrow \mp s_{\psi}$, 
$s_{\psi} \rightarrow \pm c_{\psi}$ &
\\
\hline 
IVA & 
$\theta_{23} \rightarrow - \theta_{23}$, 
$\theta_{13} \rightarrow - \theta_{13}$. & 
$\lambda_{1} \leftrightarrow \lambda_{2}$, 
$\phi \rightarrow - \phi$ & 
$\widetilde{\alpha}_{\tau e} \rightarrow - \widetilde{\alpha}_{\tau e}$, 
$\widetilde{\alpha}_{\tau \mu} \rightarrow - \widetilde{\alpha}_{\tau \mu}$ 
\\ 
 & & 
$c_{\psi} \rightarrow \mp s_{\psi}$, 
$s_{\psi} \rightarrow \pm c_{\psi}$ &
\\
\hline 
IVB & 
$\theta_{23} \rightarrow - \theta_{23}$, 
$\theta_{13} \rightarrow - \theta_{13}$, & 
$\lambda_{1} \leftrightarrow \lambda_{2}$, 
$\phi \rightarrow - \phi$ &
same as IVA
\\ 
 &
$\theta_{12} \rightarrow - \theta_{12}$, $\delta \rightarrow \delta + \pi$. 
 &
$c_{\psi} \rightarrow \pm s_{\psi}$, $s_{\psi} \rightarrow \pm c_{\psi}$ &
\\
\hline 
\end{tabular}
\end{center}
\vglue -0.4cm 
\end{table}

\subsection{The second condition}
\label{sec:2nd-condition}

The first-order terms in the SF equation~\eqref{SFeq-DMPUV} constitute the second condition which can be decomposed into the $\nu$SM, EV, and the UV parts,
\begin{eqnarray} 
&&
R \mathcal{W}_{ \nu\text{SM} }^{(1)} ( \theta_{12}, \theta_{13}, \delta, \phi, \psi; \lambda_{1}, \lambda_{2} ) 
R^{\dagger} 
= 
\mathcal{W}_{ \nu\text{SM} }^{(1)} ( \theta_{12}^{\prime}, \theta_{13}^{\prime}, \delta + \xi, \phi^{\prime}, \psi^{\prime}; \lambda_{2}, \lambda_{1} ), 
\nonumber \\
&& 
R \mathcal{W}_{ \text{EV} }^{(1)} ( \theta_{23}, \delta, \phi, \psi, \widetilde{\alpha}_{\beta \gamma}; \lambda_{1}, \lambda_{2} ) 
R^{\dagger} 
= 
\mathcal{W}_{ \text{EV} }^{(1)} ( \theta_{23}^{\prime}, \delta + \xi, \phi^{\prime}, \psi^{\prime}, \widetilde{\alpha}_{\beta \gamma}^{\prime}; \lambda_{2}, \lambda_{1} ), 
\nonumber \\
&& 
R \mathcal{Z}_{ \text{UV} }^{(1)} (\theta_{23}, \delta, \phi, \psi, \widetilde{\alpha}_{\beta \gamma})
R^{\dagger} 
= 
\mathcal{Z}_{ \text{UV} }^{(1)} (\theta_{23}^{\prime}, \delta + \xi, \phi^{\prime}, \psi^{\prime}, \widetilde{\alpha}_{\beta \gamma}^{\prime}).
\label{2nd-eq-DMPUV} 
\end{eqnarray}
The decomposability of the second condition implies, together with the common first condition in the both DMP and DMP-UV theories, that the symmetries of the DMP-UV theory cannot be larger than the eight symmetries of the $\nu$SM DMP. The question is whether all of them survive in the UV extension. 

Given the $c_{\psi} \leftrightarrow s_{\psi}$ exchange property up to sign, the rest of the job in the $\nu$SM part is to determine the sign as well as to determine whether $\theta_{ij}$ ($ij = 12, 13, 23$) flips sign or not. This is done by solving the second condition. It produces the eight DMP symmetries, IA, IB, $\cdot \cdot \cdot $, IVB, where the type A (B) means that no $\delta$ is involved ($\delta$ is involved) in the symmetry transformations, as worked out in ref.~\cite{Minakata:2021dqh}. 
See Table~\ref{tab:DMPUV-symmetry}. 
With the knowledge of the eight $\nu$SM symmetries we examine the second conditions on $\mathcal{W}_{ \text{EV} }^{(1)}$ and $\mathcal{Z}_{ \text{UV} }^{(1)}$ to know if the consistent solutions exist. In the rest of this section, we reduct the EV and UV second conditions a little further to make them ready to solve. It will be followed by the solutions of the $\mathcal{Z}_{ \text{UV} }^{(1)}$ and $\mathcal{W}_{ \text{EV} }^{(1)}$ equations in the next two sections~\ref{sec:SF-solution-UV} and \ref{sec:SF-solution-EV},  
respectively. 

\subsection{The second condition on the unitary evolution part}
\label{sec:2nd-condition-EV}

The second condition~\eqref{2nd-eq-DMPUV} with the explicit form of $\mathcal{W}_{ \text{EV} }^{(1)}$ in eq.~\eqref{W-DMP-UV} can be written, using the $H$ matrix defined in eq.~\eqref{Hmatrix-def}, as 
\begin{eqnarray} 
&&
\hspace{-8mm}
\left[
\begin{array}{ccc}
0 & 
e^{ i ( \delta + \alpha + \beta ) } H_{21} \frac{ b }{ \lambda_2 - \lambda_1 } & 
- e^{ i \alpha } H_{23} \frac{ b }{ \lambda_3 - \lambda_2 } \\
- e^{ - i ( \delta + \alpha + \beta ) } H_{12} \frac{ b }{ \lambda_2 - \lambda_1 } & 
0 & e^{ - i ( \delta + \beta) } H_{13} \frac{ b }{ \lambda_3 - \lambda_1 }  \\
e^{ - i \alpha } H_{32} \frac{ b }{ \lambda_3 - \lambda_2 } & 
- e^{ i ( \delta + \beta) } H_{31} \frac{ b }{ \lambda_3 - \lambda_1 } & 0 \\
\end{array}
\right] 
\nonumber \\
&& 
\hspace{10mm}
=
\left[
\begin{array}{ccc}
0 & - e^{ i ( \delta + \xi ) } H_{12}^{\prime} \frac{ b }{ \lambda_2 - \lambda_1 }  & 
H_{13}^{\prime} \frac{ b }{ \lambda_3 - \lambda_2 } \\
e^{ - i ( \delta + \xi ) } H_{21}^{\prime} \frac{ b }{ \lambda_2 - \lambda_1 } & 0 & 
e^{ - i ( \delta + \xi ) } H_{23}^{\prime} \frac{ b }{ \lambda_3 - \lambda_1 } \\
- H_{31}^{\prime} \frac{ b }{ \lambda_3 - \lambda_2 } & 
- e^{ i ( \delta + \xi ) } H_{32}^{\prime} \frac{ b }{ \lambda_3 - \lambda_1 } & 0 \\
\end{array}
\right].
\label{2nd-condition-EV}
\end{eqnarray} 
We notice that it can be written in the condensed form as 
\begin{eqnarray} 
&& 
H_{12}^{\prime} 
= 
- e^{ i ( \alpha + \beta - \xi ) } H_{21}, 
\nonumber \\
&& 
H_{23}^{\prime} 
= 
e^{ - i ( \beta - \xi ) } H_{13}, 
\nonumber \\
&& 
H_{13}^{\prime} 
= - e^{ i \alpha } H_{23}. 
\label{2nd-condition-EV2}
\end{eqnarray} 
The condition on $H_{ji}$ can be obtained from that on $H_{ij}$ using the hermitisity 
$H_{ji} = H_{ij}^*$. 

\subsection{The second condition on the genuine non-unitary part}
\label{sec:2nd-condition-UV}

Using eq.~\eqref{Z-sol} the second condition with $\mathcal{Z}_{ \text{UV} }^{(1)}$ in eq.~\eqref{2nd-eq-DMPUV} takes the form 
\begin{eqnarray}
&&
R \left[ V^{(0)} ( \theta_{23}, \phi, \psi, \delta) \right]^{\dagger} \widetilde{\alpha} 
V^{(0)} ( \theta_{23}, \phi, \psi, \delta) R^{\dagger} 
= 
\left[ V^{(0)} ( \theta_{23}^{\prime}, \phi^{\prime}, \psi^{\prime}, \delta + \xi) \right]^{\dagger} \widetilde{\alpha} ^{\prime} 
V^{(0)} ( \theta_{23}^{\prime}, \phi^{\prime}, \psi^{\prime}, \delta + \xi).
\nonumber \\
\label{2nd-condition-UV}
\end{eqnarray}
where $V^{(0)} ( \theta_{23}, \psi, \phi, \delta)$ is defined in eq.~\eqref{Vmatrix-0th-1st-UV}. Then, the transformed $\widetilde{\alpha}$ can be obtained in a closed form as 
\begin{eqnarray}
&&
\widetilde{\alpha} ^{\prime} 
=
V^{(0)} ( \theta_{23}^{\prime}, \phi^{\prime}, \psi^{\prime}, \delta + \xi) 
R 
\left[ V^{(0)} ( \theta_{23}, \phi, \psi, \delta) \right]^{\dagger} 
\widetilde{\alpha} 
V^{(0)} ( \theta_{23}, \phi, \psi, \delta) R^{\dagger} 
\left[ V^{(0)} ( \theta_{23}^{\prime}, \phi^{\prime}, \psi^{\prime}, \delta + \xi) \right]^{\dagger}. 
\nonumber \\
\label{2nd-condition-UV2}
\end{eqnarray}
Notice the vastly different features of the two second conditions, the one for $\mathcal{Z}_{ \text{UV} }^{(1)}$ in eq.~\eqref{2nd-condition-UV2} and the other for $\mathcal{W}_{ \text{EV} }^{(1)}$ in eq.~\eqref{2nd-condition-EV2}. It makes consistency between them highly nontrivial. 

\section{Solution to the SF equation: Genuine non-unitary part}
\label{sec:SF-solution-UV} 

As we have already pointed out, possibility of UV extension of the $\nu$SM DMP symmetries depends on whether or not the second conditions on $\mathcal{W}_{ \text{EV} }^{(1)}$ and $\mathcal{Z}_{ \text{UV} }^{(1)}$ produce {\em independently} the consistent solutions for the transformation properties of the $\widetilde{\alpha}$ parameters. Therefore, we first examine the second condition~\eqref{2nd-condition-UV2} on $\mathcal{Z}_{ \text{UV} }^{(1)}$, the genuine non-unitary part, because it will reveal an interesting feature for the SF formalism itself. But in the first place, it will tell us how $\widetilde{\alpha}$ transform under the DMP-UV symmetries. 

If we use the simplified notation $[V^{\prime} R V^{\dagger}] \equiv V^{(0)} ( \theta_{23}^{\prime}, \phi^{\prime}, \psi^{\prime}, \delta + \xi) R \left[ V^{(0)} ( \theta_{23}, \phi, \psi, \delta) \right]^{\dagger}$, eq.~\eqref{2nd-condition-UV2} can be written as $\alpha^{\prime} = [V^{\prime} R V^{\dagger}] \alpha [V^{\prime} R V^{\dagger}]^{\dagger}$. Therefore, we calculate $[V^{\prime} R V^{\dagger}]$ first. To calculate it in a transparent way, we define $C [12]$, 
\begin{eqnarray}
&&
C [12] \equiv 
\left[
\begin{array}{ccc}
c_{\psi}^{\prime} & s_{\psi}^{\prime} e^{ i (\delta + \xi) }  &  0  \\
- s_{\psi}^{\prime} e^{- i (\delta + \xi) } & c_{\psi}^{\prime} & 0 \\
0 & 0 & 1 \\
\end{array}
\right] 
\left[
\begin{array}{ccc}
0 & - e^{ i ( \delta + \alpha) } & 0 \\
e^{ - i ( \delta + \beta) } & 0 & 0 \\
0 & 0 & 1 \\
\end{array}
\right] 
\left[
\begin{array}{ccc}
c_{\psi} & - s_{\psi} e^{ i \delta}  &  0  \\
s_{\psi} e^{- i \delta} & c_{\psi} & 0 \\
0 & 0 & 1 \\
\end{array}
\right], ~~~
\label{C12-def}
\end{eqnarray}
such that the key ingredient $[V^{\prime} R V^{\dagger}]$ can be written as 
\begin{eqnarray}
&& 
[V^{\prime} R V^{\dagger}] 
\equiv 
V^{(0)} ( \theta_{23}^{\prime}, \phi^{\prime}, \psi^{\prime}, \delta + \xi) 
R 
\left[ V^{(0)} ( \theta_{23}, \phi, \psi, \delta) \right]^{\dagger}  
\nonumber \\
&=&
\left[
\begin{array}{ccc}
1 & 0 &  0  \\
0 & c_{23}^{\prime} & s_{23}^{\prime} \\
0 & - s_{23}^{\prime} & c_{23}^{\prime} \\
\end{array}
\right] 
\left[
\begin{array}{ccc}
c_{\phi}^{\prime} & 0 & s_{\phi}^{\prime} \\
0 & 1 & 0 \\
- s_{\phi}^{\prime} & 0 & c_{\phi}^{\prime} \\
\end{array}
\right] 
C [12]
\left[
\begin{array}{ccc}
c_{\phi} & 0 & - s_{\phi} \\
0 & 1 & 0 \\
s_{\phi} & 0 & c_{\phi} \\
\end{array}
\right] 
\left[
\begin{array}{ccc}
1 & 0 &  0  \\
0 & c_{23} & - s_{23} \\
0 & s_{23} & c_{23} \\
\end{array}
\right]. 
\label{VRVdagger}
\end{eqnarray}
We will see immediately below that $C [12]$ and $[V^{\prime} R V^{\dagger}]$ in eq.~\eqref{VRVdagger} are equal to each other in a quite interesting manner, see eq.~\eqref{identity}. 

\subsection{Useful identities for the rephasing matrix} 
\label{sec:identity}

We calculate $C [12]$ and $[V^{\prime} R V^{\dagger}]$ by inserting the solutions to the SF equation tabulated in Table~\ref{tab:DMPUV-symmetry}, and the phase parameters $\alpha$, $\beta$, etc. given in Table~\ref{tab:SF-solutions}. Since each solution for Symmetry X has the upper and lower $\pm$ signs there are total sixteen cases. The computed results we have obtained for them are the ones totally unexpected to us. Despite the profound dependences on the $\nu$SM parameters in $C [12]$ and $[V^{\prime} R V^{\dagger}]$, the computed results are the constants, which depend only on the symmetry type, Symmetry X with X = I, II, III, IV:\footnote{
We generically quote symmetry as ``Symmetry X'' when X applies to all the DMP-UV symmetries. When we quote X for a particular property in a more specific way, such as e.g., ``X = II, III, and IV'', it means that the property holds for the both XA and XB. For our repeated use of the phrase ``$\mathcal{O}$ transforms under the transformations of Symmetry X'', we simply say ``$\mathcal{O}$ transforms under Symmetry X'' to avoid cumbersome repetition of the words. }
\begin{eqnarray}
&&
C [12] 
= 
V^{(0)} ( \theta_{23}^{\prime}, \phi^{\prime}, \psi^{\prime}, \delta + \xi) 
R 
\left[ V^{(0)} ( \theta_{23}, \phi, \psi, \delta) \right]^{\dagger} 
= \text{Rep(X)}. 
\label{identity}
\end{eqnarray}
Rep(X) is the rephasing matrix which is introduced to characterize the transformation property of the flavor basis Hamiltonian as $H \rightarrow \text{Rep(X)} H \text{Rep(X)}^{\dagger}$ under Symmetry X~\cite{Minakata:2021dqh}. It is the key concept in the proof of the DMP symmetries as the Hamiltonian symmetries. Rep(X) is given by Rep(I) = diag (1,1,1), and\footnote{
Rep(II) = diag (-1,1,-1) in ref.~\cite{Minakata:2021dqh}, but this is equivalent to diag (1,-1,1) as in eq.~\eqref{Rep-II-III-IV}. Similarly, Rep(IV) can be written as diag (1,1,-1). The overall sign of Rep(X) does not affect the physical observables.
}
\begin{eqnarray} 
&&
\text{Rep(II)} = 
\left[
\begin{array}{ccc}
1 & 0 & 0 \\
0 & -1 & 0 \\
0 & 0 & 1 \\
\end{array}
\right],
\hspace{6mm}
\text{Rep(III)} = 
\left[
\begin{array}{ccc}
- 1 & 0 & 0 \\
0 & 1 & 0 \\
0 & 0 & 1
\end{array}
\right], 
\hspace{6mm}
\text{Rep(IV)} = 
\left[
\begin{array}{ccc}
- 1 & 0 & 0 \\
0 & -1 & 0 \\
0 & 0 & 1
\end{array}
\right].
\label{Rep-II-III-IV} 
\end{eqnarray}
In Appendix~\ref{sec:proof-IV}, a sketchy proof of the identity eq.~\eqref{identity} is given for Symmetry IV. The computations of $C [12]$ and $[V^{\prime} R V^{\dagger}]$ for the rest of symmetries X=I, II, III, are left for the interested readers. 

\subsection{Solution to the second condition: Genuine non-unitary part} 
\label{sec:solution-UV}

The solution of the second condition \eqref{2nd-condition-UV2} can readily be obtained by using the second identity in eq.~\eqref{identity}: 
\begin{eqnarray}
&&
\widetilde{\alpha} ^{\prime} 
=
\text{ Rep(X) } \widetilde{\alpha} \text{ Rep(X)}^{\dagger}, 
\label{2nd-condition-UV-sol}
\end{eqnarray}
which implies that $\widetilde{\alpha} ^{\prime} = \widetilde{\alpha}$ for Symmetry X = I, and for X = II, III, and IV, in order 
\begin{eqnarray}
&&
\widetilde{\alpha} ^{\prime} 
= 
\left[ 
\begin{array}{ccc}
\widetilde{\alpha}_{ee} & 0 & 0 \\
- \widetilde{\alpha}_{\mu e} & \widetilde{\alpha}_{\mu \mu}  & 0 \\
\widetilde{\alpha}_{\tau e}  & - \widetilde{\alpha}_{\tau \mu} & \widetilde{\alpha}_{\tau \tau} \\
\end{array}
\right], 
\hspace{5mm}
\left[ 
\begin{array}{ccc}
\widetilde{\alpha}_{ee} & 0 & 0 \\
- \widetilde{\alpha}_{\mu e} & \widetilde{\alpha}_{\mu \mu}  & 0 \\
- \widetilde{\alpha}_{\tau e}  & \widetilde{\alpha}_{\tau \mu} & \widetilde{\alpha}_{\tau \tau} \\
\end{array}
\right], 
\hspace{5mm}
\left[ 
\begin{array}{ccc}
\widetilde{\alpha}_{ee} & 0 & 0 \\
\widetilde{\alpha}_{\mu e} & \widetilde{\alpha}_{\mu \mu}  & 0 \\
- \widetilde{\alpha}_{\tau e}  & - \widetilde{\alpha}_{\tau \mu} & \widetilde{\alpha}_{\tau \tau} \\
\end{array}
\right]. 
\label{alpha-transf-II-IV}
\end{eqnarray}
The resulting transformation properties of the $\widetilde{\alpha}$ parameters and Rep(X) are summarized in Table~\ref{tab:DMPUV-symmetry} and Table~\ref{tab:SF-solutions}, respectively. 

We have reached a very interesting feature that the UV $\widetilde{\alpha}$ parameters do {\em not} transform under Symmetry IA- and IB-DMP-UV, but do transform under the rest of six DMP-UV symmetries: The Rep symmetry, as a whole, recognizes and distinguishes the $\nu$SM and the UV sectors of the theory. It suggests an intriguing possibility that the symmetry can be utilized to diagnose an extended theory which possesses the $\nu$SM and UV parts, which will be further discussed in section~\ref{sec:conclusion}. 

\subsection{The key identity and its possible topological nature} 
\label{sec:topological} 

Probably, the most important observation made in this paper is the  key identity~\eqref{identity}, abbreviated as $V^{(0)} (\Phi^{\prime}) R [V^{(0)} (\Phi) ]^{\dagger} =$ Rep(X). It has several interesting consequences. They include an innovation in the Hamiltonian proof of the symmetries and a conjecture on possible further enlargement of symmetry, which will be presented, respectively, in sections~\ref{sec:hamiltonian-symmetry} and~\ref{sec:conclusion}. 

However, the nature of the identity is highly intriguing and is not understood. While the left-hand side has full of the $\nu$SM variables dependences, the right-hand side is the constant diagonal matrix with the elements $\pm 1 = e^{ \pm i k \pi }$ ($k=0,1$). Such a coherent behavior of the sixteen (eight symmetries duplicated by the upper and lower signs) quantities is not thinkable without a particular reason. The only possibility which we are aware, to the best of our knowledge, is somehow $V^{(0)} (\Phi^{\prime}) R [V^{(0)} (\Phi) ]^{\dagger}$ has a topological nature. While its mathematical proof eludes us, since it is so natural, we suspect that a solid argument for backing up the topological origin of the identity could exist. Since the author presented an argument in favor of this possibility in analogy with the $U(1)$ charge quantization around a vortex in ref.~\cite{Minakata:2022yvs}, we do not repeat it here. 

\section{Solution to the SF equation: Unitary evolution part}
\label{sec:SF-solution-EV} 

We analyze the the second condition in eq.~\eqref{2nd-condition-EV2} for the unitary evolution part to determine the transformation property of the $H_{ij}$ parameters. After verifying the consistency with the $\widetilde{\alpha}$ parameter transformations~\eqref{alpha-transf-II-IV}, we examine the invariance of the oscillation probability under the $H_{ij}$ parameters. 

\subsection{Solution to the second condition: Unitary evolution part}
\label{sec:solution-EV} 

The solutions of the first condition depend not only on the symmetry types denoted generically as XA and XB, but also the upper and lower signs of the phase parameters $\alpha$, $\beta$, etc., as summarized in Table~\ref{tab:SF-solutions}. Using the phase parameters, one can show that the second condition~\eqref{2nd-condition-EV2} implies that $H_{ij}$ transform under Symmetry X as 
\begin{eqnarray} 
&& 
\text{Symmetry IA, IIB}: ~~~
H_{12}^{\prime} = - H_{21}, 
\hspace{8mm}
H_{23}^{\prime} = \pm H_{13}, 
\hspace{8mm}
H_{13}^{\prime} = \mp H_{23}, 
\nonumber \\
&&
\text{Symmetry IB, IIA}: ~~~
H_{12}^{\prime} = H_{21}, 
\hspace{8mm}
H_{23}^{\prime} = \pm H_{13}, 
\hspace{8mm}
H_{13}^{\prime} = \pm H_{23}, 
\nonumber \\
&&
\text{Symmetry IIIA, IVB}: ~~
H_{12}^{\prime} = H_{21}, 
\hspace{8mm}
H_{23}^{\prime} = \mp H_{13}, 
\hspace{8mm}
H_{13}^{\prime} = \mp H_{23}, 
\nonumber \\
&&
\text{Symmetry IIIB, IVA}: ~~ 
H_{12}^{\prime} = - H_{21}, 
\hspace{8mm}
H_{23}^{\prime} = \mp H_{13}, 
\hspace{8mm}
H_{13}^{\prime} = \pm H_{23}.
\label{Hij-transf}
\end{eqnarray}
where $\pm$ (or $\mp$) sign refers to the upper and lower signs in Table~\ref{tab:SF-solutions} and Table~\ref{tab:DMPUV-symmetry}, which are synchronized between them. Notice that the transformation property of $H_{ji}$ can be obtained by using the hermiticity $H_{ji} = (H_{ij})^*$. 
The pairings in eq.~\eqref{Hij-transf} may look curious because the pair IA and IIB, and also IB and IIA, differ in the property that the former (latter) does not (does) contain the transformation of the $\widetilde{\alpha}$ parameters. But, we reassure these pairings in the next section~\ref{sec:consistency}. It appears that the pairing of the two symmetries is dictated by the transformation property of $\psi$, which is the same inside the pair. 

\subsection{Consistency between the $H_{ij}$ and the $\widetilde{\alpha}$ parameter  transformations}
\label{sec:consistency} 

Now, we are left with the consistency check between the solutions of the second conditions derived from the genuine non-unitary part~\eqref{alpha-transf-II-IV} and the unitary evolution part~\eqref{Hij-transf}. The way we carry it out is to use the transformation properties of the $\nu$SM variables and the $\widetilde{\alpha}$ parameters in eq.~\eqref{alpha-transf-II-IV}, which are both in Table~\ref{tab:DMPUV-symmetry}, to obtain the $H_{ij}$ transformation properties and see if they agree with the ones in eq.~\eqref{Hij-transf}. 

Then, we immediately encounter the problem for $H_{ii}$ ($i=1,2,3$), the diagonal elements. Their transformation properties are absent in eq.~\eqref{Hij-transf}. 
It is because they are absorbed into the eigenvalues, see eq.~\eqref{ren-eigenvalues}. Note that $G_{ii}=H_{ii}$. In fact, $H_{ii}$ calculated with the above recipe transform, under all Symmetry X = IA, IB, $\cdot \cdot \cdot $, IVB, as 
\begin{eqnarray} 
&& 
H_{11} \leftrightarrow H_{22}, 
\label{H-diag-transf}
\end{eqnarray} 
i.e., $H_{11}$-$H_{22}$ exchange, and $H_{33}$ is invariant. It must be the case because we are dealing with the 1-2 state exchange symmetry $\lambda_{1} \leftrightarrow \lambda_{2}$ for which the both transformations $\lambda_{1}^{ \nu\text{SM} } \leftrightarrow \lambda_{2}^{ \nu\text{SM} }$ and $H_{11} \leftrightarrow H_{22}$ must occur simultaneously under all Symmetry X. Therefore, the consistency is met for the diagonal $H_{ii}$.

Now we discuss the off-diagonal $H_{ij}$ ($i \neq j$). To have a clearer view of the ``curious paring'' we give explicit treatments of Symmetry I and II. In Appendix~\ref{sec:H-elements} the $H_{ij}$ elements are given by using the $K_{ij}$ elements, whose latter depend only on $\theta_{23}$, $\phi$, and the $\widetilde{\alpha}$ parameters. Since none of them is involved in the transformations of Symmetry IA and IB, all the $K_{ij}$ elements are invariant under these symmetries. Then, the $H_{ij}$ elements transform under IA and IB only through the $\nu$SM parameter transformations as 
\begin{eqnarray} 
&& 
H_{12} \rightarrow - H_{21}, 
\hspace{8mm} 
H_{13} \rightarrow \mp H_{23}, 
\hspace{8mm} 
H_{23} \rightarrow \pm H_{13} ~~~~\text{(IA)},
\nonumber \\
&&
H_{12} \rightarrow H_{21}, 
\hspace{8mm} 
H_{13} \rightarrow \pm H_{23}, 
\hspace{8mm} 
H_{23} \rightarrow \pm H_{13} ~~~~\text{(IB)}. 
\label{Hij-transf-IA-IB}
\end{eqnarray} 
Notice that they reproduce the relevant lines in eq.~\eqref{Hij-transf}. It means that no $\widetilde{\alpha}$ parameter transformation is involved in Symmetry IA- and IB-DMP-UV, and only the $\nu$SM parameter transformations suffice. Thus, in consistent with the result obtained in section~\ref{sec:solution-UV}, our SF treatment reproduces the somewhat puzzling result\footnote{
Invariance of the oscillation probability in the DMP-UV perturbation theory under Symmetry IA- and IB-DMP (without UV extension) was indeed observed while the author worked on ref.~\cite{Minakata:2021nii}, but it was not mentioned because of its puzzling feature. }
that no UV variable transformation is induced in Symmetry IA and IB.\footnote{
Here is a clarifying note on our claim that the symmetry can distinguish between the $\nu$SM and the UV sectors of the theory. To make a clearer statement, we go back to the original framework in ref.~\cite{Minakata:2021nii} in which we restrict to the unrenormalized treatment of the eigenvalues $\lambda_{i} = \lambda_{i}^{ \nu\text{SM} }$ by removing the $H_{ii}$ term in eq.~\eqref{ren-eigenvalues}. Now there exists the $(b x) / 2E$ term in $P(\nu_\mu \rightarrow \nu_e)^{(1)} _{ \text{EV} }$ as given in eq.~\eqref{bx-correction}. Then, there is no mixed up between the $\nu$SM and the UV variables. In this treatment the generators for UV variables' transformations do not act on the $\nu$SM part of the theory. }

Under Symmetry IIA and IIB, the two $K_{ij}$ elements flip sign 
\begin{eqnarray} 
&& 
K_{12} \rightarrow - K_{12}, 
\hspace{10mm} 
K_{23} \rightarrow - K_{23}, 
\label{Kij-transf-II}
\end{eqnarray}
and all the other $K_{ij}$ elements are invariant. Then, one can show that the resulting $H_{ij}$ element transformations are as 
\begin{eqnarray} 
&& 
H_{12} \rightarrow H_{21}, 
\hspace{8mm} 
H_{13} \rightarrow \pm H_{23}, 
\hspace{8mm} 
H_{23} \rightarrow \pm H_{13} ~~~~\text{(IIA)},
\nonumber \\
&&
H_{12} \rightarrow - H_{21}, 
\hspace{8mm} 
H_{13} \rightarrow \mp H_{23}, 
\hspace{8mm} 
H_{23} \rightarrow \pm H_{13} ~~~~\text{(IIB)}.
\label{Hij-transf-IIA-IIB}
\end{eqnarray} 
Therefore, the pairings between IA-IIB, and IB-IIA in the first and second lines of eq.~\eqref{Hij-transf}, which we have referred as ``curious''  are reproduced.

Similarly one can work out the $H_{ij}$ transformation properties for Symmetry III and IV to confirm eq.~\eqref{Hij-transf}. Therefore, the $\widetilde{\alpha}$ parameter transformation from the second condition on the genuine non-unitary part $\mathcal{Z}_{ \text{UV} }^{(1)}$ is perfectly consistent with the $H_{ij}$ transformation property derived from that of the unitary evolution part $\mathcal{W}_{ \text{EV} }^{(1)}$. 

\subsection{Invariance of the oscillation probability}
\label{sec:invariance-P} 

The oscillation probability $P(\nu_{\mu} \rightarrow \nu_{e})^{(1)}_{ \text{EV} }$ given in Appendix~\ref{sec:EV-part} is written in terms of the $\nu$SM and $H_{ij}$ parameters without any naked $\widetilde{\alpha}$ parameters. Therefore,  showing the invariance under Symmetry X can be carried out straightforwardly for all the eight symmetries with the transformation properties of these parameters given in Table~\ref{tab:DMPUV-symmetry} and eq.~\eqref{Hij-transf}. This exercise for invariance proof, simple but slightly lengthy, is left for the interested readers. 

On the other hand, the probability $P(\nu_{\mu} \rightarrow \nu_{e})^{(1)}_{ \text{UV} }$ in Appendix~\ref{sec:UV-part} consists of the $\nu$SM and the naked $\widetilde{\alpha}$ parameters. We can use the transformation properties of these variables summarized in Table~\ref{tab:DMPUV-symmetry} to prove the invariance under the all Symmetry X.

In this paper we do not discuss explicitly the oscillation channels other than $\nu_{\mu} \rightarrow \nu_{e}$, because we will prove the Hamiltonian invariance in section~\ref{sec:hamiltonian-symmetry} which automatically applies to all the oscillation channels. 

\section{DMP-UV symmetry as a Hamiltonian symmetry}
\label{sec:hamiltonian-symmetry}

In this section we show that all the DMP-UV symmetries summarized in Table~\ref{tab:DMPUV-symmetry} are the symmetries of the flavor basis Hamiltonian~$H_{ \text{flavor} }$. In unitary case in vacuum $H_{ \text{flavor} } = U \check{H} U^{\dagger}$, where $\check{H}$ is the vacuum mass eigenstate basis Hamiltonian, and $U$ the $\nu$SM flavor mixing matrix, see eq.~\eqref{U-alpha-SOL-def}. In non-unitary case in matter, since the flavor basis $\nu$ is related to the mass eigenstate basis $\check{\nu}$ as $\nu = N \check{\nu}$, $H_{ \text{flavor} } = N \check{H} N^{\dagger}$, where the check basis Hamiltonian $\check{H}$ is defined in eq.~\eqref{evolution-check-basis}. We denote $H_{ \text{flavor} }$ constructed in this way as $H_{\text{\tiny VM}}$. 

We can construct $H_{ \text{flavor} }$ in an alternative way. To formulate the DMP-UV perturbation theory we diagonalized the dominant part of $\check{H}$ to obtain the bar basis Hamiltonian with the result presented in eq.~\eqref{barH-0th-1st} to first order. Then, we can transform back to the check basis, and then transform to the flavor basis using $\nu = N \check{\nu}$. The thereby obtained $H_{ \text{flavor} }$ is denoted as $H_{\text{\tiny Diag}}$. The subscripts in $H_{\text{\tiny VM}}$ and $H_{\text{\tiny Diag}}$ imply ``vacuum-matter'' and ``diagonalized'', respectively. Of course they are equal to each other, $H_{\text{\tiny VM}} = H_{\text{\tiny Diag}}$.\footnote{
Usage of $H_{\text{\tiny VM}}$ and $H_{\text{\tiny Diag}}$ instead of our previous notations $H_{\text{\tiny LHS}}$ and $H_{\text{\tiny RHS}}$, respectively, is to unify our notation with ref.~\cite{Minakata:2022yvs}. } 

\subsection{Transformation property of $H_{\text{\tiny VM}}$} 
\label{sec:H-LHS-transf}

Using $N= \left( 1 - \widetilde{\alpha} \right) U$ and $N N^{\dagger} = \left( 1 - \widetilde{\alpha} \right) \left( 1 - \widetilde{\alpha} \right)^{\dagger}$, $H_{\text{\tiny VM}} = N \check{H} N^{\dagger}$ can be written as (after multiplied by $2E$) 
\begin{eqnarray}
&&
\hspace{-4mm}
2E H_{\text{\tiny VM}} =
\left( 1 - \widetilde{\alpha} \right) 
\left\{ 
U (\Xi)
\left[
\begin{array}{ccc}
m^2_{1} & 0 & 0 \\
0 & m^2_{2} & 0 \\
0 & 0 & m^2_{3} \\
\end{array}
\right] 
U (\Xi)^{\dagger}  
+ 
( 1 - \widetilde{\alpha} )^{\dagger} 
\cdot 
\left[
\begin{array}{ccc}
a - b & 0 & 0 \\
0 & - b & 0 \\
0 & 0 & - b \\
\end{array}
\right] 
\cdot
\left( 1 - \widetilde{\alpha} \right) 
\right\} 
( 1 - \widetilde{\alpha} )^{\dagger},
\nonumber \\
\label{H-LHS}
\end{eqnarray}
where we have used a collective notation $\Xi$ for all the vacuum parameters involved. Here we have used a slightly different phase-redefined basis from the one in eq.~\eqref{evolution-check-basis} to make the vacuum Hamiltonian $\propto$ diag($m^2_{1}, m^2_{2}, m^2_{3}$) making it more symmetric, which however does not affect our symmetry discussion. 

We have shown in ref.~\cite{Minakata:2021dqh} that the vacuum term transforms under Symmetry X as 
\begin{eqnarray}
&&
\left\{ 
U (\Xi)
\left[
\begin{array}{ccc}
m^2_{1} & 0 & 0 \\
0 & m^2_{2} & 0 \\
0 & 0 & m^2_{3} \\
\end{array}
\right] 
U (\Xi)^{\dagger} 
\right\} 
\rightarrow
\text{ Rep(X) } 
\left\{ 
U (\Xi)
\left[
\begin{array}{ccc}
m^2_{1} & 0 & 0 \\
0 & m^2_{2} & 0 \\
0 & 0 & m^2_{3} \\
\end{array}
\right] 
U (\Xi)^{\dagger} 
\right\}
\text{ Rep(X)}^{\dagger}.
\nonumber \\
\label{Hvac-transform}
\end{eqnarray}
where Rep(X) is defined in eq.~\eqref{Rep-II-III-IV}. 
Using the transformation property of the $\widetilde{\alpha}$ parameters in eq.~\eqref{2nd-condition-UV-sol}, $\widetilde{\alpha} ^{\prime} = \text{ Rep(X) } \widetilde{\alpha} \text{ Rep(X)}^{\dagger}$, the matter term in eq.~\eqref{H-LHS} which originates from the $\nu$SM and the UV sectors of the theory obeys the same transformation property as in the vacuum term. Then, the whole $H_{\text{\tiny VM}}$ transforms under Symmetry X as 
\begin{eqnarray}
&&
H_{\text{\tiny VM}} 
\rightarrow 
\text{ Rep(X) } H_{\text{\tiny VM}} 
\text{ Rep(X)}^{\dagger}, 
\label{H-LHS-transformed}
\end{eqnarray}
which means that $H_{\text{\tiny VM}}$ is invariant under Symmetry X up to the rephasing factor Rep(X). By being the diagonal matrix with the elements $e^{ \pm i m \pi }$ ($m=0,1$), Rep(X) does not affect physical observables as it can be absorbed into the neutrino wave functions. 

We note that our success in demonstrating the transformation property of $H_{\text{\tiny VM}}$ in a transparent way as above heavily owes the $\alpha$ parametrization~\cite{Escrihuela:2015wra} of the non-unitary matrix, eq.~\eqref{alpha-matrix-def}. It entails the neat $\widetilde{\alpha}$ transformation~\eqref{2nd-condition-UV-sol}, and enabled us to have the simple and revealing form of $H_{\text{\tiny VM}}$ in eq.~\eqref{H-LHS}. 

Therefore, the invariance property~\eqref{H-LHS-transformed} ultimately comes from the fact that the transformation property of the $\widetilde{\alpha}$ parameters is determined by the identical rephasing matrix Rep(X) that governs the transformation of the $\nu$SM part of the Hamiltonian. Despite that it must be the case for proving the Hamiltonian invariance, it is remarkable to see that it indeed occurs, being enforced by the genuine UV part of the SF equation~\eqref{2nd-condition-UV2}. It indicates an intriguing interplay between the $\nu$SM and UV sectors in the theory. 
In passing, we note that we do not use the property that the matter density is uniform to obtain the invariance proof, the feature which prevails in the proof of invariance of $H_{\text{\tiny Diag}}$ in section~\ref{sec:H-RHS-transf}. 

\subsection{Transformation property of $H_{\text{\tiny Diag}}$: New method} 
\label{sec:H-RHS-transf}

In this section we discuss $H_{\text{\tiny Diag}}$ to show that it is invariant under Symmetry X-DMP-UV with the same rephasing matrix as needed for $H_{\text{\tiny VM}}$. By $H_{\text{\tiny Diag}}$ we mean the flavor-basis Hamiltonian written in terms of the diagonalized variables. 
Using the expression of the flavor-basis state by the bar-basis state 
$\nu_{\alpha} = \left[ ( 1 - \widetilde{\alpha} ) U_{23} U_{13}(\phi) U_{12} (\psi, \delta) \right] _{\alpha j} \bar{\nu}_{j}$ given in eq.~\eqref{flavor-check}, $H_{\text{\tiny Diag}}$ is given by 
\begin{eqnarray} 
H_{\text{\tiny Diag}} 
&=& 
( 1 - \widetilde{\alpha} ) 
U_{23} (\theta_{23} ) U_{13} (\phi) U_{12} (\psi, \delta) 
\bar{H} 
U^{\dagger}_{12} (\psi, \delta) U^{\dagger}_{13} (\phi) U^{\dagger}_{23} (\theta_{23}) 
( 1 - \widetilde{\alpha} )^{\dagger}. 
\label{H-RHS}
\end{eqnarray}
The bar-basis Hamiltonian $\bar{H}$ is given in eq.~\eqref{barH-0th-1st}, which ignores the second order term eq.~\eqref{barH-UV} with the $G^{(2)}$ matrix in eq.~\eqref{G-G2-def}. In this section we proceed with the bar-basis Hamiltonian~\eqref{barH-0th-1st} without the second order term to prove invariance of $H_{\text{\tiny Diag}}$ under Symmetry X. In section~\ref{sec:2nd-order} we will present a simple argument to show that our proof of invariance prevails even after we include the second order effect. 

Since we use an entirely new method to prove the invariance $H_{\text{\tiny Diag}}$, we include the $\nu$SM part as well, though its invariance has been fully discussed in ref.~\cite{Minakata:2021dqh}. From the identity~\eqref{identity} one obtains 
\begin{eqnarray}
&&
V^{(0)} ( \theta_{23}^{\prime}, \phi^{\prime}, \psi^{\prime}, \delta^{\prime} ) 
= 
\text{Rep(X)} 
V^{(0)} ( \theta_{23}, \phi, \psi, \delta) R^{\dagger}.
\label{V0-prime-V0}
\end{eqnarray}
Then, $H_{\text{\tiny Diag}}$ in eq.~\eqref{H-RHS} with use of eq.~\eqref{Vmatrix-0th-1st-UV} transforms under Symmetry X as 
\begin{eqnarray} 
&&
H_{\text{\tiny Diag}} 
= 
( 1 - \widetilde{\alpha} ) 
V^{(0)} ( \theta_{23}, \phi, \psi, \delta) 
\bar{H} ( \theta_{23}, \theta_{12}, \phi, \psi, \delta; \widetilde{\alpha}_{\beta \gamma}, \lambda_{i} ) 
\left[ V^{(0)} ( \theta_{23}, \phi, \psi, \delta) \right]^{\dagger} 
( 1 - \widetilde{\alpha} )^{\dagger} 
\nonumber \\
&\rightarrow&_{\text{\tiny Symmetry X}} 
( 1 - \widetilde{\alpha}^{\prime} )  
V^{(0)} ( \theta_{23}^{\prime}, \phi^{\prime}, \psi^{\prime}, \delta^{\prime} ) 
\bar{H} ( \theta_{23}^{\prime}, \theta_{12}^{\prime}, \phi^{\prime}, \psi^{\prime}, \delta^{\prime}; \widetilde{\alpha}_{\beta \gamma}^{\prime}, \lambda_{i}^{\prime} ) 
\left[ V^{(0)} ( \theta_{23}^{\prime}, \phi^{\prime}, \psi^{\prime}, \delta^{\prime} ) \right]^{\dagger} 
( 1 - \widetilde{\alpha}^{\prime} )^{\dagger} 
\nonumber \\
&& 
\hspace{-12mm} 
=  
\text{Rep(X)} 
( 1 - \widetilde{\alpha} ) 
V^{(0)} ( \theta_{23}, \phi, \psi, \delta) R^{\dagger} 
\bar{H} ( \theta_{23}^{\prime}, \theta_{12}^{\prime}, \phi^{\prime}, \psi^{\prime}, \delta^{\prime}; \widetilde{\alpha}_{\beta \gamma}^{\prime}, \lambda_{i}^{\prime} ) 
R 
\left[ V^{(0)} ( \theta_{23}, \phi, \psi, \delta) \right]^{\dagger} 
( 1 - \widetilde{\alpha} )^{\dagger} 
\text{Rep(X)} ^{\dagger}.
\nonumber \\
\label{H-RHS-UV}
\end{eqnarray}
Note that $R$ is the ``untransformed'' matrix. What is remarkable is that one can show by using the $H_{ij}$ transformation property in eq.~\eqref{Hij-transf} that 
\begin{eqnarray} 
&& 
R^{\dagger} 
\bar{H} ( \theta_{23}^{\prime}, \theta_{12}^{\prime}, \phi^{\prime}, \psi^{\prime}, \delta^{\prime}; \widetilde{\alpha}_{\beta \gamma}^{\prime}, \lambda_{i}^{\prime} ) 
R 
= 
\bar{H} ( \theta_{23}, \theta_{12}, \phi, \psi, \delta; \widetilde{\alpha}_{\beta \gamma}, \lambda_{i} )
\label{RGR}
\end{eqnarray} 
for all Symmetry X-DMP-UV where X= IA, IB, $\cdot \cdot \cdot $, IVB. The proof must be done for all the Symmetry X with the upper and lower signs of the solutions of the first condition.\footnote{
A careful reader might have detected an extreme similarity with the treatment in ref.~\cite{Minakata:2022yvs} for the helio-UV perturbation theory. But, we remark that despite the similarity at the equation level, the calculation needed for proof of eq.~\eqref{RGR} differs in its figure and in volume. }
The fact that eq.~\eqref{RGR} holds implies that $H_{\text{\tiny Diag}}$ transforms under Symmetry X as 
\begin{eqnarray} 
&& 
H_{\text{\tiny Diag}} 
\rightarrow 
\text{Rep(X)} 
H_{\text{\tiny Diag}} 
\text{Rep(X)} ^{\dagger}.
\label{H-RHS-UV-invariance}
\end{eqnarray}
That is, $H_{\text{\tiny Diag}}$ is invariant under Symmetry X apart from the rephasing factors $\text{Rep(X)}$ and $\text{Rep(X)} ^{\dagger}$. Notice again that Rep(X) is solely rooted in the $\nu$SM, see eq.~\eqref{identity}, but also governs the UV part of the theory. 

\subsection{Including the second-order UV effect} 
\label{sec:2nd-order} 

Now let us include the second-order UV effect into our proof of invariance by turning on $\bar{H}^{(2)}_{ \text{UV}} = - (b/2E) G^{(2)}$ in eq.~\eqref{barH-UV}. 
Recapitulating our nomenclature, once the $A^{(2)}$ matrix is given as in eq.~\eqref{A-A2-def}, one can define the $G^{(2)}$ matrix as in eq.~\eqref{G-G2-def}. Then, we can similarly define $H^{(2)}$ matrix as $G^{(2)} = D H^{(2)} D^{\dagger}$ as in eq.~\eqref{Hmatrix-def}. If one wants to obtain the explicit forms of the $H^{(2)}_{ij}$ matrix elements, one can follow eq.~\eqref{G-K-def} by replacing $A$ by $A^{(2)}$ in Appendix~\ref{sec:H-elements}. 

We first show that the transformation properties of the $A^{(2)}$ matrix is identical with that of the $A$ matrix, the both defined in eq.~\eqref{A-A2-def}. Notice that only the $\widetilde{\alpha}$ parameters transform in them. Using the $\widetilde{\alpha}$ parameter transformations~\eqref{alpha-transf-II-IV} and with the explicit expressions of the $A$ and $A^{(2)}$ matrices it is a straightforward to show that the transformation properties of the $A$ and $A^{(2)}$ matrices under Symmetry X are the same. 
It means that the transformation properties of the $G^{(2)}$ matrix under Symmetry X, where X=IA, IB, $\cdot \cdot \cdot $, IVB, is the same as that of $G$ matrix, because $G = [ V^{(0)} ( \theta_{23}, \phi, \psi, \delta) ]^{\dagger} A V^{(0)} ( \theta_{23}, \phi, \psi, \delta)$ and $G^{(2)} = [ V^{(0)} ( \theta_{23}, \phi, \psi, \delta) ]^{\dagger} A^{(2)} V^{(0)} ( \theta_{23}, \phi, \psi, \delta)$, see eq.~\eqref{G-G2-def}. 

Our last step is to show that the key equation~\eqref{RGR} for proof of invariance of the Hamiltonian is satisfied after the second order effect $- (b/2E) G^{(2)}$ is included. To prove eq.~\eqref{RGR} we have used the $H_{ij}$ transformations in eq.~\eqref{Hij-transf}. But, one can readily show that the above constructed $H^{(2)}_{ij}$ matrix elements transform under Symmetry X in the same way as $H_{ij}$. Since inclusion of the second order UV term merely changes $H_{ij}$ to $H_{ij} - H_{ij}^{(2)}$ in eq.~\eqref{RGR}, and their transformation properties are the same, the invariance proof given in section~\ref{sec:H-RHS-transf} remains valid with inclusion of the second order UV effect. 

To summarize, we have shown in this section that the flavor basis Hamiltonian $H_{ \text{flavor} }$ (the both $H_{\text{\tiny VM}}$ and $H_{\text{\tiny Diag}}$) transforms as $H_{ \text{flavor} } \rightarrow \text{Rep(X)} H_{ \text{flavor} } \text{Rep(X)} ^{\dagger}$ under all Symmetry X. This establishes the property of Symmetry X as the Hamiltonian symmetry which holds in all orders in the DMP-UV perturbation theory in all the oscillation channels. 

\subsection{The DMP-UV symmetry: Summary} 
\label{sec:summary} 

Here, we give our summary of all the eight DMP-UV symmetries, denoted as Symmetry X-DMP-UV where X = IA, IB, $\cdot \cdot \cdot $, IVB. The SF formulation and the solutions for the DMP-UV symmetry, which are valid to first order in the DMP-UV perturbation theory, are described in sections~\ref{sec:SF-DMP-UV}, \ref{sec:SF-solution-UV}, and \ref{sec:SF-solution-EV}. In Table~\ref{tab:SF-solutions}, we tabulate the solutions of the first condition~\eqref{1st-condition} and the rephasing matrix Rep(X) given in eq.~\eqref{Rep-II-III-IV} for all Symmetry X-DMP-UV. In Table~\ref{tab:DMPUV-symmetry} we give the transformation properties of the $\nu$SM and the UV $\widetilde{\alpha}$ parameters under Symmetry X. The transformations of the $H_{ij}$ matrix elements for the EV (unitary evolution) part are given in eq.~\eqref{Hij-transf} in section~\ref{sec:solution-EV}. 
Finally, the Hamiltonian proof of the DMP-UV symmetry is given in this section which guarantees that the Rep symmetry holds in all orders in the perturbation theory. 

\section{Conclusion and outlook}
\label{sec:conclusion} 

In this paper we have applied Symmetry Finder (SF) to identify the reparametrization (Rep) symmetry in the UV (unitarity violation)-extended version of the DMP perturbation theory. The extended framework of the $\nu$SM with the inclusion of non-unitary flavor mixing matrix is one of the promising ways to discuss physics beyond the $\nu$SM. The summary of the uncovered Rep symmetries in the UV-extended DMP theory, Symmetry X-DMP-UV (X=IA, IB, $\cdot \cdot \cdot $, IVB), is given in Table~\ref{tab:DMPUV-symmetry}. 
The result of this paper adds one more success example to the SF symmetry list which contains the twin 1-2 exchange symmetries~\cite{Minakata:2021dqh,Minakata:2022yvs}, and the 1-3 exchange ones with and without non-unitarity~\cite{Minakata:2022yvs,Minakata:2021goi}.

In essence the UV-extended Rep symmetry differs from the original DMP symmetry~\cite{Minakata:2021dqh} by addition of the UV $\widetilde{\alpha}$ variable transformation $\widetilde{\alpha} \rightarrow \text{ Rep(X) } \widetilde{\alpha} \text{ Rep(X)}^{\dagger}$. Rep(X) is the rephasing matrix which is introduced to describe the transformation property of the flavor-basis Hamiltonian in the $\nu$SM, see eq.~\eqref{Rep-II-III-IV} for the definition. Therefore, it is inherently the $\nu$SM concept. In this sense it is very interesting to observe that $\widetilde{\alpha}$ transforms under Symmetry X-DMP-UV in exactly the same way as the flavor-basis Hamiltonian, with or without the UV extension. It illuminates a remarkable interplay between the $\nu$SM and UV sectors probed by the Rep symmetry.

The above transformation property implies that $\widetilde{\alpha}$ does not transform under IA- and IB-DMP symmetries, as Rep(I) = 1. The Hamiltonian is then invariant under the original $\nu$SM IA- and IB-DMP symmetry transformations even though it contains the UV-originated parts. That is, the symmetry generators of IA- and IB-DMP do not feel the existence of non-unitarity. On the other hand, the remaining six symmetries X=II, III, and IV with the both A (without $\delta$) and B (with $\delta$) types do recognize the UV sector by transforming $\widetilde{\alpha}$. This feature strongly suggests that the Rep symmetry can be used as a diagnostic tool for low-energy effective theories which possesses the $\nu$SM and UV sectors. 

Did we successfully answer the question raised earlier?:~``Can one extract new physical insights from the symmetry of reparameterizing the same physics?'' Our answer is {\em Yes}. Quantum mechanics governs the both $\nu$SM and UV sectors, and hence the Rep symmetry which is rooted deep in it inevitably possesses the transformations which relate between the two sectors. In the above, we just saw the concrete example for this feature, the $\widetilde{\alpha}$ transformation property $\widetilde{\alpha} \rightarrow \text{ Rep(X) } \widetilde{\alpha} \text{ Rep(X)}^{\dagger}$, which illuminates that the UV variable transforms by the $\nu$SM one, Rep(X). It is intriguing to speculate on possible reason why $\widetilde{\alpha}$ does not transform in a strongly mixed way with the $\nu$SM variables, but only by phases Rep(X). Since UV effect at low energies originates in a new physics at high scale, coupling between UV and $\nu$SM sector must be ``weak'', and the coupling via phases would be the least efficient form. 
Thus, it is interesting to see that SF and the Rep symmetry provide us a novel way of probing the interrelation between the $\nu$SM and low-energy description of new physics in a model-independent manner. Of course, the final judgement about the validity of our picture should be made by the readers. 

Triggered by the interest in the system with Majorana phases we have reexamined the SF symmetry in vacuum, which entails the eight 1-2 exchange symmetries whose structure akin to the DMP SF symmetries. The vacuum SF symmetries have the mass exchange $m^2_{1} \leftrightarrow m^2_{2}$, whereas DMP has matter-dressed eigenvalue exchange, $\lambda_{1} \leftrightarrow \lambda_{2}$, but not the vacuum mass exchange. The feature reflects the difference between the quantum mechanical ground states in vacuum and in matter. 

\subsection{Inter-sector communications between the $\nu$SM and UV sectors}
\label{sec:inter-sector-com}

We note that the relationship between the $\nu$SM and UV sectors through the phases, in the above case through Rep(X), is not prominently new. We have piece of evidence for the $\nu$SM-UV inter-sector communications through the phases, the lepton KM phase $\delta$ and the phases of the UV $\alpha$ parameters, as explicitly worked out in refs.~\cite{Martinez-Soler:2018lcy,Martinez-Soler:2019noy,Minakata:2021nii}. It is entirely natural to expect existence of order-unity correlations between the $\nu$SM and UV phases from the viewpoint of unitarity polygon, a generalization of the unitarity triangle, in larger unitary theory~\cite{Fong:2016yyh,Fong:2017gke,Martinez-Soler:2018lcy}.\footnote{
Notice that the interactions between the $\nu$SM and UV variables always occur through the first-order term of the Hamiltonian, the last term in eq.~\eqref{barH-0th-1st}. Certainly it opens the way of communication between the two sectors in the way which manifests at low energies, as originally dictated by the high energy theory but brought to low scales when the high-energy sector is integrated out. Nonetheless, it is the first-order suppressed effect, and the physical picture behind it is not so transparent. This seems to be the case even if one solves exactly the system with non-unitarity, as done in ref.~\cite{Fong:2017gke}. } 
Notwithstanding whether the speculated topological nature of the identity is true or not, see section~\ref{sec:topological}, the $\nu$SM-UV inter-sector communication through the phases is a natural and real outcome from our study of the Rep symmetry in this paper. 

\subsection{How big is the Rep symmetry?} 
\label{sec:how-big} 

First of all, we try to address the easier question: Are the eight 1-2 state exchange symmetries in our restricted SF treatment in both DMP and SRP sufficiently large to cover the all allowed cases? We argue in the positive. 
All the symmetries identified by the SF framework~\cite{Minakata:2021dqh,Minakata:2022yvs,Minakata:2021goi}, including those in vacuum as shown in Table~\ref{tab:SF-symmetry-vacuum} in section~\ref{sec:symmetry-vacuum}, possess the structure X = I, II, III, and IV, apart from the doublings due to the types A or B (no or with $\delta$), or possibly with or without ``f'' ($s_{12}$ sign flip) indices. We suspect that the symmetry structure X = I, II, III, IV,  exhausts the candidate list for symmetries for the following reasoning: We have to have the rephasing matrix Rep(X) for the Hamiltonian proof of Symmetry X. Rep(X) must be a diagonal matrix because otherwise it alters the physical observables by changing the flavor labels. As far as the real diagonal matrices with the elements $\pm 1$ are concerned, our Rep(X) with X = I, II, III, IV in Table~\ref{tab:SF-solutions} constitute all the possible choices. The remaining possibility is the case of complex diagonal matrix Rep(X), whose existence, however, eludes us at this moment. 

Before conclude, we want to leave our speculation, an intriguing possibility that the Rep symmetry could be much larger. This is born out from stimulus of the key identity~\eqref{identity}, $V^{(0)} (\Phi^{\prime}) R [V^{(0)} (\Phi) ]^{\dagger} =$ Rep(X), and, if true, it supersedes our all SF symmetry discussions. The identity in eq.~\eqref{identity} and the transformation property of $H_{\text{\tiny Diag}}$ in eq.~\eqref{H-RHS-UV} with the equality in \eqref{RGR} that are required for the invariance proof, do not refer at least formally, which states are exchanged. We therefore suspect that even more generic state exchange symmetry exists which might be extended to $S_{3}$, for example, for the three-neutrino system. At the same time the key identity~\eqref{identity} must be elevated to match to this generalization with the suitably generalized $R$ matrix. 

Even if this is true, the formulation would become a more abstract one because the explicit construction with the PDG, or the corollary conventions such as the SOL, of the $U$ matrix would be very hard. (We recall the difficulty in formulating the 1-3 exchange symmetry which is overcome only by using the helio-perturbation theory~\cite{Minakata:2021goi}.) The topological nature of the identity might show up more naturally in this extended setting. The author believes that this possibility is worth to explore. 

\subsection{Rep symmetry in an extended class of observables with Majorana phases?} 
\label{sec:majorana-phase}

Our SF symmetry discussion in vacuum and in matter, so far, assumed its application to neutrino oscillation in which the observable CP phase is unique, the lepton KM phase $\delta$~\cite{Kobayashi:1973fv}. However, it is known that if neutrinos are Majorana particles~\cite{Majorana:1937vz} our world is enriched with the two more physical phases called the Majorana phases~\cite{Schechter:1980gr,Bilenky:1980cx,Doi:1980yb}. 
Then, an interesting question would be: Can the vacuum Rep symmetry in Table~\ref{tab:SF-symmetry-vacuum} accommodate the Majorana phases?

Let us discuss neutrinoless double beta decay~\cite{Avignone:2007fu} in the context of the vacuum SF symmetry. We note that to our knowledge this is the first time to address the neutrinoless double beta decay observable from the Rep symmetry viewpoint. In this circumstance it is imperative to use the symmetric parametrization of the $U$ matrix~\cite{Rodejohann:2011vc}. By ``symmetric parametrization'' we mean the form of the $U$ matrix as 
$U_{ \text{symmetric} } = U_{23} (\theta_{23}, \phi_{23}) U_{13} (\theta_{13}, \phi_{13}) U_{12} (\theta_{12}, \phi_{12})$, where $U_{13} (\theta_{13}, \phi_{13})$ is the 1-3 space rotation matrix in the PDG convention~\cite{ParticleDataGroup:2022pth} but with the lepton KM phase $\delta$ replaced by $\phi_{13}$, etc.~\cite{Rodejohann:2011vc}. Using $U_{ \text{symmetric} }$, the observable of neutrinoless double beta decay is given by 
\begin{eqnarray}
m_{0\nu\beta\beta} 
&= & 
\biggl | m_1 c_{13}^2 c_{12}^2 + 
m_2 c_{13}^2 s_{12}^2 e^{- 2 i \phi_{12} } + 
m_3 s_{13}^2 e^{- 2 i \phi_{13} } \biggr |.
\label{0nubb-obs}
\end{eqnarray} 
The CP phase $\delta$ can be expressed by using the three Majorana phases as~\cite{Rodejohann:2011vc} 
\begin{eqnarray} 
&& 
\delta = \phi_{13} - \phi_{12} - \phi_{23}. 
\label{delta}
\end{eqnarray} 
One can easily show that if $\phi_{ij}$ transform as $\phi_{ij} \rightarrow \phi_{ij}^{\prime}$ under transformations of Symmetry X, they leave $m_{0\nu\beta\beta}$ invariant if 
\begin{eqnarray} 
&& 
\phi_{12}^{\prime} = \pm \phi_{12}, 
\hspace{10mm}
\phi_{13}^{\prime} = \pm ( \phi_{12} - \phi_{13} ), 
\label{phi}
\end{eqnarray} 
where the both equations hold in mod.~$\pi$. 

We note that the phase $\phi_{23}$ does not come in to the neutrinoless double beta decay observable. To investigate the cases with $\phi_{23}$, with possibility of an intriguing interplay with the lepton KM phase measured by neutrino oscillation, we need a wider class of observables that contain $\nu_{\mu}$ or $\nu_{\tau}$~\cite{Frigerio:2002rd,Frigerio:2002fb,Bertuzzo:2013ew}. 
Though the discussion above is far from being a complete SF analysis with the Majorana phases, it suggests that the Rep symmetry analysis of the neutrinoless double beta decay observable $m_{0\nu\beta\beta}$ can accommodate the Majorana phases. If it were the case, meaning of the revealed transformation properties of the phases may be investigated in the light of concrete models of the double beta decay mechanism with Majorana phases.

\begin{acknowledgments} 

The author would like to thank the referee of Acta Physica Polonica B for his/her enthusiasm in suggesting an examination for a possibility of placing the neutrinoless double beta decay observable in the symmetry context. He thanks Stephen Parke for useful communications on the vacuum SF equation from which our Symmetry Finder approach has been started. 

\end{acknowledgments}

\appendix  

\section{Vacuum SF symmetry revisited} 
\label{sec:vacuum-SF} 

We revisit the problem of the Rep symmetry in vacuum. We restrict ourselves to the 1-2 state exchange symmetry which includes the mass exchange $m^2_{1} \leftrightarrow m^2_{2}$. The SF equation in vacuum can be obtained from the one for the DMP or DMP-UV perturbation theory given in eq.~\eqref{SFeq-DMPUV} by dropping all the interaction terms and sending the matter-dressed angles to the vacuum ones: 
\begin{eqnarray} 
&&
\left[
\begin{array}{ccc}
e^{ i \tau } & 0 & 0 \\
0 & e^{ i \sigma } & 0 \\
0 & 0 & 1 \\
\end{array}
\right]
\left[
\begin{array}{c}
\nu_{e} \\
\nu_{\mu} \\
\nu_{\tau} \\
\end{array}
\right] 
= 
\left[
\begin{array}{ccc}
1 & 0 &  0  \\
0 & c_{23} & s_{23} e^{ i \sigma } \\
0 & - s_{23} e^{ - i \sigma } & c_{23} \\
\end{array}
\right] 
\left[
\begin{array}{ccc}
c_{13} & 0 & s_{13} e^{ i \tau } \\
0 & 1 & 0 \\
- s_{13} e^{ - i \tau } & 0 & c_{13} \\
\end{array}
\right] 
F U_{12} (\theta_{12}, \delta) R^{\dagger} 
R 
\left[
\begin{array}{c}
\nu_{1} \\
\nu_{2} \\
\nu_{3} \\
\end{array}
\right] 
\nonumber \\
&=& 
\left[
\begin{array}{ccc}
1 & 0 &  0  \\
0 & c_{23}^{\prime} & s_{23}^{\prime} \\
0 & - s_{23}^{\prime} & c_{23}^{\prime} \\
\end{array}
\right] 
\left[
\begin{array}{ccc}
c^{\prime}_{13}  & 0 & s^{\prime}_{13} \\
0 & 1 & 0 \\
- s^{\prime}_{13} & 0 & c^{\prime}_{13} \\
\end{array}
\right] 
U_{12} ( \theta_{12}^{\prime}, \delta + \xi)  
R 
\left[
\begin{array}{c}
\nu_{1} \\
\nu_{2} \\
\nu_{3} \\
\end{array}
\right].
\label{SFeq-vacuum} 
\end{eqnarray}
Assuming the solution $s_{23}^{\prime} = s_{23} e^{ i \sigma }$ and $s_{13}^{\prime} = s_{13} e^{ i \tau }$ and equalities of the cosine of them we reach to the first condition in eq.~\eqref{1st-eq-DMPUV}, but with $\psi$ replaced by $\theta_{12}$. The solutions of $\tau$, $\sigma$, $\xi$, $\alpha$, and $\beta$ for Symmetry IA, IB, etc.  are the same as in Table~\ref{tab:SF-solutions}. Then, it is obvious that the solution to the vacuum SF equation~\eqref{SFeq-vacuum} is given by the combination of Table~\ref{tab:SF-solutions} and Table~\ref{tab:DMPUV-symmetry}, the both taking the vacuum limit. In order not to leave ambiguities in this statement, we have placed the eight 1-2 state exchange symmetries in vacuum, denoted as Symmetry X-vacuum where X=IA, IB, $\cdot \cdot \cdot $, IVB, in Table~\ref{tab:SF-symmetry-vacuum} in section~\ref{sec:symmetry-vacuum}. 

One can show that Symmetry X-vacuum is a Hamiltonian symmetry. That is, the vacuum Hamiltonian $H_{ \text{vac} } \equiv U~\text{diag} (m^2_{1}/2E, m^2_{2}/2E, m^2_{3}/2E )~U^{\dagger}$ is invariant under Symmetry X transformations up to the rephasing factor, 
\begin{eqnarray}
&&
H_{ \text{vac} } 
\rightarrow 
\text{ Rep(X) } H_{ \text{vac} } \text{ Rep(X)}^{\dagger}, 
\label{Hvac-transf}
\end{eqnarray}
where Rep(X) are given in eq.~\eqref{Rep-II-III-IV}. 
Notice that the transformation property with the use of the same Rep(X) as in the Rep symmetry in DMP is not so trivial. Equation~\eqref{Hvac-transf} may look like eq.~\eqref{Hvac-transform} for the vacuum term in $H_{\text{\tiny VM}}$ in DMP. But, eq.~\eqref{Hvac-transf} is completely different from eq.~\eqref{Hvac-transform}. $c_{12}$ and $s_{12}$ transform in eq.~\eqref{Hvac-transf}, but they {\em do not} in eq.~\eqref{Hvac-transform}. $c_{\psi}$ and $s_{\psi}$ do transform in DMP but they are absent in eq.~\eqref{Hvac-transform}. 
In this sense the transformation property in eq.~\eqref{Hvac-transf} with the use of the same Rep(X) is, in fact, remarkable. 

We remark here that an explicit verification of invariance of the vacuum oscillation probability $P(\nu_\mu \rightarrow \nu_e)$ and $P(\nu_\mu \rightarrow \nu_\tau)$, for example, under the transformations of Symmetry X is very simple. 
$\theta_{12}$ comes in with the forms $c^2_{12} s^2_{12}$ which are invariant under $c_{12} \leftrightarrow \pm s_{12}$, or separately as $c^2_{12}$ and $s^2_{12}$ but synchronized with $\lambda_{1} \leftrightarrow \lambda_{2}$. Similarly, $s_{23}$ and $s_{13}$ appear as squared, except for in the $\delta$ dependent terms, which may provide the unique source for the trouble. 
However, one can show by explicit calculation that the $\delta$ dependent terms can be written in the forms involving $J_{rc} \sin \delta$, $J_{rc} \cos \delta$, $J_{r} \sin \delta$, or $J_{r} \cos \delta$, which are odd under all Symmetry X. The minus sign is cancelled either by the explicit minus sign or the triple-sine combination in the T-odd term. Here, we have introduced the new notations 
\begin{eqnarray}
&&
J_{rc} \equiv c_{23} s_{23} c^2_{13} s_{13} c_{12} s_{12},  
\nonumber \\ 
&& 
J_{r} \equiv c_{23} s_{23} s_{13} c_{12} s_{12}. 
\label{Jr-Jrc-def}
\end{eqnarray}
In $P(\nu_\mu \rightarrow \nu_e)$, the Jarlskog combination $J_{rc}$~\cite{Jarlskog:1985ht} suffices, but in $P(\nu_\mu \rightarrow \nu_\tau)$ we need $J_{r}$ as well for the $\cos \delta$ terms~\cite{Asano:2011nj}. 

\section{$H$ matrix elements in first order: Summary}
\label{sec:H-elements} 

In ref.~\cite{Minakata:2021nii}, we have defined the $G$ matrix, the same one as in eq.~\eqref{G-G2-def}, in a two-step form 
\begin{eqnarray} 
G &=& 
U_{12} (\psi, \delta)^{\dagger} K U_{12} (\psi, \delta), 
\nonumber \\
K &=&
U^\dagger_{13}(\phi) U_{23}^{\dagger} (\theta_{23}) 
A 
U_{23} (\theta_{23}) U_{13}(\phi), 
\label{G-K-def}
\end{eqnarray} 
where $A$ is defined in eq.~\eqref{A-A2-def}. We follow the same style in this paper, except that we use the $H$ matrix elements to display the $G$ matrix elements, $G = D H D^{\dagger}$, $D \equiv \text{diag} (e^{ i \delta}, 1, e^{ i \delta} )$ as defined in eq.~\eqref{Hmatrix-def}.

The explicit expressions of the $H$ matrix elements through the $K$ matrix elements are given by 
\begin{eqnarray} 
&& 
H_{11} 
=
c^2_\psi K_{11} + s^2_\psi K_{22} 
- c_\psi s_\psi \left( e^{ i \delta} K_{21} + e^{ - i \delta} K_{12} \right), 
\nonumber \\
&& 
H_{12} 
=
\left[ c_\psi s_\psi \left( K_{11} - K_{22} \right) 
+ \left( c^2_\psi e^{ - i \delta} K_{12} - s^2_\psi e^{ i \delta} K_{21} \right) \right], 
\nonumber \\
&& 
H_{13} 
= 
\left( c_\psi K_{13} - s_\psi e^{ i \delta} K_{23} \right), 
\nonumber \\
&& 
H_{21} 
= 
\left[ c_\psi s_\psi \left( K_{11} - K_{22} \right) 
+ \left( c^2_\psi e^{ i \delta} K_{21} - s^2_\psi e^{ - i \delta} K_{12} \right) \right], 
\nonumber \\
&& 
H_{22} 
= 
s^2_\psi K_{11} + c^2_\psi K_{22} 
+ c_\psi s_\psi \left( e^{ i \delta} K_{21} + e^{ - i \delta} K_{12} \right), 
\nonumber \\
&& 
H_{23} 
= 
\left( s_\psi K_{13} + c_\psi e^{ i \delta} K_{23} \right), 
\nonumber \\
&& 
H_{31} 
= 
\left( c_\psi K_{31} - s_\psi e^{ - i \delta} K_{32} \right), 
\nonumber \\
&& 
H_{32} 
= 
\left( s_\psi K_{31} + c_\psi e^{ - i \delta} K_{32} \right), 
\nonumber \\
&& 
H_{33} 
= 
K_{33}. 
\label{Hij-elements}
\end{eqnarray}
The $K$ matrix elements are given by 
\begin{eqnarray} 
K_{11} &=& 
2 c^2_{\phi} \widetilde{\alpha}_{ee} \left( 1 - \frac{ a }{ b } \right) 
+ 2 s^2_{\phi} 
\left[
s_{23}^2 \widetilde{\alpha}_{\mu \mu} + c_{23}^2 \widetilde{\alpha}_{\tau \tau} + c_{23} s_{23} \mbox{Re} \left( \widetilde{\alpha}_{\tau \mu} \right) 
\right], 
\nonumber \\
&-& 
2 c_{\phi} s_{\phi} 
\mbox{Re} \left( s_{23} \widetilde{\alpha}_{\mu e} + c_{23} \widetilde{\alpha}_{\tau e} \right) 
\nonumber \\
K_{12} &=& 
c_{\phi} \left( c_{23} \widetilde{\alpha}_{\mu e}^* - s_{23} \widetilde{\alpha}_{\tau e}^* \right) 
- s_{\phi} 
\left[ 2 c_{23} s_{23} ( \widetilde{\alpha}_{\mu \mu} - \widetilde{\alpha}_{\tau \tau} ) + c_{23}^2 \widetilde{\alpha}_{\tau \mu} - s_{23}^2 \widetilde{\alpha}_{\tau \mu}^* \right] 
= \left( K_{21} \right)^*, 
\nonumber \\
K_{13} &=& 
2 c_{\phi} s_{\phi} 
\left[
\widetilde{\alpha}_{ee} \left( 1 - \frac{ a }{ b } \right) 
- \left( s_{23}^2 \widetilde{\alpha}_{\mu \mu} + c_{23}^2 \widetilde{\alpha}_{\tau \tau} \right) 
\right]
\nonumber \\
&+& 
c^2_{\phi} \left( s_{23} \widetilde{\alpha}_{\mu e}^* + c_{23} \widetilde{\alpha}_{\tau e}^* \right) 
- s^2_{\phi} \left( s_{23} \widetilde{\alpha}_{\mu e} + c_{23} \widetilde{\alpha}_{\tau e} \right) 
- 2 c_{23} s_{23} c_{\phi} s_{\phi} 
\mbox{Re} \left( \widetilde{\alpha}_{\tau \mu} \right) 
= \left( K_{31} \right)^*, 
\nonumber \\
K_{22} &=& 
2 \left[ 
c_{23}^2 \widetilde{\alpha}_{\mu \mu} + s_{23}^2 \widetilde{\alpha}_{\tau \tau} - c_{23} s_{23} \mbox{Re} \left( \widetilde{\alpha}_{\tau \mu} \right) 
\right], 
\nonumber \\
K_{23} &=& 
s_{\phi} \left( c_{23} \widetilde{\alpha}_{\mu e} - s_{23} \widetilde{\alpha}_{\tau e} \right) 
+ c_{\phi} 
\left[ 2 c_{23} s_{23} ( \widetilde{\alpha}_{\mu \mu} - \widetilde{\alpha}_{\tau \tau} ) + c_{23}^2 \widetilde{\alpha}_{\tau \mu}^* - s_{23}^2 \widetilde{\alpha}_{\tau \mu} \right] 
= \left( K_{32} \right)^*, 
\nonumber \\
K_{33} &=& 
2 s^2_{\phi} \widetilde{\alpha}_{ee} \left( 1 - \frac{ a }{ b } \right) 
+ 2 c^2_{\phi} 
\left[
s_{23}^2 \widetilde{\alpha}_{\mu \mu} + c_{23}^2 \widetilde{\alpha}_{\tau \tau} + c_{23} s_{23} \mbox{Re} \left( \widetilde{\alpha}_{\tau \mu} \right) 
\right] 
\nonumber \\
&+&
2 c_{\phi} s_{\phi} \mbox{Re} 
\left( s_{23} \widetilde{\alpha}_{\mu e} + c_{23} \widetilde{\alpha}_{\tau e} \right). 
\label{Kij-elements}
\end{eqnarray}
Notice that the $K$ matrix elements are free from the $\nu$SM vacuum mixing angles apart from $\theta_{23}$ and contain no $\delta$ in the SOL convention. $K_{ji} = K_{ij}^*$ and $K_{ii}$ are real. 

\section{First order correction in the wave function in quantum mechanics}
\label{sec:QM-1st-order} 

In ref.~\cite{Minakata:2015gra}, the numerator of eq.~\eqref{nu-bar-first-order} is written as $\bar{H}^{(1)}_{ji}$, not $\left( \bar{H}^{(1)}_{ji} \right)^*$ which is equivalent to eq.~\eqref{nu-bar-first-order}. The complex conjugate can be disregarded in ref.~\cite{Minakata:2015gra} because the $\bar{H}^{(1)}$ elements are real due to the usage of the ATM convention of the $U$ matrix. But, in our case and in general it must be kept. Though this point must be obvious from eq.~(26) in ref.~\cite{Minakata:1998bf}, we want to write a pedagogical note here not to leave  ambiguity on this point.

The non-degenerate stationary state perturbation theory formulated for the ket state reads 
\begin{equation}
| \nu_{i} \rangle^{(1)} = | \nu_i \rangle^{(0)} 
+ \sum_{j\neq i} \frac{ ( H_{1} )_{ij} }{E_i^{(0)}-E_j^{(0)}}
| \nu_{j} \rangle^{(0)},
\end{equation}
with $E_i^{(0)}$ being the zeroth-order eigenvalues and $H_{1}$ perturbed Hamiltonian. In terms of bra state it can be written as 
\begin{eqnarray} 
\langle \nu_{i} | ^{(1)}
&=& \langle  \nu_{i} | ^{(0)} 
+ \sum_{j\neq i} \langle  \nu_{j} | ^{(0)} \left( H_{1} ^{\dagger} \right) _{ji}
\frac{ 1 }{E_i^{(0)}-E_j^{(0)}} 
\nonumber \\
&=&
\langle  \nu_{i} | ^{(0)} 
+ \sum_{j\neq i} \langle  \nu_{j} | ^{(0)} 
\frac{ ( H_{1} )_{ij}  }{E_i^{(0)}-E_j^{(0)}}.
\label{bra-correction}
\end{eqnarray}
Since the wave function we have used in this paper corresponds to the bra state, we obtain eq.~\eqref{nu-bar-first-order}. 
In fact, if the complex conjugation is missed, we obtain a wrong expression of $W_{ \nu\text{SM} }$ in eq.~\eqref{W-DMP-UV} which is different from the correct one by $\delta \rightarrow - \delta$. Notice that our expression in eq.~\eqref{W-DMP-UV} is consistent with the one in ref.~\cite{Minakata:2021dqh}, which is obtained by the rephasing from the ATM convention result. 

\section{The probabilities $P(\nu_{\mu} \rightarrow \nu_{e})^{(1)}_{ \text{EV} }$ and 
$P(\nu_{\mu} \rightarrow \nu_{e})^{(1)}_{ \text{UV} }$} 
\label{sec:UV-probabilities} 

\subsection{Unitary evolution part $P(\nu_{\mu} \rightarrow \nu_{e})^{(1)} _{ \text{EV} }$}
\label{sec:EV-part}

The unitary evolution part of the oscillation probability $P(\nu_{\mu} \rightarrow \nu_{e})^{(1)} _{ \text{EV} }$ is given in a form separated to the T-even and T-odd parts: 
$P(\nu_{\mu} \rightarrow \nu_{e})^{(1)} _{ \text{EV} } 
= P(\nu_{\mu} \rightarrow \nu_{e})^{(1)} _{ \text{EV} } \vert_{ \text{T-even} } 
+ P(\nu_{\mu} \rightarrow \nu_{e})^{(1)} _{ \text{EV} } \vert_{ \text{T-odd} }$. 
\begin{eqnarray}
&&
P(\nu_{\mu} \rightarrow \nu_{e})^{(1)} _{ \text{EV} } \vert_{ \text{T-even} }
= \biggl\{
2 c^2_{\phi} 
\left[ \left( c^2_{23} - s^2_{23} s^2_{\phi} \right) \sin 4\psi 
+ \sin 2\theta_{23} s_{\phi} \cos \delta \cos 4\psi \right] 
\mbox{Re} \left( H_{12} \right) 
\nonumber \\
&-&
2 \sin 2\theta_{23} c^2_{\phi} s_{\phi} \cos 2\psi \sin \delta 
\mbox{Im} \left( H_{12} \right) 
\biggr\} 
\left( \frac{ b }{ \lambda_2 - \lambda_1 } \right) 
\sin^2 \frac{( \lambda_2 - \lambda_1) x}{4E} 
\nonumber \\
&-& 
2 
\biggl\{ 
c_{\psi} \left[ c_{\phi} s_{\phi} \sin 2\psi 
\left( c^2_{23} + s^2_{23} \cos 2\phi \right) 
+ \sin 2\theta_{23} c_{\phi} \cos \delta 
\left( s^2_{\phi} c^2_{\psi} + s^2_{\psi} \cos 2\phi \right) \right] 
\mbox{Re} \left( H_{23} \right) 
\nonumber \\
&+& 
\sin 2\theta_{23} c_{\phi} c_{\psi} \left( s^2_{\phi} + c^2_{\phi} s^2_{\psi} \right) 
\sin \delta \mbox{Im} \left( H_{23} \right) 
\biggr\} 
\left( \frac{ b }{ \lambda_3 - \lambda_2 } \right) 
\sin^2 \frac{( \lambda_2 - \lambda_1) x}{4E} 
\nonumber \\
&-& 
2 \biggl\{
s_{\psi} \left[ c_{\phi} s_{\phi} \sin 2\psi 
\left( c^2_{23} + s^2_{23} \cos 2\phi \right) 
- \sin 2\theta_{23} c_{\phi} \cos \delta 
\left( s^2_{\phi} s^2_{\psi}  + c^2_{\psi} \cos 2\phi \right) 
\right] 
\mbox{Re} \left( H_{13} \right) 
\nonumber \\
&-& 
\sin 2\theta_{23} c_{\phi} s_{\psi} \sin \delta 
\left( s^2_{\phi} + c^2_{\phi} c^2_{\psi} \right) 
\mbox{Im} \left( H_{13} \right) 
\biggr\} 
\left( \frac{ b }{ \lambda_3 - \lambda_1 } \right) 
\sin^2 \frac{( \lambda_2 - \lambda_1) x}{4E} 
\nonumber \\
&& 
\hspace{-10mm}
- 4 s_{23} c^2_{\phi} s_{\phi} 
\biggl\{ 
\left[ c_{23} \cos 2\psi \cos \delta  
- s_{23} s_{\phi} \sin 2\psi \right] \mbox{Re} \left( H_{12} \right) 
- c_{23} \sin \delta \mbox{Im} \left( H_{12} \right) 
\biggr\} 
\left( \frac{ b }{ \lambda_2 - \lambda_1 } \right) 
\sin^2 \frac{( \lambda_3 - \lambda_2) x}{4E} 
\nonumber \\
&& 
\hspace{-10mm}
+ 4 s_{23} c^2_{\phi} s_{\phi} 
\biggl\{
\left[ c_{23} \cos 2\psi \cos \delta  
- s_{23} s_{\phi} \sin 2\psi \right]  
\mbox{Re} \left( H_{12} \right)  
- c_{23} \sin \delta \mbox{Im} \left( H_{12} \right) 
\biggr\} 
\left( \frac{ b }{ \lambda_2 - \lambda_1 } \right) 
\sin^2 \frac{( \lambda_3 - \lambda_1) x}{4E} 
\nonumber \\
&-& 
2 c_{\phi} \biggl\{ 
\left[ 
2 s_{\phi} s_{\psi} 
\left\{ c^2_{23} c^2_{\psi} - s^2_{23} \cos 2\phi ( 1 + s^2_{\psi} ) \right\} 
- \sin 2\theta_{23} c_{\psi} \cos \delta 
\left\{ s^2_{\phi} ( 1 + s^2_{\psi} ) - s^2_{\psi} \cos 2\phi \right\} 
\right] 
\mbox{Re} \left( H_{23} \right) 
\nonumber \\
&-&
\sin 2\theta_{23} c_{\psi} \sin \delta ( s^2_{\phi} - c^2_{\phi} s^2_{\psi} ) 
\mbox{Im} \left( H_{23} \right) 
\biggr\} 
\left( \frac{ b }{ \lambda_3 - \lambda_2 } \right) 
\sin^2 \frac{( \lambda_3 - \lambda_2) x}{4E} 
\nonumber \\
&-& 
2 c_{\phi} \biggl\{
\left[ 
2 s_{\phi} c_{\psi} 
\left\{ c^2_{23} s^2_{\psi} - s^2_{23} \cos 2\phi ( 1 + c^2_{\psi} ) \right\} 
+ \sin 2\theta_{23} s_{\psi} \cos \delta 
\left\{ s^2_{\phi} ( 1 + c^2_{\psi} ) - c^2_{\psi} \cos 2\phi \right\} 
\right] 
\mbox{Re} \left( H_{13} \right) 
\nonumber \\
&+& 
\sin 2\theta_{23} s_{\psi} \sin \delta ( s^2_{\phi} - c^2_{\phi} c^2_{\psi} ) 
\mbox{Im} \left( H_{13} \right) 
\biggr\} 
\left( \frac{ b }{ \lambda_3 - \lambda_1 } \right) 
\sin^2 \frac{( \lambda_3 - \lambda_1) x}{4E} 
\nonumber \\
&-& 
2 c_{\phi} s_{\psi} \biggl\{ 
\left[ - 2 \left( c^2_{23} + s^2_{23} \cos 2\phi \right) 
s_{\phi} c_{\psi} s_{\psi} 
+ \sin 2\theta_{23} \cos \delta 
\left( s^2_{\phi} s^2_{\psi} + \cos 2\phi c^2_{\psi} \right) \right] 
\mbox{Re} \left( H_{13} \right) 
\nonumber \\
&+& 
\sin 2\theta_{23} \sin \delta 
\left( s^2_{\phi} + c^2_{\phi} c^2_{\psi} \right) 
\mbox{Im} \left( H_{13} \right) 
\biggr\} 
\left( \frac{ b }{ \lambda_3 - \lambda_1 } \right) 
\sin^2 \frac{( \lambda_3 - \lambda_2) x}{4E} 
\nonumber \\
&+& 
2 c_{\phi} c_{\psi} \biggl\{ 
\left[ 
2 \left( c^2_{23} + s^2_{23} \cos 2\phi \right) s_{\phi} c_{\psi} s_{\psi} 
+ \sin 2\theta_{23} \cos \delta 
\left( s^2_{\phi} c^2_{\psi} + s^2_{\psi} \cos 2\phi \right) 
\right] 
\mbox{Re} \left( H_{23} \right) 
\nonumber \\
&+& 
\sin 2\theta_{23} \sin \delta 
\left( s^2_{\phi} + c^2_{\phi} s^2_{\psi} \right) 
\mbox{Im} \left( H_{23} \right) 
\biggr\} 
\left( \frac{ b }{ \lambda_3 - \lambda_2 } \right) 
\sin^2 \frac{( \lambda_3 - \lambda_1) x}{4E}. 
\label{P-mue-EV-Teven}
\end{eqnarray}
\begin{eqnarray} 
&&
P(\nu_{\mu} \rightarrow \nu_{e})^{(1)} _{ \text{EV} } \vert_{ \text{T-odd} } 
= 8 \biggl[ 
- c_{23} s_{23} c^2_{\phi} s_{\phi} 
\left[ \cos \delta \mbox{Im} \left( H_{12} \right) 
+ \cos 2\psi \sin \delta \mbox{Re} \left( H_{12} \right) \right] 
\left( \frac{ b }{ \lambda_2 - \lambda_1 } \right) 
\nonumber \\
&& 
\hspace{-16mm}
+ c_{\phi} c_\psi \biggl\{ 
\left[ - \cos 2\theta_{23} s_{\phi} c_{\psi} s_{\psi} 
+ c_{23} s_{23} \cos \delta \left( s^2_\psi - s^2_{\phi} c^2_\psi \right) \right] 
\mbox{Im} \left( H_{23} \right) 
+ c_{23} s_{23} \sin \delta 
\left( s^2_{\phi} - c^2_{\phi} s^2_{\psi} \right) 
\mbox{Re} \left( H_{23} \right) 
\biggr\} 
\left( \frac{ b }{ \lambda_3 - \lambda_2 } \right) 
\nonumber \\
&& 
\hspace{-16mm}
+ c_{\phi} s_\psi 
\biggl\{ 
\left[ \cos 2\theta_{23} s_{\phi} c_\psi s_\psi 
+ c_{23} s_{23} \cos \delta 
\left( c^2_\psi - s^2_{\phi} s^2_\psi \right) \right] 
\mbox{Im} \left( H_{13} \right) 
+ c_{23} s_{23} \sin \delta 
\left( s^2_{\phi} - c^2_{\phi} c^2_{\psi} \right) 
\mbox{Re} \left( H_{13} \right) 
\biggr\} 
\left( \frac{ b }{ \lambda_3 - \lambda_1 } \right) 
\biggr]
\nonumber \\
&\times& 
\sin \frac{( \lambda_2 - \lambda_1) x}{4E} 
\sin \frac{( \lambda_3 - \lambda_1) x}{4E} 
\sin \frac{( \lambda_3 - \lambda_2) x}{4E}. 
\label{P-mue-EV-Todd}
\end{eqnarray}

\subsection{Genuine non-unitary part $P(\nu_{\mu} \rightarrow \nu_{e})^{(1)} _{ \text{UV} }$}
\label{sec:UV-part}

The genuine non-unitary part of the oscillation probability reads 
\begin{eqnarray} 
&&
P(\nu_\mu \rightarrow \nu_e)^{(1)}_{ \text{UV} }
= 
- 2 \left( \widetilde{\alpha}_{ee} + \widetilde{\alpha}_{\mu \mu} \right) 
\nonumber \\
&\times& 
\biggl[ 
s^2_{23} \sin^2 2\phi 
\biggl\{ 
s^2_{\psi} \sin^2 \frac{( \lambda_3 - \lambda_2) x}{4E} 
+ c^2_{\psi} \sin^2 \frac{( \lambda_3 - \lambda_1) x}{4E} 
\biggr\} 
+ c^2_{\phi} \sin^2 2\psi \left( c^2_{23} - s^2_{23} s^2_{\phi} \right) 
\sin^2 \frac{( \lambda_2 - \lambda_1) x}{4E} 
\nonumber \\
&+& 
4 J_{mr} \cos \delta 
\biggl\{ 
- \sin^2 \frac{( \lambda_3 - \lambda_2) x}{4E} 
+ \sin^2 \frac{( \lambda_3 - \lambda_1) x}{4E} 
+ \cos 2\psi \sin^2 \frac{( \lambda_2 - \lambda_1) x}{4E} 
\biggr\} 
\nonumber \\
&-& 
8 J_{mr} \sin \delta 
\sin \frac{( \lambda_2 - \lambda_1) x}{4E} 
\sin \frac{( \lambda_3 - \lambda_1) x}{4E} 
\sin \frac{( \lambda_3 - \lambda_2) x}{4E} 
\biggr] 
\nonumber \\
&+& 
2 c_{23} c_{\phi} \sin 2\psi 
\mbox{Re} \left( \widetilde{\alpha}_{\mu e} e^{ i \delta} \right) 
\biggl[
s^2_{\phi} \biggl\{ 
\sin^2 \frac{( \lambda_3 - \lambda_2) x}{4E} 
- \sin^2 \frac{( \lambda_3 - \lambda_1) x}{4E} 
\biggr\} 
+ c^2_{\phi} \cos 2\psi 
\sin^2 \frac{( \lambda_2 - \lambda_1) x}{4E} 
\biggr] 
\nonumber \\
&+& 
c_{23} c_{\phi} \sin 2\psi 
\mbox{Im} \left( \widetilde{\alpha}_{\mu e} e^{ i \delta} \right) 
\biggl[ 
s^2_{\phi} \biggl\{
\sin \frac{( \lambda_3 - \lambda_2) x}{2E} 
- \sin \frac{( \lambda_3 - \lambda_1) x}{2E} 
\biggr\} 
- c^2_{\phi} 
\sin \frac{( \lambda_2 - \lambda_1) x}{2E} 
\biggr] 
\nonumber \\
&+& 
s_{23} \sin 2\phi 
\left[ \cos \delta \mbox{Re} \left( \widetilde{\alpha}_{\mu e} e^{ i \delta} \right) 
+ \sin \delta \mbox{Im} \left( \widetilde{\alpha}_{\mu e} e^{ i \delta} \right) \right] 
\nonumber \\
&\times&
\biggl[ 
2 \cos 2\phi 
\biggl\{ 
s^2_{\psi} \sin^2 \frac{( \lambda_3 - \lambda_2) x}{4E} 
+ c^2_{\psi} \sin^2 \frac{( \lambda_3 - \lambda_1) x}{4E} 
\biggr\} 
- c^2_{\phi} \sin^2 2\psi 
\sin^2 \frac{( \lambda_2 - \lambda_1) x}{4E} 
\biggr]
\nonumber \\
&+& 
s_{23} \sin 2\phi 
\left[ \sin \delta \mbox{Re} \left( \widetilde{\alpha}_{\mu e} e^{ i \delta} \right) 
- \cos \delta \mbox{Im} \left( \widetilde{\alpha}_{\mu e} e^{ i \delta} \right) \right] 
\biggl\{ 
s^2_{\psi} 
\sin \frac{( \lambda_3 - \lambda_2) x}{2E} 
+ c^2_{\psi} 
\sin \frac{( \lambda_3 - \lambda_1) x}{2E} 
\biggr\}, 
\label{P-mue-UV}
\end{eqnarray} 
where $J_{mr} \equiv c_{23} s_{23} c^2_{\phi} s_{\phi} c_{\psi} s_{\psi}$ is the Jarlskog factor~\cite{Jarlskog:1985ht} in matter. 

\section{$(b x) / 2E$ term in $P(\nu_\mu \rightarrow \nu_e)^{(1)} _{ \text{EV} }$}
\label{sec:bx-term} 

The expressions of the probabilities $P(\nu_\mu \rightarrow \nu_e)^{(1)} _{ \text{EV} }$ and $P(\nu_\mu \rightarrow \nu_e)^{(1)} _{ \text{UV} }$ given in Appendix~\ref{sec:UV-probabilities} are based on the renormalized treatment of the eigenvalues $\lambda_{i} = \lambda_{i}^{ \nu\text{SM} } + b H_{ii}$ in eq.~\eqref{ren-eigenvalues}. To compare the present expression of $P(\nu_\mu \rightarrow \nu_e)^{(1)} _{ \text{EV} }$ to the one from the un-renormalized treatment of ref.~\cite{Minakata:2021nii}, we calculate the first-order corrections that are produced when we expand the eigenvalues $\lambda_{i}$ in $P(\nu_\mu \rightarrow \nu_e)^{(0)} _{ \nu\text{SM} }$ given for example in ref.~\cite{Minakata:2020oxb}. Notice that this is the only source of the $(b x) / 2E$ term in first order in perturbation. In the other parts of the theory the eigenvalue expansion merely induces the second order terms. 

We briefly describe the computation for it with use of the $S$ matrix method because it is easier. We expand the eigenvalues to first order 
\begin{eqnarray} 
&&
e^{ - i \frac{ \lambda_{i} }{2E} x } 
= 
e^{ - i h_{i} x } 
+ \left( - i \frac{bx}{2E} \right) H_{ii} e^{ - i h_{i} x },
\label{lambda-expand}
\end{eqnarray} 
where we have defined 
\begin{eqnarray} 
&&
h_{i} \equiv 
\frac{ \lambda_{i}^{ \nu\text{SM} } }{2E} ~~~~~~~(i=1,2,3).
\label{hi-def}
\end{eqnarray} 
Then, the flavor basis $S$ matrix element can be expanded to first order as 
\begin{eqnarray} 
&&
S^{(0)}_{e \mu} \left( \lambda_{i} \right) 
= 
S^{(0)}_{e \mu} \left(\lambda_{i}^{ \nu\text{SM} } \right)
+ S^{(1)}_{e \mu} \left( H_{ii} \right), 
\label{S-emu-expand}
\end{eqnarray} 
where 
\begin{eqnarray} 
&&
S^{(0)}_{e \mu} \left(\lambda_{i}^{ \nu\text{SM} } \right) 
=
c_{23} c_{\phi} e^{ i \delta} c_\psi s_\psi 
\left( e^{ - i h_{2} x } - e^{ - i h_{1} x } \right) 
+ s_{23} c_{\phi} s_{\phi} 
\left[ e^{ - i h_{3} x } - \left( c^2_\psi e^{ - i h_{1} x } + s^2_\psi e^{ - i h_{2} x } \right) \right], 
\nonumber \\
&&
S^{(1)}_{e \mu} \left( H_{ii} \right) 
=
\left( - i \frac{bx}{2E} \right) 
\biggl\{ 
c_{23} c_{\phi} e^{ i \delta} c_\psi s_\psi 
\left( H_{22} e^{ - i h_{2} x } - H_{11} e^{ - i h_{1} x } \right) 
\nonumber \\
&&
\hspace{30mm}
+ s_{23} c_{\phi} s_{\phi} 
\left( H_{33} e^{ - i h_{3} x } - c^2_\psi H_{11} e^{ - i h_{1} x } - s^2_\psi H_{22} e^{ - i h_{2} x } \right) 
\biggr\}.
\label{S-emu-expand-result}
\end{eqnarray}
The interference between $S^{(0)}_{e \mu} \left(\lambda_{i}^{ \nu\text{SM} } \right)$ and $S^{(1)}_{e \mu} \left( H_{ii} \right)$ produces the required first-order correction, $(b x) / 2E$ term in $P(\nu_\mu \rightarrow \nu_e)^{(1)} _{ \text{EV} }$, as 
\begin{eqnarray} 
&& 
2 \mbox{Re} \left[ 
\left\{ S^{(0)}_{e \mu} \left( \lambda_{i}^{ \nu\text{SM} } \right) \right\}^* 
S^{(1)}_{e \mu} \left( H_{ii} \right) \right] 
\nonumber \\
&=& 
2 \left( \frac{bx}{2E} \right) 
\biggl\{ 
\left( H_{22} - H_{11} \right) 
\nonumber \\
&\times& 
\left[ 
2 J_{mr} \sin \delta \sin^2 \frac{ ( h_{2} - h_{1} ) x }{2} 
+ \left\{ c^2_{\phi} c^2_\psi s^2_\psi 
\left( c^2_{23} - s^2_{23} s^2_{\phi} \right) 
+ J_{mr} \cos \delta \cos 2\psi \right\} \sin ( h_{2} - h_{1} ) x 
\right] 
\nonumber \\
&+& 
\left( H_{33} - H_{22} \right) 
\left[ 
2 J_{mr} \sin \delta \sin^2 \frac{ ( h_{3} - h_{2} ) x }{2} 
+ \left( s^2_{23} c^2_{\phi} s^2_{\phi} s^2_\psi - J_{mr} \cos \delta \right) 
\sin ( h_{3} - h_{2} ) x 
\right] 
\nonumber \\
&+& 
\left( H_{33} - H_{11} \right) 
\left[ 
- 2 J_{mr} \sin \delta \sin^2 \frac{ ( h_{3} - h_{1} ) x }{2} 
+ \left( s^2_{23} c^2_{\phi} s^2_{\phi} c^2_\psi + J_{mr} \cos \delta \right) 
\sin ( h_{3} - h_{1} ) x 
\right] 
\biggr\}. 
\nonumber \\
\label{bx-correction}
\end{eqnarray}
Using the expression of $H_{ii}$ in terms of the $K_{ij}$ variables as given in Appendix~\ref{sec:H-elements}, it can be shown that eq.~\eqref{bx-correction} precisely reproduces eq.~(4.9) in ref.~\cite{Minakata:2021nii}. 

\section{An explicit proof of the identity eq.~\eqref{identity} for Symmetry IV}
\label{sec:proof-IV}

It is easy to prove the first equality in eq.~\eqref{identity}, $C [12]=$ Rep(X), by explicit calculation of $C [12]$ for all the eight DMP symmetries with use of the values of the parameters $\alpha$, $\beta$ etc. for each symmetry, which have the upper and lower signs as given in Table~\ref{tab:SF-solutions}. To show how the second equality holds, we examine the cases of symmetries IVA and IVB and compute $[V^{\prime} R V^{\dagger}]$. Inserting $C [12]=$Rep(IV) in eq.~\eqref{VRVdagger} we obtain 
\begin{eqnarray} 
&& 
V^{(0)} ( \theta_{23}^{\prime}, \phi^{\prime}, \psi^{\prime}, \delta^{\prime} ) 
R 
\left[ V^{(0)} ( \theta_{23}, \phi, \psi, \delta) \right]^{\dagger} 
\nonumber \\
&=& 
\left[
\begin{array}{ccc}
1 & 0 &  0  \\
0 & c_{23}^{\prime} & s_{23}^{\prime} \\
0 & - s_{23}^{\prime} & c_{23}^{\prime} \\
\end{array}
\right] 
\left[
\begin{array}{ccc}
c_{\phi}^{\prime} & 0 & s_{\phi}^{\prime} \\
0 & 1 & 0 \\
- s_{\phi}^{\prime} & 0 & c_{\phi}^{\prime} \\
\end{array}
\right] 
\left[
\begin{array}{ccc}
- 1 & 0 & 0 \\
0 & -1 & 0 \\
0 & 0 & 1 \\
\end{array}
\right] 
\left[
\begin{array}{ccc}
c_{\phi} & 0 & - s_{\phi} \\
0 & 1 & 0 \\
s_{\phi} & 0 & c_{\phi} \\
\end{array}
\right] 
\left[
\begin{array}{ccc}
1 & 0 &  0  \\
0 & c_{23} & - s_{23} \\
0 & s_{23} & c_{23} \\
\end{array}
\right] 
= 
\left[
\begin{array}{ccc}
-1 & 0 & 0 \\
0 & -1 & 0 \\
0 & 0 & 1 \\
\end{array}
\right], 
\nonumber \\
\label{equality-IX}
\end{eqnarray}
thanks to the properties that 
\begin{eqnarray} 
&& 
\hspace{-10mm}
\left[
\begin{array}{ccc}
c_{\phi}^{\prime} & 0 & s_{\phi}^{\prime} \\
0 & 1 & 0 \\
- s_{\phi}^{\prime} & 0 & c_{\phi}^{\prime} \\
\end{array}
\right] 
\left[
\begin{array}{ccc}
-1 & 0 & 0 \\
0 & -1 & 0 \\
0 & 0 & 1 \\
\end{array}
\right] 
\left[
\begin{array}{ccc}
c_{\phi} & 0 & - s_{\phi} \\
0 & 1 & 0 \\
s_{\phi} & 0 & c_{\phi} \\
\end{array}
\right] 
= 
\left[
\begin{array}{ccc}
1 & 0 &  0  \\
0 & c_{23}^{\prime} & s_{23}^{\prime} \\
0 & - s_{23}^{\prime} & c_{23}^{\prime} \\
\end{array}
\right] 
\left[
\begin{array}{ccc}
-1 & 0 & 0 \\
0 & -1 & 0 \\
0 & 0 & 1 \\
\end{array}
\right] 
\left[
\begin{array}{ccc}
1 & 0 &  0  \\
0 & c_{23} & - s_{23} \\
0 & s_{23} & c_{23} \\
\end{array}
\right] 
= 
\left[
\begin{array}{ccc}
-1 & 0 & 0 \\
0 & -1 & 0 \\
0 & 0 & 1 \\
\end{array}
\right]. 
\nonumber \\
\end{eqnarray}
Notice that $s_{\phi}^{\prime} = - s_{\phi}$ and $s_{23}^{\prime} = - s_{23}$ in Symmetry IVA and IVB. What happens is that when Rep(IV) move to the left to get out to the front, it remedies the transformed parameters into the un-transformed parameters through passage, which occurs for both $s_{\phi}$ and $s_{23}$ in Symmetry IV. 
Similarly, one can prove the second equality in eq.~\eqref{identity} for all the remaining symmetries X=I, II, and III. For Symmetry I, Rep(I) = 1, and no transformation on $s_{23}$ and $s_{\phi}$ is needed. For Symmetry II (III) the ``sign remedy'' occurs only for $s_{23}$ ($s_{\phi}$).


\begin{thebibliography}{99}

\bibitem{Itzykson:1980rh}
C.~Itzykson and J.~B.~Zuber, ``Quantum Field Theory,''
McGraw-Hill, 1980,
ISBN 978-0-486-44568-7

\bibitem{Coleman:1985rnk}
S.~Coleman,
``Aspects of Symmetry: Selected Erice Lectures,'' \\
https:// doi.org/10.1017/CBO9780511565045 

\bibitem{Minakata:2021dqh}
H.~Minakata,
``Symmetry finder: A method for hunting symmetry in neutrino oscillation,''
Phys. Rev. D \textbf{104} (2021) no.7, 075024
doi:10.1103/PhysRevD.104.075024
[arXiv:2106.11472 [hep-ph]].

\bibitem{Minakata:2022yvs}
H.~Minakata,
``Symmetry in Neutrino Oscillation in Matter: New Picture and the \ensuremath{\nu}SM\textendash{}Non-Unitarity Interplay,''
Symmetry \textbf{14} (2022) no.12, 2581
doi:10.3390/sym14122581
[arXiv:2210.09453 [hep-ph]].

\bibitem{Minakata:2021goi}
H.~Minakata,
``Symmetry Finder applied to the 1\textendash{}3 mass eigenstate exchange symmetry,''
Eur. Phys. J. C \textbf{81} (2021) no.11, 1021
doi:10.1140/epjc/s10052-021-09810-5
[arXiv:2107.12086 [hep-ph]]. 

\bibitem{Denton:2016wmg}
P.~B.~Denton, H.~Minakata and S.~J.~Parke,
``Compact Perturbative Expressions For Neutrino Oscillations in Matter,''
JHEP \textbf{06} (2016), 051
doi:10.1007/JHEP06(2016)051
[arXiv:1604.08167 [hep-ph]].

\bibitem{Martinez-Soler:2019nhb}
I.~Martinez-Soler and H.~Minakata,
``Perturbing Neutrino Oscillations Around the Solar Resonance,''
PTEP \textbf{2019} (2019) no.7, 073B07
doi:10.1093/ptep/ptz067
[arXiv:1904.07853 [hep-ph]]. 

\bibitem{Minakata:2022pyr}
H.~Minakata,
``Comparative Study of the 1\textendash{}2 Exchange Symmetries in Neutrino Frameworks with Global and Local Validities,''
Acta Phys. Polon. B \textbf{54} (2023) no.4, 3
doi:10.5506/APhysPolB.54.4-A3
[arXiv:2212.06320 [hep-ph]]. 

\bibitem{Minakata:2015gra}
H.~Minakata and S.~J.~Parke,
``Simple and Compact Expressions for Neutrino Oscillation Probabilities in Matter,''
JHEP \textbf{01} (2016), 180
doi:10.1007/JHEP01(2016)180
[arXiv:1505.01826 [hep-ph]]. 

\bibitem{Wolfenstein:1977ue}
L.~Wolfenstein,
``Neutrino Oscillations in Matter,''
Phys. Rev. D \textbf{17} (1978), 2369-2374
doi:10.1103/PhysRevD.17.2369 

\bibitem{Parke:2018shx}
S.~Parke,
``Theoretical Aspects of the Quantum Neutrino,''
doi:10.1142/9789811207402\_0008
[arXiv:1801.09643 [hep-ph]]. 

\bibitem{Antusch:2006vwa}
S.~Antusch, C.~Biggio, E.~Fernandez-Martinez, M.~B.~Gavela and J.~Lopez-Pavon,
``Unitarity of the Leptonic Mixing Matrix,''
JHEP \textbf{10} (2006), 084
doi:10.1088/1126-6708/2006/10/084
[arXiv:hep-ph/0607020 [hep-ph]].

\bibitem{Escrihuela:2015wra}
F.~J.~Escrihuela, D.~V.~Forero, O.~G.~Miranda, M.~Tortola and J.~W.~F.~Valle,
``On the description of nonunitary neutrino mixing,''
Phys. Rev. D \textbf{92} (2015) no.5, 053009
[erratum: Phys. Rev. D \textbf{93} (2016) no.11, 119905]
doi:10.1103/PhysRevD.92.053009
[arXiv:1503.08879 [hep-ph]].

\bibitem{Blennow:2016jkn}
M.~Blennow, P.~Coloma, E.~Fernandez-Martinez, J.~Hernandez-Garcia and J.~Lopez-Pavon,
``Non-Unitarity, sterile neutrinos, and Non-Standard neutrino Interactions,''
JHEP \textbf{04} (2017), 153
doi:10.1007/JHEP04(2017)153
[arXiv:1609.08637 [hep-ph]]. 

\bibitem{Fong:2016yyh}
C.~S.~Fong, H.~Minakata and H.~Nunokawa,
``A framework for testing leptonic unitarity by neutrino oscillation experiments,''
JHEP \textbf{02} (2017), 114
doi:10.1007/JHEP02(2017)114
[arXiv:1609.08623 [hep-ph]].

\bibitem{Fong:2017gke}
C.~S.~Fong, H.~Minakata and H.~Nunokawa,
``Non-unitary evolution of neutrinos in matter and the leptonic unitarity test,''
JHEP \textbf{02} (2019), 015
doi:10.1007/JHEP02(2019)015
[arXiv:1712.02798 [hep-ph]]. 

\bibitem{Minakata:2021nii}
H.~Minakata,
``Toward diagnosing neutrino non-unitarity through CP phase correlations,''
PTEP \textbf{2022} (2022) no.6, 063B03
doi:10.1093/ptep/ptac078
[arXiv:2112.06178 [hep-ph]]. 

\bibitem{Agarwalla:2013tza}
S.~K.~Agarwalla, Y.~Kao and T.~Takeuchi,
``Analytical approximation of the neutrino oscillation matter effects at large $\theta_{13}$,''
JHEP \textbf{04} (2014), 047
doi:10.1007/JHEP04(2014)047
[arXiv:1302.6773 [hep-ph]]. 


\bibitem{Parke:2019vbs}
G.~Barenboim, P.~B.~Denton, S.~J.~Parke and C.~A.~Ternes,
``Neutrino Oscillation Probabilities through the Looking Glass,''
  Phys.\ Lett.\ B {\bf 791} (2019) 351
  doi:10.1016/j.physletb.2019.03.002
  [arXiv:1902.00517 [hep-ph]]. 

\bibitem{Minakata:2020oxb}
H.~Minakata,
``Neutrino amplitude decomposition in matter,''
Phys. Rev. D \textbf{103} (2021) no.5, 053004
doi:10.1103/PhysRevD.103.053004
[arXiv:2011.08415 [hep-ph]]. 

\bibitem{Parke:2019jyu}
S.~J.~Parke and X.~Zhang,
``Compact Perturbative Expressions for Oscillations with Sterile Neutrinos in Matter,''
Phys. Rev. D \textbf{101} (2020) no.5, 056005
doi:10.1103/PhysRevD.101.056005
[arXiv:1905.01356 [hep-ph]].

\bibitem{Martinez-Soler:2018lcy}
I.~Martinez-Soler and H.~Minakata,
``Standard versus Non-Standard CP Phases in Neutrino Oscillation in Matter with Non-Unitarity,''
PTEP \textbf{2020} (2020) no.6, 063B01
doi:10.1093/ptep/ptaa062
[arXiv:1806.10152 [hep-ph]].


\bibitem{Fogli:1996nn}
G.~L.~Fogli, E.~Lisi, D.~Montanino and G.~Scioscia,
``Three flavor atmospheric neutrino anomaly,''
Phys. Rev. D \textbf{55} (1997), 4385-4404
doi:10.1103/PhysRevD.55.4385
[arXiv:hep-ph/9607251 [hep-ph]].

\bibitem{deGouvea:2000pqg}
A.~de Gouvea, A.~Friedland and H.~Murayama,
``The Dark side of the solar neutrino parameter space,''
Phys. Lett. B \textbf{490} (2000), 125-130
doi:10.1016/S0370-2693(00)00989-8
[arXiv:hep-ph/0002064 [hep-ph]].

\bibitem{Fogli:2001wi}
G.~L.~Fogli, E.~Lisi and A.~Palazzo,
``Quasi energy independent solar neutrino transitions,''
Phys. Rev. D \textbf{65} (2002), 073019
doi:10.1103/PhysRevD.65.073019
[arXiv:hep-ph/0105080 [hep-ph]].

\bibitem{Altarelli:2010gt}
G.~Altarelli and F.~Feruglio,
``Discrete Flavor Symmetries and Models of Neutrino Mixing,''
Rev. Mod. Phys. \textbf{82} (2010), 2701-2729
doi:10.1103/RevModPhys.82.2701
[arXiv:1002.0211 [hep-ph]].

\bibitem{Fogli:1996pv}
G.~L.~Fogli and E.~Lisi,
``Tests of three flavor mixing in long baseline neutrino oscillation experiments,''
Phys. Rev. D \textbf{54} (1996), 3667-3670
doi:10.1103/PhysRevD.54.3667
[arXiv:hep-ph/9604415 [hep-ph]].

\bibitem{Minakata:2001qm}
H.~Minakata and H.~Nunokawa,
``Exploring neutrino mixing with low-energy superbeams,''
JHEP \textbf{10} (2001), 001
doi:10.1088/1126-6708/2001/10/001
[arXiv:hep-ph/0108085 [hep-ph]].

\bibitem{Minakata:2010zn}
H.~Minakata and S.~Uchinami,
``Parameter Degeneracy in Neutrino Oscillation -- Solution Network and Structural Overview --,''
JHEP \textbf{04} (2010), 111
doi:10.1007/JHEP04(2010)111
[arXiv:1001.4219 [hep-ph]].

\bibitem{Coloma:2016gei}
P.~Coloma and T.~Schwetz,
``Generalized mass ordering degeneracy in neutrino oscillation experiments,''
Phys. Rev. D \textbf{94} (2016) no.5, 055005
[erratum: Phys. Rev. D \textbf{95} (2017) no.7, 079903]
doi:10.1103/PhysRevD.94.055005
[arXiv:1604.05772 [hep-ph]]. 

\bibitem{Gluza:2001de}
J.~Gluza and M.~Zralek,
``Parameters' domain in three flavor neutrino oscillations,''
Phys. Lett. B \textbf{517} (2001), 158-166
doi:10.1016/S0370-2693(01)00962-5
[arXiv:hep-ph/0106283 [hep-ph]].

\bibitem{deGouvea:2008nm}
A.~de Gouvea and J.~Jenkins,
``The Physical Range of Majorana Neutrino Mixing Parameters,''
Phys. Rev. D \textbf{78} (2008), 053003
doi:10.1103/PhysRevD.78.053003
[arXiv:0804.3627 [hep-ph]].

\bibitem{Zhou:2016luk}
S.~Zhou,
``Symmetric formulation of neutrino oscillations in matter and its intrinsic connection to renormalization-group equations,''
J. Phys. G \textbf{44} (2017) no.4, 044006
doi:10.1088/1361-6471/aa5fd9
[arXiv:1612.03537 [hep-ph]]. 

\bibitem{Minkowski:1977sc}
P.~Minkowski,
``$\mu \to e\gamma$ at a Rate of One Out of $10^{9}$ Muon Decays?,''
Phys. Lett. B \textbf{67} (1977), 421-428
doi:10.1016/0370-2693(77)90435-X

\bibitem{Yanagida:1979as}
T.~Yanagida,
``Horizontal gauge symmetry and masses of neutrinos,''
Conf. Proc. C \textbf{7902131} (1979), 95-99
KEK-79-18-95.

\bibitem{Gell-Mann:1979vob}
M.~Gell-Mann, P.~Ramond and R.~Slansky,
``Complex Spinors and Unified Theories,''
Conf. Proc. C \textbf{790927} (1979), 315-321
[arXiv:1306.4669 [hep-th]].

\bibitem{Glashow:1979nm}
S.~L.~Glashow,
``The Future of Elementary Particle Physics,''
NATO Sci. Ser. B \textbf{61} (1980), 687
doi:10.1007/978-1-4684-7197-7\_15

\bibitem{Mohapatra:1979ia}
R.~N.~Mohapatra and G.~Senjanovic,
``Neutrino Mass and Spontaneous Parity Nonconservation,''
Phys. Rev. Lett. \textbf{44} (1980), 912
doi:10.1103/PhysRevLett.44.912


\bibitem{Miranda:2016wdr}
O.~G.~Miranda, M.~Tortola and J.~W.~F.~Valle,
``New ambiguity in probing CP violation in neutrino oscillations,''
Phys. Rev. Lett. \textbf{117} (2016) no.6, 061804
doi:10.1103/PhysRevLett.117.061804
[arXiv:1604.05690 [hep-ph]]. 

\bibitem{Ge:2016xya}
S.~F.~Ge, P.~Pasquini, M.~Tortola and J.~W.~F.~Valle,
``Measuring the leptonic CP phase in neutrino oscillations with nonunitary mixing,''
Phys. Rev. D \textbf{95} (2017) no.3, 033005
doi:10.1103/PhysRevD.95.033005
[arXiv:1605.01670 [hep-ph]].

\bibitem{Dutta:2016vcc}
D.~Dutta and P.~Ghoshal,
``Probing CP violation with T2K, NO$\nu$A and DUNE in the presence of non-unitarity,''
JHEP \textbf{09} (2016), 110
doi:10.1007/JHEP09(2016)110
[arXiv:1607.02500 [hep-ph]].

\bibitem{Abe:2017jit}
Y.~Abe, Y.~Asano, N.~Haba and T.~Yamada,
``Heavy neutrino mixing in the T2HK, the T2HKK and an extension of the T2HK with a detector at Oki Islands,''
Eur. Phys. J. C \textbf{77} (2017) no.12, 851
doi:10.1140/epjc/s10052-017-5294-7
[arXiv:1705.03818 [hep-ph]].

\bibitem{Rout:2017udo}
J.~Rout, M.~Masud and P.~Mehta,
``Can we probe intrinsic CP and T violations and nonunitarity at long baseline accelerator experiments?,''
Phys. Rev. D \textbf{95} (2017) no.7, 075035
doi:10.1103/PhysRevD.95.075035
[arXiv:1702.02163 [hep-ph]].

\bibitem{Li:2018jgd}
Y.~F.~Li, Z.~z.~Xing and J.~y.~Zhu,
``Indirect unitarity violation entangled with matter effects in reactor antineutrino oscillations,''
Phys. Lett. B \textbf{782} (2018), 578-588
doi:10.1016/j.physletb.2018.05.079
[arXiv:1802.04964 [hep-ph]].

\bibitem{Martinez-Soler:2019noy}
I.~Martinez-Soler and H.~Minakata,
``Physics of parameter correlations around the solar-scale enhancement in neutrino theory with unitarity violation,''
PTEP \textbf{2020} (2020) no.11, 113B01
doi:10.1093/ptep/ptaa112
[arXiv:1908.04855 [hep-ph]]. 


\bibitem{Fernandez-Martinez:2007iaa}
E.~Fernandez-Martinez, M.~B.~Gavela, J.~Lopez-Pavon and O.~Yasuda,
``CP-violation from non-unitary leptonic mixing,''
Phys. Lett. B \textbf{649} (2007), 427-435
doi:10.1016/j.physletb.2007.03.069
[arXiv:hep-ph/0703098 [hep-ph]].

\bibitem{Goswami:2008mi}
S.~Goswami and T.~Ota,
``Testing non-unitarity of neutrino mixing matrices at neutrino factories,''
Phys. Rev. D \textbf{78} (2008), 033012
doi:10.1103/PhysRevD.78.033012
[arXiv:0802.1434 [hep-ph]].

\bibitem{Antusch:2009pm}
S.~Antusch, M.~Blennow, E.~Fernandez-Martinez and J.~Lopez-Pavon,
``Probing non-unitary mixing and CP-violation at a Neutrino Factory,''
Phys. Rev. D \textbf{80} (2009), 033002
doi:10.1103/PhysRevD.80.033002
[arXiv:0903.3986 [hep-ph]].

\bibitem{Antusch:2009gn}
S.~Antusch, S.~Blanchet, M.~Blennow and E.~Fernandez-Martinez,
``Non-unitary Leptonic Mixing and Leptogenesis,''
JHEP \textbf{01} (2010), 017
doi:10.1007/JHEP01(2010)017
[arXiv:0910.5957 [hep-ph]].

\bibitem{Antusch:2014woa}
S.~Antusch and O.~Fischer,
``Non-unitarity of the leptonic mixing matrix: Present bounds and future sensitivities,''
JHEP \textbf{10} (2014), 094
doi:10.1007/JHEP10(2014)094
[arXiv:1407.6607 [hep-ph]].

\bibitem{Fernandez-Martinez:2016lgt}
E.~Fernandez-Martinez, J.~Hernandez-Garcia and J.~Lopez-Pavon,
``Global constraints on heavy neutrino mixing,''
JHEP \textbf{08} (2016), 033
doi:10.1007/JHEP08(2016)033
[arXiv:1605.08774 [hep-ph]].
  
\bibitem{Parke:2015goa}
S.~Parke and M.~Ross-Lonergan,
``Unitarity and the three flavor neutrino mixing matrix,''
Phys. Rev. D \textbf{93} (2016) no.11, 113009
doi:10.1103/PhysRevD.93.113009
[arXiv:1508.05095 [hep-ph]].

\bibitem{Ellis:2020hus}
S.~A.~R.~Ellis, K.~J.~Kelly and S.~W.~Li,
``Current and Future Neutrino Oscillation Constraints on Leptonic Unitarity,''
JHEP \textbf{12} (2020), 068
doi:10.1007/JHEP12(2020)068
[arXiv:2008.01088 [hep-ph]].

\bibitem{Coloma:2021uhq}
P.~Coloma, J.~L\'opez-Pav\'on, S.~Rosauro-Alcaraz and S.~Urrea,
``New physics from oscillations at the DUNE near detector, and the role of systematic uncertainties,''
JHEP \textbf{08} (2021), 065
doi:10.1007/JHEP08(2021)065
[arXiv:2105.11466 [hep-ph]].




\bibitem{Ohlsson:2012kf}
T.~Ohlsson,
``Status of non-standard neutrino interactions,''
Rept. Prog. Phys. \textbf{76} (2013), 044201
doi:10.1088/0034-4885/76/4/044201
[arXiv:1209.2710 [hep-ph]].

\bibitem{Miranda:2015dra}
O.~G.~Miranda and H.~Nunokawa,
``Non standard neutrino interactions: current status and future prospects,''
New J. Phys. \textbf{17} (2015) no.9, 095002
doi:10.1088/1367-2630/17/9/095002
[arXiv:1505.06254 [hep-ph]].

\bibitem{Farzan:2017xzy}
Y.~Farzan and M.~Tortola,
``Neutrino oscillations and Non-Standard Interactions,''
Front. in Phys. \textbf{6} (2018), 10
doi:10.3389/fphy.2018.00010
[arXiv:1710.09360 [hep-ph]]. 

\bibitem{Proceedings:2019qno}
P.~S.~Bhupal Dev, K.~S.~Babu, P.~B.~Denton, P.~A.~N.~Machado, C.~A.~Arg\"uelles, J.~L.~Barrow, S.~S.~Chatterjee, M.~C.~Chen, A.~de Gouv\^ea and B.~Dutta, \textit{et al.}
``Neutrino Non-Standard Interactions: A Status Report,''
SciPost Phys. Proc. \textbf{2} (2019), 001
doi:10.21468/SciPostPhysProc.2.001
[arXiv:1907.00991 [hep-ph]].

\bibitem{Davidson:2003ha}
S.~Davidson, C.~Pena-Garay, N.~Rius and A.~Santamaria,
``Present and future bounds on nonstandard neutrino interactions,''
JHEP \textbf{03} (2003), 011
doi:10.1088/1126-6708/2003/03/011
[arXiv:hep-ph/0302093 [hep-ph]].

\bibitem{Antusch:2008tz}
S.~Antusch, J.~P.~Baumann and E.~Fernandez-Martinez,
``Non-Standard Neutrino Interactions with Matter from Physics Beyond the Standard Model,''
Nucl. Phys. B \textbf{810} (2009), 369-388
doi:10.1016/j.nuclphysb.2008.11.018
[arXiv:0807.1003 [hep-ph]].

\bibitem{Biggio:2009nt}
C.~Biggio, M.~Blennow and E.~Fernandez-Martinez,
``General bounds on non-standard neutrino interactions,''
JHEP \textbf{08} (2009), 090
doi:10.1088/1126-6708/2009/08/090
[arXiv:0907.0097 [hep-ph]].

\bibitem{Esteban:2018ppq}
I.~Esteban, M.~C.~Gonzalez-Garcia, M.~Maltoni, I.~Martinez-Soler and J.~Salvado,
``Updated constraints on non-standard interactions from global analysis of oscillation data,''
JHEP \textbf{08} (2018), 180
doi:10.1007/JHEP08(2018)180
[arXiv:1805.04530 [hep-ph]]. 



\bibitem{Maki:1962mu}
Z.~Maki, M.~Nakagawa and S.~Sakata,
``Remarks on the unified model of elementary particles,''
Prog. Theor. Phys. \textbf{28} (1962), 870-880
doi:10.1143/PTP.28.870

\bibitem{Minakata:1998bf}
H.~Minakata and H.~Nunokawa,
``CP violation versus matter effect in long baseline neutrino oscillation experiments,''
Phys. Rev. D \textbf{57} (1998), 4403-4417
doi:10.1103/PhysRevD.57.4403
[arXiv:hep-ph/9705208 [hep-ph]].

\bibitem{Zaglauer:1988gz}
H.~W.~Zaglauer and K.~H.~Schwarzer,
``The Mixing Angles in Matter for Three Generations of Neutrinos and the Msw Mechanism,''
Z. Phys. C \textbf{40} (1988), 273
doi:10.1007/BF01555889

\bibitem{Mikheyev:1985zog}
S.~P.~Mikheyev and A.~Y.~Smirnov,
``Resonance Amplification of Oscillations in Matter and Spectroscopy of Solar Neutrinos,''
Sov. J. Nucl. Phys. \textbf{42} (1985), 913-917

\bibitem{Barger:1980tf}
V.~D.~Barger, K.~Whisnant, S.~Pakvasa and R.~J.~N.~Phillips,
``Matter Effects on Three-Neutrino Oscillations,''
Phys. Rev. D \textbf{22} (1980), 2718
doi:10.1103/PhysRevD.22.2718 

\bibitem{Smirnov:2016xzf}
A.~Y.~Smirnov,
``Solar neutrinos: Oscillations or No-oscillations?,''
[arXiv:1609.02386 [hep-ph]].

\bibitem{Schechter:1980gr}
J.~Schechter and J.~W.~F.~Valle,
``Neutrino Masses in SU(2) x U(1) Theories,''
Phys. Rev. D \textbf{22} (1980), 2227
doi:10.1103/PhysRevD.22.2227

\bibitem{Okubo:1961jc}
S.~Okubo,
``Note on unitary symmetry in strong interactions,''
Prog. Theor. Phys. \textbf{27} (1962), 949-966
doi:10.1143/PTP.27.949

\bibitem{ParticleDataGroup:2022pth}
R.~L.~Workman \textit{et al.} [Particle Data Group],
``Review of Particle Physics,''
PTEP \textbf{2022} (2022), 083C01
doi:10.1093/ptep/ptac097

\bibitem{Kobayashi:1973fv}
M.~Kobayashi and T.~Maskawa,
``CP Violation in the Renormalizable Theory of Weak Interaction,''
Prog. Theor. Phys. \textbf{49} (1973), 652-657
doi:10.1143/PTP.49.652 

\bibitem{Nunokawa:2005nx}
H.~Nunokawa, S.~J.~Parke and R.~Zukanovich Funchal,
``Another possible way to determine the neutrino mass hierarchy,''
Phys. Rev. D \textbf{72} (2005), 013009
doi:10.1103/PhysRevD.72.013009
[arXiv:hep-ph/0503283 [hep-ph]]. 

\bibitem{Majorana:1937vz}
E.~Majorana,
``Teoria simmetrica dell\textquoteright{}elettrone e del positrone,''
Nuovo Cim. \textbf{14} (1937), 171-184
doi:10.1007/BF02961314

\bibitem{Bilenky:1980cx}
S.~M.~Bilenky, J.~Hosek and S.~T.~Petcov,
``On Oscillations of Neutrinos with Dirac and Majorana Masses,''
Phys. Lett. B \textbf{94} (1980), 495-498
doi:10.1016/0370-2693(80)90927-2

\bibitem{Doi:1980yb}
M.~Doi, T.~Kotani, H.~Nishiura, K.~Okuda and E.~Takasugi,
``CP Violation in Majorana Neutrinos,''
Phys. Lett. B \textbf{102} (1981), 323-326
doi:10.1016/0370-2693(81)90627-4

\bibitem{Avignone:2007fu}
F.~T.~Avignone, III, S.~R.~Elliott and J.~Engel,
``Double Beta Decay, Majorana Neutrinos, and Neutrino Mass,''
Rev. Mod. Phys. \textbf{80} (2008), 481-516
doi:10.1103/RevModPhys.80.481
[arXiv:0708.1033 [nucl-ex]].

\bibitem{Rodejohann:2011vc}
W.~Rodejohann and J.~W.~F.~Valle,
``Symmetrical Parametrizations of the Lepton Mixing Matrix,''
Phys. Rev. D \textbf{84} (2011), 073011
doi:10.1103/PhysRevD.84.073011
[arXiv:1108.3484 [hep-ph]].


\bibitem{Frigerio:2002rd}
M.~Frigerio and A.~Y.~Smirnov,
``Structure of neutrino mass matrix and CP violation,''
Nucl. Phys. B \textbf{640} (2002), 233-282
doi:10.1016/S0550-3213(02)00570-9
[arXiv:hep-ph/0202247 [hep-ph]].

\bibitem{Frigerio:2002fb}
M.~Frigerio and A.~Y.~Smirnov,
``Neutrino mass matrix: Inverted hierarchy and CP violation,''
Phys. Rev. D \textbf{67} (2003), 013007
doi:10.1103/PhysRevD.67.013007
[arXiv:hep-ph/0207366 [hep-ph]].

\bibitem{Bertuzzo:2013ew}
E.~Bertuzzo, P.~A.~N.~Machado and R.~Z.~Funchal,
``Neutrino Mass Matrix Textures: A Data-driven Approach,''
JHEP \textbf{06} (2013), 097
doi:10.1007/JHEP06(2013)097
[arXiv:1302.0653 [hep-ph]].

\bibitem{Jarlskog:1985ht}
C.~Jarlskog,
``Commutator of the Quark Mass Matrices in the Standard Electroweak Model and a Measure of Maximal CP Violation,''
Phys. Rev. Lett. \textbf{55} (1985), 1039
doi:10.1103/PhysRevLett.55.1039

\bibitem{Asano:2011nj}
K.~Asano and H.~Minakata,
``Large-Theta(13) Perturbation Theory of Neutrino Oscillation for Long-Baseline Experiments,''
JHEP \textbf{06} (2011), 022
doi:10.1007/JHEP06(2011)022
[arXiv:1103.4387 [hep-ph]].


\end{thebibliography}
\end{document}